\documentclass[12pt,aps,prd,a4paper]{revtex4}

\usepackage{graphicx}
\usepackage{bm}
\usepackage{amsfonts}
\usepackage{amsmath}

\usepackage[russian,ngerman,english]{babel}

\newcommand{\muas}[0]{\hbox{\rm $\mu$as}}

\newcommand{\ve}[1]{\mbox{\boldmath$#1$}}

\let\oldbibitem\bibitem
\renewcommand\bibitem[2][]{\oldbibitem{#2}}

\begin{document}

\title{Total light deflection in the gravitational field of an axisymmetric body at rest with full mass and spin multipole structure}

\author{Sven Zschocke}

\affiliation{Institute of Planetary Geodesy - Lohrmann Observatory, TUD Dresden University of Technology, 
Helmholtzstrasse 10, D-01069 Dresden, Germany}

\begin{abstract}
The tangent vector of the light trajectory at future infinity and the angle of total light deflection in the gravitational field of an isolated axisymmetric body 
at rest with full set of mass-multipoles and spin-multipoles is determined in harmonic coordinates in the 1PN and 1.5PN approximation of the post-Newtonian (PN) scheme. 
It is found that the evaluation of the tangent vector and of the angle of total light deflection caused by mass-multipoles and spin-multipoles 
leads directly and in a compelling way to Chebyshev polynomials of first and second kind, respectively. 
This fact allows to determine the upper limits of the total light deflection, which are strictly valid in the 1PN and 1.5PN approximation. 
They represent a criterion to identify those multipoles which contribute significantly to the total light deflection for a given astrometric accuracy.  
These upper limits are used to determine the total light deflection in the gravitational field of the Sun and giant planets of the solar system. It is found that the first 
few mass-multipoles with $l \le 10$ and the first few spin-multipoles with $l \le 3$ are sufficient for an accuracy on the nano-arcsecond level 
in astrometric angular measurements. 
\end{abstract}

\pacs{95.10.Jk, 95.10.Ce, 95.30.Sf, 04.25.Nx, 04.80.Cc}

\begin{center}


\end{center}

\maketitle 

\newpage

\tableofcontents 

\newpage

\section{Introduction}\label{Section0}

The precision in angular measurements of celestial objects has reached the micro-arcsecond (\muas) scale of accuracy. This level 
of precision requires a relativistic modeling of light trajectories in the curved space-time of the solar system, where not only 
the mass-monopole structure of the solar system bodies needs to be taken into account, but also the mass-quadrupole structure 
of some giant planets. In particular, the space-based astrometry mission {\it Gaia} of the European Space Agency (ESA) measures 
angular distances between stellar objects with an accuracy of about $5\,\muas$ \cite{Gaia1,Gaia2}, which allows to detect the 
effect of light deflection caused by the mass-quadrupole structure of the giant planets Jupiter and Saturn. Near future astrometry 
missions are aiming at the sub-micro-arcsecond (sub-\muas) level in angular measurements \cite{Proceeding_Sub_Microarcsecond_Astrometry,article_sub_micro_4}, 
where it becomes possible to detect not only the light deflection effect caused by higher mass-multipoles but perhaps also spin-multipoles  
of solar system bodies. In fact, there are several astrometry missions proposed to ESA, which are aiming at such unprecedented  
level of astrometric precision, like the astrometry missions {\it Gaia NIR} \cite{Gaia_NIR}, {\it Theia} \cite{Theia}, {\it Astrod} \cite{Astrod1,Astrod2},
{\it Lator} \cite{Lator1,Lator2}, {\it Odyssey} \cite{Odyssey}, {\it Sagas} \cite{Sagas}, or {\it TIPO} \cite{TIPO}.  

A precise relativistic modeling of light trajectories in the curved space-time of the solar system is required, in order to interpret such astrometric 
measurements correctly. In the flat space-time, assumed to be covered by Minkowskian four-coordinates $\left(x^0,x^1,x^2,x^3\right)$, a light signal 
which is emitted at some initial time $t_0$ by some light source located at $\ve{x}_0$ in spatial direction $\ve{\sigma}$, propagates along a straight line, 
\begin{eqnarray} 
	\ve{x}_{\rm N}\left(t\right) &=& \ve{x}_0 + c \left(t - t_0\right) \ve{\sigma}\;,
	\label{Introduction_5}
\end{eqnarray} 

\noindent
where sublabel ${\rm N}$ stands for the unperturbed light ray according to Newtonian theory, while  
according to general theory of relativity, a massive body causes space-time to curve and light is no longer propagating along a straight line, but 
along geodesics of the curved space-time. For weak gravitational fields, like in the solar system, the light trajectory is then given by 
\begin{eqnarray}
	\ve{x}\left(t\right) &=& \ve{x}_0 + c \left(t - t_0\right) \ve{\sigma} + \Delta \ve{x}\left(t\right)\;,
        \label{Introduction_10}
\end{eqnarray}

\noindent 
where $\Delta \ve{x}\left(t\right)$ denote the corrections to the unperturbed light ray. The light trajectory (\ref{Introduction_10}) is curvilinear 
and the angle of total light deflection, denoted by $\delta\left(\ve{\sigma}, \ve{\nu}\right)$, is the angle between the three-vectors $\ve{\sigma}$ and $\ve{\nu}$, 
which are the unit tangent vectors along the light trajectory at past and future infinity, respectively, defined by 
\begin{eqnarray}
        \ve{\sigma} &=& \frac{\dot{\ve{x}}\left(t\right)}{c}\,\bigg|_{t \rightarrow - \infty} \;,  
        \label{Introduction_15}
	\\
	\ve{\nu} &=& \frac{\dot{\ve{x}}\left(t\right)}{c}\,\bigg|_{t \rightarrow + \infty} \;, 
        \label{Introduction_20}
\end{eqnarray}
 
\noindent 
where a dot means differentiation with respect to coordinate time. 
The total light deflection is an upper limit for bending of light by the massive body, in the sense that if the light source or the observer are located at finite 
spatial distances from the body, then the light deflection angle would become smaller than the total light deflection angle. 

For weak gravitational fields and slow motions of matter one may apply the post-Newtonian (PN) approximation (weak-field slow-motion approximation). In the 1.5PN approximation, 
the unit tangent vector of light ray at future infinity, caused by a spherically symmetric body at rest in uniform rotational motion with period $T$, 
is given by \cite{Einstein,MTW,Poisson_Will,Kopeikin_Efroimsky_Kaplan,Klioner1991}  
\begin{eqnarray}
	\ve{\nu} &=& \ve{\sigma} + \ve{\nu}_{\rm 1PN}^{M_0} + \ve{\nu}_{\rm 1.5PN}^{S_1} + {\cal O}\left(c^{-4}\right) . 
        \label{Introduction_Tangent_Vector}
\end{eqnarray}

\noindent 
The mass-monopole term reads (e.g. Eq.~(16) in \cite{Klioner1991}) 
\begin{eqnarray}
	\ve{\nu}_{\rm 1PN}^{M_0} &=& - \frac{4 G M}{c^2 d_{\sigma}}\,\frac{\ve{d}_{\sigma}}{d_{\sigma}}\,, 
        \label{Introduction_Monopole}
\end{eqnarray}

\noindent
where $G$ is the Newtonian gravitational constant, $M$ is the mass of the body, $c$ is the speed of light, and
$\ve{d}_{\sigma}$ is the impact vector of the unperturbed light ray and its absolute value, $d_{\sigma}$,
is the impact parameter. In case of the Sun, the total light deflection (\ref{Introduction_Monopole}) amounts to $1.75\,{\rm arcsecond}$ for grazing rays, an effect which belongs
to the most famous predictions of the general theory of relativity in 1915 (cf. text below Eq.~(75) in \cite{Einstein}).

The spin-dipole term reads (Eq.~(60) in \cite{Klioner1991} or Eq.~(72) in \cite{Kopeikin_Mashhoon})
\begin{eqnarray}
	\ve{\nu}_{\rm 1.5PN}^{S_1} &=& - \frac{4 G}{c^3\left(d_{\sigma}\right)^2} 
        \bigg[2\,\frac{\left(\ve{\sigma}\times\ve{d}_{\sigma}\right) \cdot \ve{S}}{d_{\sigma}}\,\frac{\ve{d}_{\sigma}}{d_{\sigma}} +
        \left(\ve{\sigma} \times \ve{S}\right)\bigg],  
        \label{Introduction_Spin}
\end{eqnarray}

\noindent 
where the intrinsic angular momentum in (\ref{Introduction_Spin}) is given by 
\begin{eqnarray}
        \ve{S} &=& \kappa^2\,M\,\Omega\,P^2\,\ve{e}_3\;,
        \label{Introduction_Spin_Dipole}
\end{eqnarray}

\noindent
where $\kappa^2$ is the dimensionless moment of inertia, $\Omega = 2 \pi / T$ is the angular velocity of the body, $P$ its equatorial radius, and 
$\ve{e}_3$ is the unit vector along the rotational axis. 
The effect of total light deflection caused by the rotational motion (\ref{Introduction_Spin}) in case of the Sun amounts to $0.7\,{\rm micro}$-${\rm arcsecond}$ for grazing rays, 
an effect which becomes detectable at the sub-micro-arcsecond level of astrometric precision.

The expression in (\ref{Introduction_Tangent_Vector}) with (\ref{Introduction_Monopole}) and (\ref{Introduction_Spin}) determines the total light deflection 
in the gravitational field of a spherically symmetric body in uniform rotation. In reality, however, massive bodies are not spherically symmetric, but can be of 
arbitrary shape, inner structure, oscillations, and rotational motions of inner currents. The gravitational fields of such gravitational systems are determined 
with the aid of multipole expansion, which is a series expansion of the metric tensor in terms of mass-multipoles $M_L$ (mass-monopole, mass-quadrupole, mass-octupole, etc.) 
and spin-multipoles $S_L$ (spin-dipole, spin-hexapole, spin-decapole, etc.) \cite{MTW,Poisson_Will,Kopeikin_Efroimsky_Kaplan}. The multipole expansion of the gravitational fields implies 
a corresponding multipole expansion of the exact light trajectory in (\ref{Introduction_10}). Accordingly, the expressions (\ref{Introduction_Monopole}) and (\ref{Introduction_Spin}) 
represent only the first terms of an infinite multipole series of the unit tangent vector $\ve{\nu}$ in (\ref{Introduction_20}) (e.g. Eq.~(11) in \cite{Klioner1991}),
\begin{eqnarray}
        \ve{\nu} &=& \ve{\sigma} + \sum\limits_{l=0}^{\infty} \ve{\nu}_{\rm 1PN}^{M_L} + \sum\limits_{l=1}^{\infty} \ve{\nu}_{\rm 1.5PN}^{S_L} + {\cal O}\left(c^{-4}\right),  
        \label{Introduction_nu}
\end{eqnarray}

\noindent 
as well as of the angle of total light deflection, $\delta\left(\ve{\sigma},\ve{\nu}\right) = \arcsin |\ve{\sigma} \times \ve{\nu}|$,  
\footnote{Let us notice here, that these terms in (\ref{Introduction_Multipoles}) can be positive or negative, because they are not angles,
but individual terms which contribute to the angle of total light deflection, which is a positive real number per definition.},  
\begin{eqnarray}
	\delta\left(\ve{\sigma},\ve{\nu}\right) &=& 
	\sum\limits_{l=0}^{\infty} \delta\!\left(\!\ve{\sigma},\ve{\nu}_{\rm 1PN}^{M_L}\!\right) 
	+ \sum\limits_{l=1}^{\infty} \delta\!\left(\!\ve{\sigma},\ve{\nu}_{\rm 1.5PN}^{S_L}\!\right) + {\cal O}\left(c^{-4}\right),
	\label{Introduction_Multipoles}
\end{eqnarray}

\noindent  
where the terms of second post-Newtonian order (2PN) have been neglected in (\ref{Introduction_nu}) and (\ref{Introduction_Multipoles}); see also the final comment in the 
summary Section below Eq.~(\ref{Summary_S_L_Grazing}). The 1PN and 1.5PN expressions for the unit tangent vector in (\ref{Introduction_nu}) and for the angle of 
total light deflection (\ref{Introduction_Multipoles}) have been derived for the first time in \cite{Kopeikin1997} for the most general case of one body at rest 
with full mass-multipole and spin-multipole structure. 

In order to determine the unit tangent vector in (\ref{Introduction_nu}) as well as the angle of total light deflection in (\ref{Introduction_Multipoles}) and their upper limits 
one may consider the model of an axisymmetric body in uniform rotational motion around its axis of symmetry. In this investigation it has been found, that the calculation of both 
these observables in (\ref{Introduction_nu}) and (\ref{Introduction_Multipoles}) leads directly to Chebyshev polynomials of first and second kind, 
$T_l$ and $U_l$ \cite{Arfken_Weber,Abramowitz_Stegun}. This remarkable feature allows two important things: (a) it allows to derive the expressions for the 
unit tangent vector at future infinity in a straightforward manner and (b) it allows for a straightforward determination of the upper limit of the 
angle of total light deflection, because the upper limits of Chebyshev polynomials are given by 
\begin{eqnarray}
	\left|T_l\right| \le 1 \quad {\rm and} \quad \left|U_{l-1}\right| \le l \;.
	\label{Introduction_Chebyshev_polynomials_2}
\end{eqnarray}

\noindent 
The angle of total light deflection represents the upper limit for the bending of light by some massive body. Therefore, the upper limits derived in this investigation allow for a 
straightforward determination of the maximal effect of light deflection for realistic solar system bodies. Furthermore, they can be reached by realistic astrometric configurations. 
In this sense they are just {\it the upper limits}.

The manuscript is organized as follows: 
In Section~\ref{Section1} the multipole decomposition of the metric for an isolated body as well as the geodesic equation for light rays is given. Furthermore, the approach 
presented in \cite{Kopeikin1997} about how to derive the light trajectory in 1PN and 1.5PN approximation is considered. The formal expressions for the total light deflection 
in 1PN and 1.5PN approximation are given in Section~\ref{Section2}. The mass-multipoles and spin-multipoles for an axisymmetric body in uniform rotational motion are 
presented in Section~\ref{Section3}. In Section~\ref{Section4} the total light deflection caused by mass-multipoles and spin-multipoles of an axisymmetric body are calculated and 
it is demonstrated that the total light deflection is given by Chebyshev polynomials. Furthermore, their upper limits are determined. 
Numerical results for the total light deflection at the Sun and the giant planets of the solar system are shown in Section~\ref{Section_Light_Deflection}. 
A summary is given in Section~\ref{Summary}. The notations, conventions and details of the calculations are shifted into a set of several appendices.

\section{Metric tensor and geodesic equation}\label{Section1} 

\subsection{Metric tensor and geodesic equation}

Let the space-time be covered by harmonic four-coordinates, $x^{\mu} = \left(x^0,x^1,x^2,x^3\right)$, where $x^0 = ct$ is the speed of light times coordinate time, 
while $x^1,x^2,x^3$ are the spatial coordinates, and the origin of spatial axes is assumed to be located at the center-of-mass of the body. 
The metric tensor of the curved space-time in the exterior of the massive body is assumed to be time-independent. In the post-Newtonian scheme the metric tensor can be 
series-expanded in terms of inverse powers of the speed of light, which in the time-independent case is given by \cite{MTW,Poisson_Will,Kopeikin_Efroimsky_Kaplan}, 
\begin{eqnarray}
        g_{\alpha\beta}\left(\ve{x}\right) &=& \eta_{\alpha\beta} + h_{\alpha\beta}^{\left(2\right)}\left(\ve{x}\right)  
        + h_{\alpha\beta}^{\left(3\right)}\left(\ve{x}\right) + {\cal O}(c^{-4}),   
        \label{PN_Expansion_1PN_15PN}
\end{eqnarray}

\noindent
where the non-vanishing metric perturbations, in canonical gauge, are \cite{Thorne,Blanchet_Damour1,Multipole_Damour_2,Kopeikin_Efroimsky_Kaplan} 
\footnote{The metric perturbations in Eqs.~(\ref{Metric_00}) - (\ref{Metric_ij}) can be obtained from Eqs.~(5.32.a) - (5.32.c) in \cite{Multipole_Damour_2} by means 
of the relation between metric perturbation and metric density perturbation as given, for instance, by relation (3.5.10) in \cite{Kopeikin_Efroimsky_Kaplan}. 
One may also consider the post-Newtonian expansion of Eq.~(2) in \cite{Zschocke_Soffel} in the stationary case.}
\begin{eqnarray}
        h_{00}^{\left(2\right)}\left(\ve{x}\right) &=& \frac{2}{c^2} \sum\limits_{l=0}^{\infty} \frac{\left(-1\right)^l}{l!}\,\hat{M}_L\, 
        \hat{\partial}_L \frac{1}{r}\;,
        \label{Metric_00}
        \\
        h_{0i}^{\left(3\right)}\left(\ve{x}\right) &=& \frac{4}{c^3} \sum\limits_{l=1}^{\infty} \frac{\left(-1\right)^l\,l}{\left(l+1\right)!} \, 
        \epsilon_{iab}\,\hat{S}_{b L-1}\, \hat{\partial}_{a L-1} \frac{1}{r}\;,  
        \label{Metric_0i}
        \\
        h_{ij}^{\left(2\right)}\left(\ve{x}\right) &=& \frac{2}{c^2}\,\delta_{ij}  
        \sum\limits_{l=0}^{\infty} \frac{\left(-1\right)^l}{l!}\,\hat{M}_L\,\hat{\partial}_L \frac{1}{r}\;, 
        \label{Metric_ij}
        \end{eqnarray}

\noindent
where $r = \left|\ve{x}\right|$ and 
\begin{eqnarray}
	\hat{\partial}_L &=& {\rm STF}_{i_1 \dots i_l}\,\frac{\partial}{\partial x^{i_1}} \dots \frac{\partial}{\partial x^{i_l}}\;,  
        \label{Partial_Derivatives}
        \end{eqnarray}

\noindent 
where the hat in $\hat{\partial}_L$ indicates the symmetric tracefree (STF) operation with respect to the indices $L = i_1 \dots i_l$, 
which makes a Cartesian tensor symmetric and tracefree with respect to all spatial indices; 
for details see Appendix~\ref{Appendix_STF}. The mass-multipoles and spin-multipoles for a stationary source of matter are given by 
(cf. Eqs.~(5.38) and (5.41) in \cite{Multipole_Damour_2})
\begin{eqnarray}
	\hat{M}_L &=& \int d^3 x \; \hat{x}_L\;\Sigma + {\cal O}\left(c^{-4}\right), 
\label{Mass_Multipoles}
\\
        \hat{S}_L &=& \int d^3 x \;\epsilon_{j k < i_l}\,\hat{x}_{L-1 >}\;x^j\;\Sigma^k + {\cal O}\left(c^{-4}\right), 
\label{Spin_Multipoles}
\end{eqnarray}

\noindent
where the integration runs over the volume of the body and the notation 
$\Sigma = \left(T^{00} + T^{kk}\right)/c^2$ and $\Sigma^k = T^{0k}/c$ has been adopted, with $T^{\alpha\beta}$ being the stress-energy tensor 
of the body. 

In flat (Minkowskian) space-time, a light signal which is emitted at some initial time, $t_0$, into some three-direction, $\ve{\sigma}$, propagates along a straight trajectory,  
\begin{eqnarray}
        \ve{x}_{\rm N} &=& \ve{x}_0 + c \left(t - t_0\right) \ve{\sigma}\;. 
        \label{Unperturbed_Lightray_2}
\end{eqnarray}

\noindent
In curvilinear space-time the light trajectory is determined by the geodesic equation, which in 1.5PN approximation reads \cite{Brumberg1991,KopeikinSchaefer1999_Gwinn_Eubanks} 
\begin{eqnarray}
\frac{\ddot{x}^i \left(t\right)}{c^2} &=& \frac{\partial h_{00}^{(2)}}{\partial x^i} 
- 2\,\frac{\partial h_{00}^{(2)}}{\partial x^j}\,\sigma^i \sigma^j 
- \frac{\partial h_{0i}^{(3)}}{\partial x^j}\,\sigma^j 
+ \frac{\partial h_{0j}^{(3)}}{\partial x^i}\,\sigma^j 
- \frac{\partial h_{0j}^{(3)}}{\partial x^k}\,\sigma^i \sigma^j \sigma^k
\label{Geodesic_Equation_15PN}
\end{eqnarray}

\noindent
up to terms of the order ${\cal O}\left(c^{-4}\right)$. 
The double-dot on the left-hand side in (\ref{Geodesic_Equation_15PN}) means twice of the total differential with respect to the coordinate time, and 
$h_{ij}^{\left(2\right)} = h_{00}^{\left(2\right)}\,\delta_{ij}$ has been taken into account. The geodesic equation (\ref{Geodesic_Equation_15PN}) is a differential 
equation of second order. Thus, a unique solution of (\ref{Geodesic_Equation_15PN}) requires initial-boundary conditions 
\cite{Brumberg1991,Kopeikin1997,KopeikinSchaefer1999_Gwinn_Eubanks,Klioner1991,KlionerKopeikin1992,Zschocke_1PN,Zschocke_15PN}: the spatial position of light source $\ve{x}_0$ 
and the unit-direction $\ve{\sigma}$ of light ray at past infinity,
\begin{eqnarray} 
        \ve{x}_0 \; &=& \; \ve{x}\left(t\right) \bigg|_{t = t_0}\;,
        \label{Initial_B}
        \\
        \ve{\sigma} &=& \frac{\dot{\ve{x}}\left(t\right)}{c}\,\bigg|_{t \rightarrow - \infty} \;,  
        \label{vector_sigma}
\end{eqnarray}

\noindent
with $\ve{\sigma} \cdot \ve{\sigma} = 1$. The geodesic equation (\ref{Geodesic_Equation_15PN}) can be solved by iteration, and the solution of first and second integration  
reads formally 
\begin{eqnarray}
	\frac{\dot{\ve{x}}\left(t\right)}{c} &=& \ve{\sigma} 
	+ \sum\limits_{l=0}^{\infty} \frac{\Delta \dot{\ve{x}}^{M_L}_{\rm 1PN}\left(t\right)}{c} 
        + \sum\limits_{l=1}^{\infty} \frac{\Delta \dot{\ve{x}}^{S_L}_{\rm 1.5PN}\left(t\right)}{c}\,,  
        \label{First_Interation_1PN}
        \\
	\ve{x}\left(t\right) &=& \ve{x}_{\rm N}\left(t\right) + \sum\limits_{l=0}^{\infty} \Delta\ve{x}^{M_L}_{\rm 1PN}\left(t, t_0\right) 
        + \sum\limits_{l=1}^{\infty} \Delta\ve{x}^{S_L}_{\rm 1.5PN}\left(t, t_0\right),  
        \label{Second_Interation_1PN}
\end{eqnarray}

\noindent 
up to terms of the order ${\cal O}\left(c^{-4}\right)$. 
The metric (\ref{PN_Expansion_1PN_15PN}) is valid in the entire space, while in the geodesic equation (\ref{Geodesic_Equation_15PN}) the metric components have 
to be taken at the concrete spatial position of the unperturbed light ray (\ref{Unperturbed_Lightray_2}) at coordinate time $t$. This fact implies that one has, 
first of all, to perform the differentiations in the metric components (\ref{Metric_00}) - (\ref{Metric_ij}) as well as in the geodesic equation 
(\ref{Geodesic_Equation_15PN}), and afterwards one may replace the spatial variable $\ve{x}$ by the unperturbed light ray $\ve{x}_{\rm N}\left(t\right)$. 
Then, in the very final step, one may perform the integration of geodesic equation. However, that standard procedure leads to cumbersome expressions already 
for the very first few multipoles. Therefore, advanced integration methods have been developed in \cite{Kopeikin1997}, which allows one to integrate 
(\ref{Geodesic_Equation_15PN}) exactly and which will be considered in what follows.

\subsection{Metric tensor and geodesic equation in terms of new variables $\left(c\tau, \ve{\xi}\right)$}

In the approach in \cite{Kopeikin1997} advanced integration methods were introduced, basing on the new parameters,
\begin{eqnarray}
        c \tau &=& \ve{\sigma} \cdot \ve{x}_{\rm N}\;,
        \label{Parameter1}
        \\
        \xi^i &=& P^i_j\,x_{\rm N}^j\;,  
        \label{Parameter2}
\end{eqnarray}

\noindent
where the unperturbed light ray is given by Eq.~(\ref{Unperturbed_Lightray_2}) and the operator 
\begin{eqnarray}
        P^{ij} &=& \delta^{ij} - \sigma^i \sigma^j
        \label{Projection_Operator}
\end{eqnarray}

\noindent
is a projection operator onto the plane perpendicular to vector $\ve{\sigma}$. Clearly, by inserting the unperturbed light ray (\ref{Unperturbed_Lightray_2}) and the projector 
(\ref{Projection_Operator}) into (\ref{Parameter2}), one may identify the auxiliary variable $\ve{\xi}$ as impact vector $\ve{d}_{\sigma}$ of the unperturbed light ray, which will  
later been introduced by Eq.~(\ref{impact_vector}). Here, we prefer to keep the notation of \cite{Kopeikin1997} and shall distinguish between these both three-vectors. 
Like the spatial and time variable, $\ve{x}$ and $t$, also these new variables, $\ve{\xi}$ and $c \tau$, are independent of each 
other. The unperturbed light ray $\ve{x}_{\rm N}$ and its absolute value $r_{\rm N} = \left|\ve{x}_{\rm N}\right|$ can be parameterized in terms of these new variables and take the form 
\begin{eqnarray}
	\ve{x}_{\rm N} &=& \ve{\xi} + c \tau\,\ve{\sigma}\;, 
	\label{Parameter3}
	\\
	r_{\rm N} &=& \sqrt{\xi^2 + c^2 \tau^2}\;, 
	\label{Parameter4}
\end{eqnarray}

\noindent 
where in (\ref{Parameter4}) it has been used that the three-vector $\ve{\xi}$ is laying in the two-dimensional space perpendicular to the three-vector $\ve{\sigma}$.  
Therefore, that vector has only two independent components and the partial derivatives in this plane are given by 
(cf. Eq.~(23) in \cite{Kopeikin1997}, see also Eq.~(11.2.12) in \cite{Book_Soffel_Han})  
\begin{eqnarray}
        \frac{\partial \xi^i}{\partial \xi^j} &=& P^i_j = P^{ij} = P_{ij}\;.
\label{Differentiation_5}
\end{eqnarray}

\noindent
Then, the spatial derivatives in (\ref{Metric_00}) - (\ref{Metric_ij}), when transformed in terms of these new variables, 
are given by (cf. Eq.~(20) in \cite{Kopeikin1997})
\begin{eqnarray}
        \frac{\partial}{\partial x^i} &=& \frac{\partial}{\partial \xi^i} + \sigma_i\,\frac{\partial}{\partial c \tau}\;, 
        \label{spatial_derivative_1}
\end{eqnarray}

\noindent
where $\ve{\xi}$ means a three-vector in the two-dimensional plane perpendicular to $\ve{\sigma}$. In practical calculations it is, however, 
often convenient to treat the spatial components of this vector $\ve{\xi}$ as formally independent and, 
therefore, a subsequent projection onto this two-dimensional plane by means of $P^{ij}$ is necessary 
(cf. text above Eq.~(31) in \cite{KopeikinSchaefer1999_Gwinn_Eubanks}). Accordingly, one gets the familiar result (see also Eq.~(11.2.13) in \cite{Book_Soffel_Han}) 
\begin{eqnarray}
        \frac{\partial \xi^i}{\partial \xi^j} &=& \delta^i_j = \delta^{ij} = \delta_{ij}\;, 
\label{Differentiation_10}
\end{eqnarray}

\noindent
with subsequent projection into the two-dimensional plane perpendicular to three-vector $\ve{\sigma}$, 
\begin{eqnarray}
        P^k_j\,\frac{\partial \xi^i}{\partial \xi^k} &=& P^i_j = P^{ij} = P_{ij}\;. 
\label{Differentiation_15}
\end{eqnarray}

\noindent 
That means, for the spatial derivatives, when transformed into derivatives expressed in terms of these new variables, one obtains 
\begin{eqnarray}
        \frac{\partial}{\partial x^i} &=& P_i^j\,\frac{\partial}{\partial \xi^j} + \sigma_i\,\frac{\partial}{\partial c \tau}\;. 
        \label{spatial_derivative_2}
\end{eqnarray}

\noindent
This relation coincides with Eq.~(33) in \cite{KopeikinSchaefer1999_Gwinn_Eubanks} in case of time-independent functions. 
It is notice that (\ref{spatial_derivative_2}) in combination with (\ref{Differentiation_10}) is identical with 
(\ref{spatial_derivative_1}) in combination with (\ref{Differentiation_5}). 

Here, we will prefer this procedure in (\ref{spatial_derivative_2}), that means we will consider the spatial components of $\ve{\xi}$ as three independent components 
of the three-vector $\ve{\xi}$ with a subsequent projection into the two-dimensional plane perpendicular to the three-vector $\ve{\sigma}$. Then, using (\ref{spatial_derivative_2}) 
and the binomial theorem \cite{Arfken_Weber,Abramowitz_Stegun}, 
\begin{eqnarray}
	\left(a + b\right)^l &=& \sum\limits_{p=0}^l {l \choose p} a^{l-p}\,b^{p}\;,  
        \label{binomial_theorem_1}
\end{eqnarray}

\noindent
where 
\begin{eqnarray}
        {l \choose p} &=& \frac{l!}{p! \left(l - p\right)!}
        \label{binomial_coefficients}
\end{eqnarray}

\noindent
are the binomial coefficients, 
one finds for $l$ partial derivatives (\ref{Partial_Derivatives}) expressed in terms of these new variables 
\begin{eqnarray} 
\hat{\partial}_{L} &=& {\rm STF}_{i_1 \dots i_l}\;\sum\limits_{p=0}^{l} \frac{l!}{p! \left(l-p\right)!}\, 
\sigma_{i_1}\,...\,\sigma_{i_p}\, 
P_{i_{p+1}}^{j_{p+1}}\, \dots \,P_{i_l}^{j_l}\, 
\frac{\partial}{\partial \xi^{j_{p+1}}}\, \dots \, 
\frac{\partial}{\partial \xi^{j_{l}}}\,\left(\frac{\partial}{\partial c\tau}\right)^p \;,  
\label{Transformation_Derivative_3}
\end{eqnarray}

\noindent 
where the hat in $\hat{\partial}_L$ indicates the STF operation with respect to the indices $L = i_1 \dots i_l$ 
\footnote{The differential operator in Eq.~(\ref{Transformation_Derivative_3}) is just the STF version of expression (101) in \cite{Zschocke_1PN}. It also coincides 
with the summand $q=0$ in Eq.~(94) in \cite{Zschocke_15PN}. The same differential operator is given by Eq.~(11.2.23) in \cite{Book_Soffel_Han}. Let us also notice 
that the differential operator in Eq.~(\ref{Transformation_Derivative_3}) resembles Eq.~(24) in \cite{Kopeikin1997}, where the only difference is just the projector 
as described in the main text between Eqs.~(\ref{Projection_Operator}) and (\ref{Transformation_Derivative_3}).}.
The metric tensor (\ref{PN_Expansion_1PN_15PN}) with (\ref{Metric_00}) - (\ref{Metric_ij}) in terms of these new parameters $\left(c\tau, \ve{\xi}\right)$ reads
\begin{eqnarray}
        g_{\alpha\beta}\left(c\tau, \ve{\xi}\right) &=& \eta_{\alpha\beta} + h_{\alpha\beta}^{\left(2\right)}\left(c\tau, \ve{\xi}\right)  
        + h_{\alpha\beta}^{\left(3\right)}\left(c\tau, \ve{\xi}\right) + {\cal O}(c^{-4}), 
        \label{PN_Expansion_1PN_15PN_New}
\end{eqnarray}

\noindent
where non-vanishing metric perturbations in terms of these new variables are given by \cite{Kopeikin1997}
\footnote{One may obtain Eqs.~(\ref{Metric_00_New}) - (\ref{Metric_ij_New}) from Eqs.~(69) - (72) in \cite{Zschocke_15PN} for a massive body at rest in stationary case.}
\begin{eqnarray}
        h_{00}^{\left(2\right)}\left(c\tau, \ve{\xi}\right) &=& \frac{2}{c^2} \sum\limits_{l=0}^{\infty} \frac{\left(-1\right)^l}{l!}\,\hat{M}_L\, 
        \hat{\partial}_L \frac{1}{r_{\rm N}}\;,
        \label{Metric_00_New}
        \\
        h_{0i}^{\left(3\right)}\left(c\tau, \ve{\xi}\right) &=& \frac{4}{c^3} \sum\limits_{l=1}^{\infty} \frac{\left(-1\right)^l\,l}{\left(l+1\right)!} \, 
        \epsilon_{iab}\,\hat{S}_{b L-1}\, \hat{\partial}_{a L-1} \frac{1}{r_{\rm N}}\;,  
        \label{Metric_0i_New}
        \\
        h_{ij}^{\left(2\right)}\left(c\tau, \ve{\xi}\right) &=& \frac{2}{c^2}\,\delta_{ij}  
        \sum\limits_{l=0}^{\infty} \frac{\left(-1\right)^l}{l!}\,\hat{M}_L\,\hat{\partial}_L \frac{1}{r_{\rm N}}\;.   
        \label{Metric_ij_New}
        \end{eqnarray}

\noindent
Accordingly, the metric tensor (\ref{PN_Expansion_1PN_15PN_New}) depends both on spatial variable $\ve{\xi}$ as well as on time-variable $c\tau$, besides 
that the metric describes the stationary curved space-time caused by the massive body at rest with time-independent mass-multipoles and time-independent 
spin-multipoles. The geodesic equation (\ref{Geodesic_Equation_15PN}) expressed in terms of these new variables $c \tau$ and $\ve{\xi}$ reads \cite{KopeikinSchaefer1999_Gwinn_Eubanks} 
\begin{eqnarray}
\frac{\ddot{x}^i \left(\tau\right)}{c^2} =
P^{ij} \frac{\partial h_{00}^{(2)}}{\partial \xi^j}
- \frac{\partial h_{00}^{(2)}}{\partial c \tau}\,\sigma^i 
- \frac{\partial h_{0i}^{(3)}}{\partial c \tau} 
+ P^{ik} \frac{\partial h_{0j}^{(3)}}{\partial \xi^k}\,\sigma^j 
\label{Geodesic_Equation_15PN_New}
\end{eqnarray}

\noindent
up to terms of the order ${\cal O}\left(c^{-4}\right)$. 
The double-dot on the left-hand side in (\ref{Geodesic_Equation_15PN_New}) means twice of the total differential with respect to the new variable
$\tau$. We also note that the signature of the metric has been chosen such that covariant and contravariant spatial indices are equal, while covariant 
and contravariant time indices differ by a sign in front. For instance: $\xi^i = \xi_i$ and $h_{0i} = h_0^i$, but $h_{0i} = - h_i^0$. 

The decisive advantage of (\ref{Geodesic_Equation_15PN_New}) compared to (\ref{Geodesic_Equation_15PN}) is that the metric components (\ref{Metric_00_New}) - (\ref{Metric_ij_New}) 
are given in terms of the unperturbed light ray and, therefore, the integration of the geodesic equation (\ref{Geodesic_Equation_15PN_New}) can immediately be performed with respect 
to the integration variable $c \tau$. After that integration of the geodesic equation all differentiations can be computed, which still need to be performed according to the metric 
components in (\ref{Metric_00_New}) - (\ref{Metric_ij_New}) as they have to be implemented into the geodesic equation (\ref{Geodesic_Equation_15PN_New}). 
That solution for the first and second integration of (\ref{Geodesic_Equation_15PN_New}) reads formally 
\begin{eqnarray}
        \frac{\dot{\ve{x}}\left(\tau\right)}{c} &=& \ve{\sigma} 
	+ \sum\limits_{l=0}^{\infty} \frac{\Delta \dot{\ve{x}}^{M_L}_{\rm 1PN}\left(\tau\right)}{c} 
	+ \sum\limits_{l=1}^{\infty} \frac{\Delta \dot{\ve{x}}^{S_L}_{\rm 1.5PN}\left(\tau\right)}{c}\;,  
        \label{First_Interation_1PN_New}
        \\
	\ve{x}\left(\tau\right) &=& \ve{x}_{\rm N}\left(\tau\right) 
	+ \sum\limits_{l=0}^{\infty} \Delta\ve{x}^{M_L}_{\rm 1PN}\left(\tau, \tau_0\right) 
        + \sum\limits_{l=1}^{\infty} \Delta\ve{x}^{S_L}_{\rm 1.5PN}\left(\tau, \tau_0\right), 
        \label{Second_Interation_1PN_New}
\end{eqnarray}

\noindent 
up to terms of the order ${\cal O}\left(c^{-4}\right)$; the unperturbed light ray $\ve{x}_{\rm N}\left(\tau\right)$ in (\ref{Second_Interation_1PN_New}) is given by 
Eq.~(\ref{Parameter3}). The explicit expressions of (\ref{First_Interation_1PN_New}) and (\ref{Second_Interation_1PN_New}) are given by Eqs.~(32), (34), (37) in \cite{Kopeikin1997}
and Eqs.~(33), (36), (38) in \cite{Kopeikin1997}, respectively. The solutions in (\ref{First_Interation_1PN_New}) and (\ref{Second_Interation_1PN_New}) are still given 
in terms of the auxiliary variables $c\tau$ and $\ve{\xi}$. After performing all differentiations with respect to these auxiliary variables, one obtains the solutions of the 
first and second integration of geodesic equation in terms of the four-coordinates $\left(ct, \ve{x}\right)$, that means the solutions in (\ref{First_Interation_1PN}) 
and (\ref{Second_Interation_1PN}), respectively, just by replacing the auxiliary parameter $c\tau$ by $\ve{\sigma} \cdot \ve{x}_{\rm N}$ and $\ve{\xi}$ by the impact vector 
$\ve{d}_{\sigma}$ of the unperturbed light ray, which is defined by 
\begin{eqnarray}
        \ve{d}_{\sigma} &=& \ve{\sigma} \times \left(\ve{x}_0 \times \ve{\sigma}\right).
        \label{impact_vector}
\end{eqnarray}

\begin{widetext}
\begin{figure}[!ht]
\begin{center}
\includegraphics[scale=0.25]{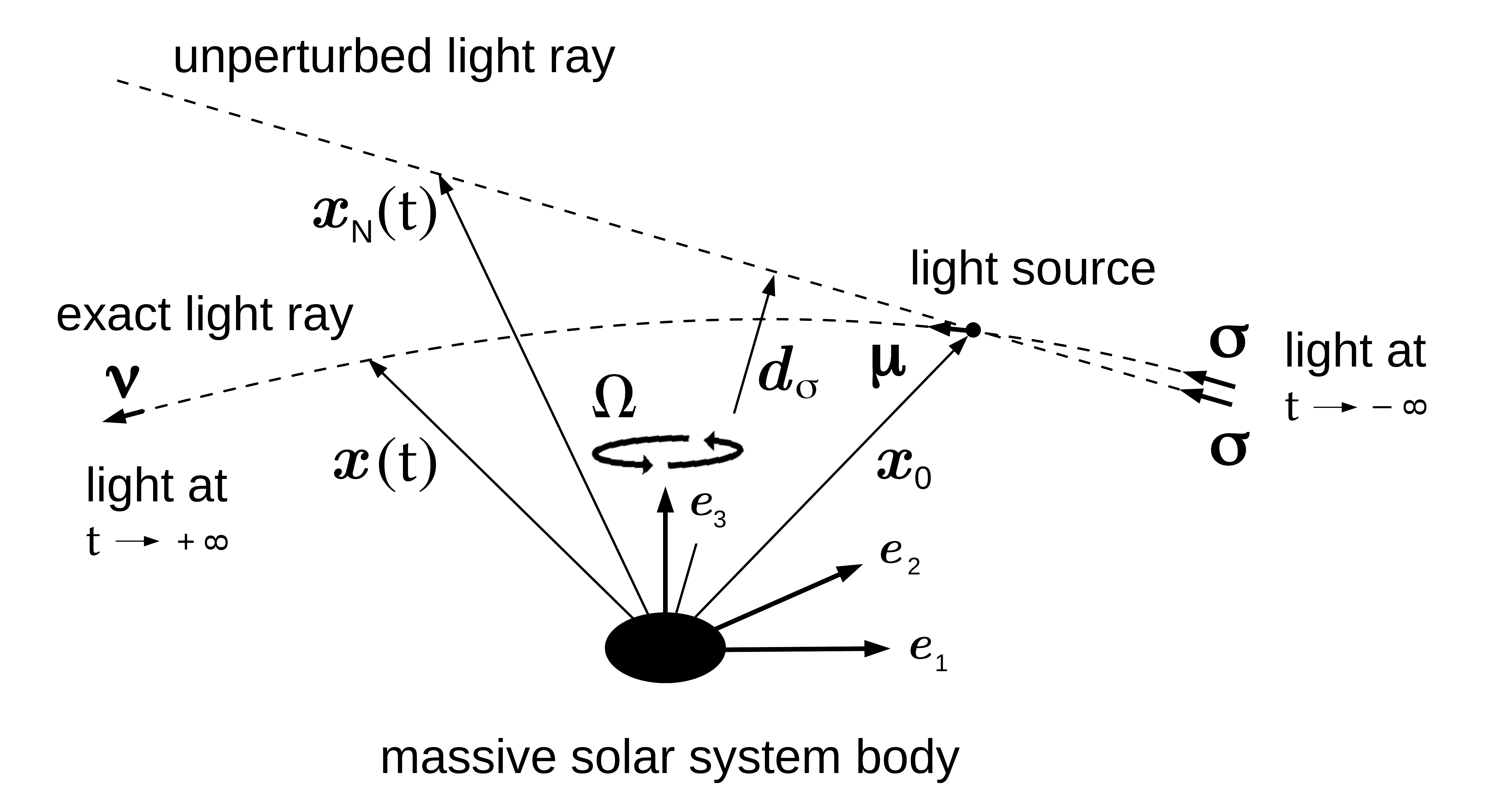}
\end{center}
\caption{
        A geometrical representation of the propagation of a light signal through the gravitational field of a massive body at rest.
	The origin of the spatial coordinates $\left(x^1,x^2,x^3\right)$ is assumed to be located at the center of mass of the body. 
	The spatial axes of the coordinate system are aligned with the unit vectors $\ve{e}_1, \ve{e}_2, \ve{e}_3$ of the principal axes 
	of the massive body. The body is in uniform rotational motion around its symmetry axis $\ve{e}_3$ with angular velocity $\Omega$. 
        The light signal is emitted by the light source at $\ve{x}_0$ in the direction of the unit-vector $\ve{\mu}$ and propagates along the exact
        light trajectory $\ve{x}\left(t\right)$.
        The unit tangent vectors $\ve{\sigma}$ and $\ve{\nu}$ of the light trajectory at past infinity and future infinity are
        defined by Eqs.~(\ref{vector_sigma}) and (\ref{vector_nu}), respectively.
        The unperturbed light ray $\ve{x}_{\rm N}\left(t\right)$ is given by
        Eq.~(\ref{Unperturbed_Lightray_2}) and propagates in the direction of $\ve{\sigma}$ along a straight line through the position
        of the light source at $\ve{x}_0$. The impact vector $\ve{d}_{\sigma}$ of the unperturbed light ray is given by Eq.~(\ref{impact_vector}).}
\label{Diagram}
\end{figure}
\end{widetext}

\noindent
The impact vector is a three-vector which points from the origin of coordinate system (center of mass of the body) towards the unperturbed light ray at their closest distance,  
as elucidated in Figure~\ref{Diagram}. The absolute value of the impact vector, $d_{\sigma} = |\ve{d}_{\sigma}|$, is denoted as impact parameter.

\section{Total light deflection in case of a body with arbitrary shape and in arbitrary rotational motions}\label{Section2}

\subsection{The tangent vector of light ray at future infinity}

The unit tangent vector along the light trajectory at future infinity is defined by   
\begin{eqnarray}
        \ve{\nu} &=& \frac{\dot{\ve{x}}\left(t\right)}{c}\,\bigg|_{t \rightarrow + \infty} \;, 
        \label{vector_nu}
\end{eqnarray}

\noindent
with $\ve{\nu} \cdot \ve{\nu} = 1$. 
The coordinate velocity of the light signal, $\dot{\ve{x}}\left(\tau\right)$, has been obtained  
for the case of a massive body at rest with full mass and spin multipole structure in \cite{Kopeikin1997} and for the case of a slowly 
moving massive body with full mass and spin multipole structure in \cite{Zschocke_1PN,Zschocke_15PN}, which agrees with \cite{Kopeikin1997} 
in case of a body at rest. In the limit $\tau \rightarrow + \infty$ one obtains from Eq.~(30) with (34) and (37) in \cite{Kopeikin1997} in 1.5PN approximation 
(e.g. Eq.~(11) in \cite{Klioner1991}),  
\begin{eqnarray}
	\ve{\nu} &=& \ve{\sigma} + \sum\limits_{l=0}^{\infty} \ve{\nu}_{\rm 1PN}^{M_L} + \sum\limits_{l=1}^{\infty} \ve{\nu}_{\rm 1.5PN}^{S_L} + {\cal O}\left(c^{-4}\right),  
        \label{vector_mu_M_L_S_L}
\end{eqnarray}
	
\noindent
where the individual terms in the sum (\ref{vector_mu_M_L_S_L}) are 
\begin{eqnarray}
	\ve{\nu}_{\rm 1PN}^{M_L} &=& - \frac{4 G}{c^2} \frac{\left(-1\right)^l}{l!} \, \hat{M}_L \,\hat{\partial}_L\,
	\frac{\ve{\xi}}{\left|\ve{\xi}\right|^2} \;,
        \label{vector_mu_M_L}
	\\
	\ve{\nu}_{\rm 1.5PN}^{S_L} &=& - \frac{8 G}{c^3}\,\sigma^c\,\epsilon_{abc} \frac{\left(-1\right)^l\,l}{\left(l+1\right)!}\,
	\hat{S}_{bL-1} \,\hat{\partial}_{aL-1}\,\frac{\ve{\xi}}{\left|\ve{\xi}\right|^2} \;.
        \label{vector_mu_S_L}
\end{eqnarray}

\noindent
It can be shown that (\ref{vector_mu_M_L_S_L}) is a unit vector up to higher order terms: $\ve{\nu} \cdot \ve{\nu} = 1 + {\cal O}\left(c^{-4}\right)$. 
Let us notice again that after performing the differentiation in (\ref{vector_mu_M_L}) and (\ref{vector_mu_S_L}) with respect to $\ve{\xi}$ this auxiliary vector 
is replaced by the impact vector $\ve{d}_{\sigma}$ and that the 'hat' implies STF with respect to all indices, e.g.: 
$\hat{S}_{bL-1} = {\rm STF}_{bL-1}\,S_{bL-1}$ and $\hat{\partial}_{aL-1} = {\rm STF}_{aL-1}\,\partial_{aL-1}$.  
These expressions for the total light deflection in (\ref{vector_mu_M_L}) and (\ref{vector_mu_S_L}) can also be deduced from Eqs.~(110) and (113) in \cite{Zschocke_15PN} in the limit of
a massive body at rest as well as in the limit $\tau \rightarrow + \infty$. It is noticed that the mass-dipole term (i.e. $l=1$ in (\ref{vector_mu_M_L})) vanishes if the origin of
spatial axes is located at the center-of-mass of the massive body. 

The expressions for the three-vector of total light deflection in (\ref{vector_mu_M_L}) and (\ref{vector_mu_S_L}) can be rewritten in a considerably simpler form by using the relation  
\begin{eqnarray}
        && \hat{\partial}_L\,\frac{\xi^i}{|\ve{\xi}|^2} = \hat{\partial}_L\,P^{ij}\,\frac{\partial}{\partial \xi^j} \ln |\ve{\xi}| 
= P^{ij}\,\frac{\partial}{\partial \xi^j}\,\hat{\partial}_L\,\ln |\ve{\xi}| \,,
\label{Relation_A}
\end{eqnarray}

\noindent
which is valid for any $l \ge 1$. It is noticed that the projector in (\ref{Relation_A}) has to be implemented, because $\ve{\xi}$ on the l.h.s. of (\ref{Relation_A}) 
is a three-vector in the plane perpendicular to three-vector $\ve{\sigma}$, while in the derivative in the first term on the r.h.s. of (\ref{Relation_A}) the vector 
$\ve{\xi}$ is treated as spatial vector with three independent components, hence afterwards it has to be projected back into the plane perpendicular to three-vector $\ve{\sigma}$,
in line with the statements made between Eqs.~(\ref{Differentiation_10}) and (\ref{spatial_derivative_2}).
By inserting (\ref{Relation_A}) into (\ref{vector_mu_M_L}) and (\ref{vector_mu_S_L}) one obtains for the spatial components  
\begin{eqnarray}
        {\nu}_{\rm 1PN}^{i\,M_L} &=& - \frac{4 G}{c^2} P^{ij} \frac{\partial}{\partial \xi^j}\,\frac{\left(-1\right)^l}{l!} \, \hat{M}_L \,\hat{\partial}_L\,\ln |\ve{\xi}|, 
        \label{vector_mu_M_L_Simpler_Form}
        \\
        {\nu}_{\rm 1.5PN}^{i\,S_L} &=& - \frac{8 G}{c^3} P^{ij} \frac{\partial}{\partial \xi^j}\,\sigma^c\,\epsilon_{i_lbc} \frac{\left(-1\right)^l\,l}{\left(l+1\right)!}\,
        \hat{S}_{bL-1} \,\hat{\partial}_{L}\,\ln |\ve{\xi}|\,, 
        \label{vector_mu_S_L_Simpler_Form}
\end{eqnarray}

\noindent 
where from (\ref{vector_mu_S_L}) to (\ref{vector_mu_S_L_Simpler_Form}) the summation index has been renamed, $a \rightarrow i_l$, in favor of a simpler notation 
for the differential operator $\hat{\partial}_{i_l L-1} = \hat{\partial}_{L}$. 
From (\ref{vector_mu_M_L_Simpler_Form}) and (\ref{vector_mu_S_L_Simpler_Form}) follows immediately that $\ve{\sigma} \cdot \ve{\nu}_{\rm 1PN}^{M_L} = 0$ as well as
$\ve{\sigma} \cdot \ve{\nu}_{\rm 1.5PN}^{S_L} = 0$, hence $\ve{\nu}$ is, in fact, a unit-vector, that means $\ve{\nu} \cdot \ve{\nu} = 1$ up to terms beyond 1.5PN approximation.

In order to determine the three-vector of total light deflection (\ref{vector_mu_M_L_S_L}) 
with (\ref{vector_mu_M_L_Simpler_Form}) and (\ref{vector_mu_S_L_Simpler_Form}), we need the term
\begin{eqnarray}
        \hat{\partial}_{L} \, \ln \left|\ve{\xi}\right| &=& {\rm STF}_{i_1 \dots i_l}\;
        P_{i_{1}}^{j_{1}}\, \dots \,P_{i_l}^{j_l}\,
\frac{\partial}{\partial \xi^{j_{1}}}\, \dots \,
\frac{\partial}{\partial \xi^{j_{l}}}\; \ln \left|\ve{\xi}\right|,  
\label{term_partial_ln}
\end{eqnarray}

\noindent
where the differential operator $\hat{\partial}_{L}$ has been defined by Eq.~(\ref{Transformation_Derivative_3}) and it has been taken into account that one
only needs the term $p=0$ in (\ref{Transformation_Derivative_3}), because $\ln \left|\ve{\xi}\right|$ is independent of $\tau$. The expression (\ref{term_partial_ln})
has been calculated in Appendix \ref{Appendix2} and reads (cf. Eq.~(\ref{Appendix_Relation_C_120}) in Appendix \ref{Appendix2}):
\begin{eqnarray}
        \hat{\partial}_{L}\,\ln \left|\ve{\xi}\right| &=& \frac{\left(-1\right)^{l+1}}{\left| \ve{\xi}\right|^l}\;{\rm STF}_{i_1 \dots i_l}\;  
        \sum\limits_{n=0}^{[l/2]} G_n^l\;P_{i_1 i_2}\, \dots \, P_{i_{2 n - 1} i_{2 n}}\; 
        \frac{\xi_{i_{2 n + 1}}\,\dots\,\xi_{i_{l}}}{\left|\ve{\xi}\right|^{l-2n}}\;,  
        \label{Relation_C}
\end{eqnarray}

\noindent
which is valid for any natural number $l \ge 1$ and the more explicit form for the tensor $n_{i_1 \dots i_l} = \xi_{i_1} \dots \xi_{i_l} / \left|\ve{\xi}\right|^l$ has been inserted;
the notation $\left[l/2\right]$ means the largest integer less than or equal to $l/2$ (see also Eqs.~(\ref{even_l}) and (\ref{odd_l})).
The expression in (\ref{Relation_C}) is considerably simpler that the expression in (\ref{term_partial_ln}), because the differentiations with respect to the auxiliary variable
$\ve{\xi}$ have been performed. Accordingly, one is allowed to replace the auxiliary variable $\ve{\xi}$ by the impact vector $\ve{d}_{\sigma}$.
Here, such a replacement is postponed for awhile in favor of simpler notation for the moment being.
The scalar coefficients in (\ref{Relation_C}) are given by (cf. Eq.~(\ref{Appendix_Relation_G_l_n}) in Appendix~\ref{Appendix2})
\begin{eqnarray}
        G^l_n &=& \left(-1\right)^{n}\,2^{l - 2 n - 1}\,\frac{l!}{n!}\,\frac{\left(l - n - 1\right)!}{\left(l - 2 n\right)!}\;.  
        \label{Relation_D}
\end{eqnarray}

\noindent 
Inserting (\ref{Relation_C}) into (\ref{vector_mu_M_L_Simpler_Form}) and (\ref{vector_mu_S_L_Simpler_Form}) yields 
\begin{eqnarray}
        && \hspace{-0.5cm} 
	{\nu}_{\rm 1PN}^{i\,M_L} = \frac{4 G}{c^2} P^{ij} \frac{\partial}{\partial \xi^j}\,\frac{1}{l!} \, \hat{M}_L\,\frac{1}{\left| \ve{\xi}\right|^l} \, 
	{\rm STF}_{i_1 \dots i_l} \sum\limits_{n=0}^{[l/2]} G_n^l\;P_{i_1 i_2}\, \dots \, P_{i_{2 n - 1} i_{2 n}}\, 
        \frac{\xi_{i_{2 n + 1}}\,\dots\,\xi_{i_{l}}}{\left|\ve{\xi}\right|^{l-2n}}\,,
        \label{vector_mu_M_L_Final_Form}
        \\
	&& \hspace{-0.5cm} 
	{\nu}_{\rm 1.5PN}^{i\,S_L} = \frac{8 G}{c^3} P^{ij} \frac{\partial}{\partial \xi^j}\,\sigma^c\,\epsilon_{i_lbc} \frac{l}{\left(l+1\right)!}\,\frac{1}{\left| \ve{\xi}\right|^l} 
        \hat{S}_{bL-1}\,  
	{\rm STF}_{i_1 \dots i_l} \sum\limits_{n=0}^{[l/2]} G_n^l\;P_{i_1 i_2}\, \dots \, P_{i_{2 n - 1} i_{2 n}}\, 
        \frac{\xi_{i_{2 n + 1}}\,\dots\,\xi_{i_{l}}}{\left|\ve{\xi}\right|^{l-2n}}\,,  
	\nonumber\\
        \label{vector_mu_S_L_Final_Form}
\end{eqnarray}

\noindent 
which is the final form for the components of the unit tangent vector (\ref{vector_mu_M_L_S_L}) of the light trajectory at future infinity. Because these 
terms in (\ref{vector_mu_M_L_Final_Form}) and (\ref{vector_mu_S_L_Final_Form}) are perpendicular to vector $\ve{\sigma}$, one concludes immediately 
that the tangent vector (\ref{vector_mu_M_L_S_L}) is, in fact, a unit vector.

\subsection{The angle of total light deflection}

The total light deflection angle is defined as angle between the unit tangent vector along the light trajectory at past and future infinity,
\begin{eqnarray}
	\delta\left(\ve{\sigma}, \ve{\nu}\right) &=& \arcsin |\ve{\sigma} \times \ve{\nu}|, 
	\label{Definition_total_light_deflection}
\end{eqnarray}

\noindent 
where $\ve{\sigma}$ was defined by (\ref{vector_sigma}), while the unit-vector $\ve{\nu}$ is defined by (\ref{vector_nu}). 
This angle (\ref{Definition_total_light_deflection}) follows from (\ref{vector_mu_M_L_S_L}) to be (e.g. Eq.~(12) in \cite{Klioner1991}), 
\begin{eqnarray}
	\delta\left(\ve{\sigma},\ve{\nu}\right) &=& 
	\left| \sum\limits_{l=0}^{\infty} \ve{\sigma} \times \ve{\nu}_{\rm 1PN}^{M_L} + \sum\limits_{l=1}^{\infty} \ve{\sigma} \times \ve{\nu}_{\rm 1.5PN}^{S_L} \right| 
	+ {\cal O}\left(c^{-4}\right), 
	\label{total_light_deflection_angle_1}
\end{eqnarray}

\noindent
where $\arcsin x = x + {\cal O} \left(x^3\right)$ for $x \ll 1$ has been used. 
The effect of the monopole field is several orders of magnitude larger than the higher mass-multipole or spin-multipole terms. 
Accordingly, by taking the monopole term out of the total sum in (\ref{total_light_deflection_angle_1}), then 
performing a series expansion of (\ref{total_light_deflection_angle_1}) and keeping only terms which are linear in the multipoles, one obtains the total light deflection 
(\ref{total_light_deflection_angle_1}) in the following form (see also Eqs.~(14) and (20) in \cite{Klioner1991} as well as Eqs.~(44) and (46) - (48) in \cite{Kopeikin1997}), 
\begin{eqnarray}
	\delta\left(\ve{\sigma},\ve{\nu}\right) &=& \sum\limits_{l=0}^{\infty} \delta\!\left(\ve{\sigma},\ve{\nu}_{\rm 1PN}^{M_L}\right) 
	+ \sum\limits_{l=1}^{\infty} \delta\!\left(\ve{\sigma},\ve{\nu}_{\rm 1.5PN}^{S_L}\right) + {\cal O}\left(c^{-4}\right), 
        \label{total_light_deflection}
\end{eqnarray}

\noindent
where the individual multipole terms are (cf. Eqs.~(20) and (61) in \cite{Klioner1991} as well as text above Eq.~(44) and above Eq.~(50) in \cite{Kopeikin1997}),  
\begin{eqnarray}
	\delta\left(\ve{\sigma},\ve{\nu}_{\rm 1PN}^{M_L}\right) &=& - \ve{\nu}_{\rm 1PN}^{M_L} \cdot \frac{\ve{d}_{\sigma}}{d_{\sigma}} \;,  
         \label{total_light_deflection_M_L}
         \\
	 \delta\left(\ve{\sigma},\ve{\nu}_{\rm 1.5PN}^{S_L}\right) &=& - \ve{\nu}_{\rm 1.5PN}^{S_L} \cdot \frac{\ve{d}_{\sigma}}{d_{\sigma}} \;. 
         \label{total_light_deflection_S_L}
\end{eqnarray}

\noindent
In order to get (\ref{total_light_deflection_M_L}) and (\ref{total_light_deflection_S_L}), also the relation
$|\ve{\sigma} \times \ve{\nu}| = |\ve{\sigma} \times \left(\ve{\nu} \times \ve{\sigma}\right)|$ has been used. 
The expression in (\ref{total_light_deflection_M_L}) and (\ref{total_light_deflection_S_L}) contain all terms which are linear in the multipoles, while terms  
are neglected, which are products of mass and spin multipoles. Clearly, in actual astrometric observations the total light deflection (\ref{total_light_deflection})
with (\ref{total_light_deflection_M_L}) and (\ref{total_light_deflection_S_L}) is the relevant quantity, rather than $|\ve{\sigma} \times \ve{\nu}_{\rm 1PN}^{M_L}|$
or $|\ve{\sigma} \times \ve{\nu}_{\rm 1.5PN}^{S_L}|$ \cite{Klioner1991,Kopeikin1997}. 

When the spatial components of the unit tangent vector (\ref{vector_mu_M_L_Simpler_Form}) and (\ref{vector_mu_S_L_Simpler_Form}) are inserted into 
(\ref{total_light_deflection_M_L}) and (\ref{total_light_deflection_S_L}), one encounters the following term, 
\begin{eqnarray}
\xi^i\,P^{ij}\,\frac{\partial}{\partial \xi^j}\,\hat{\partial}_L\,\ln |\ve{\xi}| = \left(-l\right)\,\hat{\partial}_L\,\ln |\ve{\xi}|\,,
\label{Relation_B}
\end{eqnarray}

\noindent
which is valid for $l \ge 1$. This relation can be shown by calculating the left-hand side of (\ref{Relation_B}), where $\hat{\partial}_L\,\ln |\ve{\xi}|$ is given by
the expression in (\ref{Relation_C}) and taking into account that $\displaystyle \xi^i \frac{\partial}{\partial \xi^i}\,\xi_{i_{2n+1}} \dots \xi_{i_l}/|\ve{\xi}|^{l-2n} = 0$.
Then, using relation (\ref{Relation_B}), one obtains for the mass-multipole and spin-multipole terms in (\ref{total_light_deflection_M_L}) and (\ref{total_light_deflection_S_L}) 
the following expressions, 
\begin{eqnarray}
	\delta\left(\ve{\sigma}, \ve{\nu}_{\rm 1PN}^{M_L}\right) &=& 
	- \frac{4 G}{c^2}\,\frac{1}{\left|\ve{\xi}\right|} \frac{\left(-1\right)^l}{\left(l-1\right)!} \,
        \hat{M}_{L} \, \hat{\partial}_{L}\;\ln \left|\ve{\xi}\right| ,
        \label{light_deflection_mass}
	\\
	\delta\left(\!\ve{\sigma}, \ve{\nu}_{\rm 1.5PN}^{S_L}\!\right) &=& 
        - \frac{8 G}{c^3} \frac{1}{\left|\ve{\xi}\right|} \epsilon_{abc} \sigma^c \frac{\left(-1\right)^l l^2}{\left(l+1\right)!} 
        \hat{S}_{b L-1} \hat{\partial}_{a L-1} \ln \left|\ve{\xi}\right|, 
        \label{light_deflection_spin}
\end{eqnarray}

\noindent
which are valid for $l \ge 1$; see also Eqs.~(47) and (48) in \cite{Kopeikin1997} 
(for the overall sign of the spin multipole terms see also footnote $3$ in \cite{Ciufolini}). 
The expressions in (\ref{light_deflection_mass}) and (\ref{light_deflection_spin}) are further treated 
by inserting (\ref{Relation_C}) into (\ref{light_deflection_mass}) and (\ref{light_deflection_spin}), which yields the 
total light deflection caused by the mass-multipole and spin-multipole structure of the body,  
\begin{eqnarray}
        \delta\left(\ve{\sigma}, \ve{\nu}_{\rm 1PN}^{M_L}\right) &=& 
	\frac{4 G}{c^2}\,\frac{1}{\left|\ve{\xi}\right|^{l+1}}\,\frac{\hat{M}_{L}}{\left(l-1\right)!} 
	\;{\rm STF}_{i_1 \dots i_l}\;  
        \sum\limits_{n=0}^{[l/2]} G_n^l\,P_{i_1 i_2}\, \dots \, P_{i_{2 n - 1} i_{2 n}}\,
	\frac{\xi_{i_{2 n + 1}}\,\dots\,\xi_{i_{l}}}{\left|\ve{\xi}\right|^{l-2n}}\;, 
        \label{light_deflection_mass_final}
        \\
        \delta\left(\ve{\sigma}, \ve{\nu}_{\rm 1.5PN}^{S_L}\right) &=& 
	\frac{8 G}{c^3} \frac{1}{\left|\ve{\xi}\right|^{l+1}} \epsilon_{i_l bc}\,\sigma^c\,\frac{l}{l+1}\,\frac{\hat{S}_{b L-1}}{\left(l-1\right)!}\,
	{\rm STF}_{i_1 \dots i_l} 
        \sum\limits_{n=0}^{[l/2]} G_n^l\,P_{i_1 i_2}\, \dots \, P_{i_{2 n - 1} i_{2 n}}\,
        \frac{\xi_{i_{2 n + 1}}\,\dots\,\xi_{i_{l}}}{\left|\ve{\xi}\right|^{l-2n}} ,
	\nonumber\\ 
        \label{light_deflection_spin_final}
\end{eqnarray}

\noindent
which are valid for any natural number $l \ge 1$. The mass-multipoles, $\hat{M}_L$, and the spin-multipoles, $\hat{S}_L$, for a body being of arbitrary shape, 
inner structure, rotational motions and inner currents, are given by Eqs.~(\ref{Mass_Multipoles}) and (\ref{Spin_Multipoles}), respectively. Accordingly, 
the expressions in (\ref{light_deflection_mass_final}) and (\ref{light_deflection_spin_final}) represent the total light deflection in the gravitational field of a 
body being of arbitrary shape, inner structure, as well as rotational motions of inner currents. The term in (\ref{light_deflection_mass_final}) is a scalar, while 
(\ref{light_deflection_spin_final}) is a pseudo-scalar, a fact which implies that they are invariant under arbitrary rotations of the spatial axes of the coordinate system. 

In order to calculate the numerical values of the light deflection vectors in (\ref{vector_mu_M_L_Final_Form}) and (\ref{vector_mu_S_L_Final_Form}) and the light deflection angles in 
(\ref{light_deflection_mass_final}) and (\ref{light_deflection_spin_final}), one needs the explicit expressions for the multipoles in (\ref{Mass_Multipoles}) and (\ref{Spin_Multipoles}). 
This will be the subject in the subsequent Section.

\section{The multipoles for an axisymmetric body}\label{Section3}

In order to calculate the total light deflection in (\ref{light_deflection_mass_final}) and in (\ref{light_deflection_spin_final}) one has to determine the 
mass-multipoles $\hat{M}_L$ and spin-multipoles $\hat{S}_L$ as defined by Eqs.~(\ref{Mass_Multipoles}) and (\ref{Spin_Multipoles}), respectively. 
For that one has to consider a concrete model for the massive solar system bodies. 
To a good approximation, especially for the Sun and the planets, a massive solar system body can be described by a  
rigid axisymmetric structure with radial-dependent mass-density, having the shape 
\begin{eqnarray}
	\frac{\left(x^1\right)^2}{A^2} + \frac{\left(x^2\right)^2}{B^2} + \frac{\left(x^3\right)^2}{C^2} &=& 1\;,
	\label{axisymmetry}
\end{eqnarray}

\noindent
where $A \neq B \neq C$ are the principal semi-axes of the body; we note that for an axisymmetric ellipsoid we would have $A = B$.  
Furthermore, the body can be assumed to be in uniform rotational motion around the symmetry axis, $\ve{e}_3$, of the body.  
If the coordinate system is chosen such that the rotational axis of the massive body is aligned with the $x^3$-axis of the coordinate system, $\ve{e}_3 = \left(0, 0, 1\right)$,  
then, the mass-multipoles (\ref{Mass_Multipoles}) and spin-multipoles (\ref{Spin_Multipoles}) are given by
\begin{eqnarray}
	\hat{M}_0 &=& - M \left(P\right)^0\,J_0\;, 
	\label{M}
	\\ 
	\hat{M}_L &=& - M \left(P\right)^l\,J_l\,\delta^{\;3}_{<{i_1}}\,\dots\,\delta^3_{{i_l}>} \;,
	\label{M_L}
        \\ 
        \hat{S}_{a} &=& - \kappa^2\,M\,\Omega\left(P\right)^2\,J_0\,\delta^3_{a}\;, 
        \label{S}
        \\
	\hat{S}_L &=& - M\,\Omega \left(P\right)^{l+1} \, J_{l-1}\,\frac{l+1}{l+4}\,\delta^{\;3}_{<{i_1}} \; \dots \;\delta^3_{{i_l}>} \;, 
        \label{S_L}
\end{eqnarray}

\noindent 
where (\ref{M_L}) is valid for any natural number of $l \ge 2$, while (\ref{S_L}) is valid for natural number of $l \ge 3$. 
In order to show these expressions in (\ref{M_L}) and (\ref{S_L}) one may apply the very same steps as presented in Appendix~B in \cite{Zschocke_Time_Delay_2PN}, 
except that we assume an axisymmetric body $A \neq B \neq C$, while in \cite{Zschocke_Time_Delay_2PN} an axisymmetric oblate ellipsoid $A=B$ has been considered. 

One comment should be in order about the multipoles in (\ref{M}) - (\ref{S_L}), which are valid for a coordinate system, $\left(x^1,x^2,x^3\right)$, 
where the symmetry axis of the massive body is aligned with the $x^3$-axis. Let us assume another coordinate system, $\left(x^{\prime\,1},x^{\prime\,2},x^{\prime\,3}\right)$, 
and both systems are having the same origin of their spatial coordinates. Let us further assume that these systems are related by a rotation of their spatial axes 
\cite{MTW,Kopeikin_Efroimsky_Kaplan,Poisson_Will,Arfken_Weber}, 
\begin{eqnarray}
	x^{\prime\,a} &=& R^a_b\,x^{b}\;,
        \label{Rotation_Coordinate_System}
\end{eqnarray}

\noindent 
where the orthogonal matrix of rotation $R^a_b$ can be parameterized, for instance, by three Euler angles, and is given, for example, by Eq.~(3.94) in \cite{Arfken_Weber}. 
Then, the multipoles in (\ref{M}) - (\ref{S_L}) in coordinate system $\{x^a\}$ and the multipoles in coordinate system $\{x^{\prime\,a}\}$ are related to each other 
by the standard transformation of tensors in three-space \cite{MTW,Kopeikin_Efroimsky_Kaplan,Poisson_Will,Arfken_Weber}, 
\begin{eqnarray}
	\hat{M}^{\prime}_{i_1 \dots i_l} &=& \hat{M}_{j_1 \dots j_l}\,R^{j_1}_{i_1}\,\dots R^{j_l}_{i_l}\;, 
        \label{Rotation_Mass_Multipoles}
	\\
	\hat{S}^{\prime}_{i_1 \dots i_l} &=& \hat{S}_{j_1 \dots j_l}\,R^{j_1}_{i_1}\,\dots R^{j_l}_{i_l}\;.
        \label{Rotation_Spin_Multipoles}
\end{eqnarray}

\noindent 
These relations allow to switch the multipoles from one coordinate system to the other. An explicit example of relation (\ref{Rotation_Mass_Multipoles}) is given for the 
mass-quadrupole by Eqs.~(48) - (53) in \cite{Klioner2003b}. A rotation of the spatial axes (\ref{Rotation_Coordinate_System}) would cause a change of the spatial components 
of all multipole-tensors and three-vectors. However, as mentioned above, the total light deflection terms (\ref{light_deflection_mass_final}) and (\ref{light_deflection_spin_final}) 
are invariant under arbitrary rotations of the spatial axes of the coordinate system, because they are scalars and pseudo-scalars, respectively. Therefore, without loss of generality, 
one may chose the coordinate system $\left(x^1,x^2,x^3\right)$ such that the $x^3$-axis is aligned with the axis of symmetry, $\ve{e}_3$, of the massive body. 

In (\ref{M}) - (\ref{S_L}) the Newtonian mass of the body is $M$, $P$ is the equatorial radius of the body, $\Omega$ is the angular velocity of the rotating body, 
and $J_l$ are the zonal harmonic coefficients of index $l$ defined by (cf. Eq.~(17) in \cite{Teyssandier1} or Eq.~(1.143) with (1.112) and (1.139) in \cite{Poisson_Will})
\begin{eqnarray}
        J_l &=& - \frac{1}{M\,\left(P\right)^l} \int d^3 x\,r^l\,\Sigma\,P_l\left(\cos \theta\right),
        \label{zonal_harmonic_coefficients}
\end{eqnarray}

\noindent
where $\Sigma$ is the mass-energy density of the massive body (cf. text below Eq.~(\ref{Spin_Multipoles})), $P_l$ are the Legendre polynomials, while the angle $\theta$ is the 
colatitude. 

The zonal harmonic coefficients in (\ref{zonal_harmonic_coefficients}) are defined for an axisymmetric body, $A \neq B \neq C$, hence they are non-zero for any natural number of $l$, 
while in case of an axisymmetric oblate ellipsoid $A = B$ they are non-zero only for even values of $l$. In case of a spherically symmetric body they vanish, except for $l=0$. 
The values of $J_l$ defined in (\ref{zonal_harmonic_coefficients}) are model-depend in the sense that they depend on assumptions made for the
mass distribution in the interior of the solar system bodies. Therefore, it is preferable to use actual zonal harmonics which are deduced from real measurements
of the gravitational fields of the Sun and giant planets. Their numerical values are given in Table~\ref{Table1} for the giant planets of the solar system.

The parameter $\kappa^2$ in (\ref{S}) is defined by \cite{Ellipticity} (see also Eqs.~(B60) - (B62) in \cite{Zschocke_Time_Delay_2PN}) 
\begin{eqnarray}
         \kappa^2 &=& \frac{I}{M\,P^2}\;, 
        \label{kappa}
\end{eqnarray}

\noindent
where $I$ is the moment of inertia of the real solar system body under consideration, which is related to the body's angular momentum via $|\ve{S}| = I\,\Omega$. 
As stated above, for a spherically symmetric body with uniform density $\kappa^2 = 2/5$ (cf. Eq.~(1.20) in \cite{Ellipticity}), 
while for real solar system bodies $\kappa^2 < 2/5$ because the mass densities are increasing towards the center of the massive bodies. 
The values of $\kappa^2$ are given in Table~\ref{Table1} for the Sun and giant planets of the solar system bodies. 

The STF tensor $\delta^{\;3}_{<{i_1}} \; \dots \; \delta^3_{{i_l}>} = {\rm STF}_{i_1 \dots i_l}\;\delta^{3}_{i_1} \dots \delta^{3}_{i_l}$ in (\ref{M_L})
and (\ref{S_L}) denotes products of Kronecker symbols which are symmetric and traceless with respect to indices $i_1 \dots i_l$.
They are given by the formula (cf. Eq.~(A20a) in \cite{Blanchet_Damour1}, Eq.~(B34) in \cite{Zschocke_Time_Delay_2PN}, or Eq.~(1.155) in \cite{Poisson_Will}):
\begin{eqnarray}
	\delta^{\;3}_{<{i_1}} \; \dots \; \delta^{3}_{{i_l}>} &=& \sum\limits_{p=0}^{[l/2]} H^l_p\,  
        \delta_{\{ i_1 i_2}\,\dots\, \delta_{i_{2p - 1} i_{2p}} \,\delta^{3}_{i_{2p + 1}}\,\dots\, \delta^{3}_{i_l \}}
\label{delta}
\end{eqnarray}

\noindent 
for any natural number $l \ge 1$ and the scalar coefficients are given by   
\begin{eqnarray}
        H^l_p &=& \left(-1\right)^p\, 
        \frac{\left(2 l - 2 p - 1\right)!!}{\left(2 l - 1 \right)!!}\,. 
\label{Coefficients_G_l_p}
\end{eqnarray}

\noindent
These multipoles (\ref{M_L}) and (\ref{S_L}) are in agreement with the resolutions of the International Astronomical Union (IAU) \cite{IAU_Resolution1}. That agreement is shown in 
some detail in Appendix~B in \cite{Zschocke_Time_Delay_2PN} for the mass-quadrupole as well as for the spin-hexapole in case of a rigid axisymmetric body with uniform mass-density. 
For explicit examples of mass-quadrupole, mass-octupole, and spin-hexapole it is referred to Eqs.~(B36), (B37), and (B59) in \cite{Zschocke_Time_Delay_2PN}, respectively.

\section{Total light deflection in case of an axisymmetric body in uniform rotation}\label{Section4}

In this Section the three-vector of total light deflection of a light signal in the gravitational field of a body with full mass-multipole and spin-multipole structure 
are given and the upper limits are determined. In what follows it will be shown the three-vector of total light deflection and the angle of the total light deflection 
are naturally given in terms of Chebyshev polynomials. 

\subsection{Chebyshev polynomials} 

In this Section we will briefly review the Chebyshev polynomials and their relations which are relevant for our considerations. 

There are Chebyshev polynomials of first and second kind, 
$T_l\left(x\right)$ and $U_l\left(x\right)$, which form a sequence of orthogonal polynomials \cite{Arfken_Weber,Abramowitz_Stegun}.
The power representation of Chebyshev polynomials of first kind reads (cf. Eqs.~(13.67) and (13.88a) in \cite{Arfken_Weber})
\begin{eqnarray}
        T_0 \left(x\right)  &=& 1 \;,
        \label{Chebyshev_Polynomials_0}
        \\
        T_l \left(x\right) &=& \frac{l}{2} \sum \limits_{n=0}^{[l/2]} \frac{\left(-1\right)^n}{n!} \,\frac{\left(l - n - 1\right)!}{\left(l - 2 n\right)!}
        \,\left(2 x\right)^{l - 2 n}\,,
        \label{Chebyshev_Polynomials_1}
\end{eqnarray}

\noindent
where $l \ge 1$. 
The power representation of Chebyshev polynomials of second kind reads (cf. Eq.~(13.88b) in \cite{Arfken_Weber}) 
\begin{eqnarray}
        U_l \left(x\right) &=& \sum \limits_{n=0}^{[l/2]} \frac{\left(-1\right)^n}{n!} \,\frac{\left(l - n\right)!}{\left(l - 2 n\right)!}
        \,\left(2 x\right)^{l - 2 n}\;, 
        \label{Chebyshev_Polynomials_2}
\end{eqnarray}

\noindent
where $l \ge 0$. 
The argument of the Chebyshev polynomials is a real number of the interval  
\begin{eqnarray}
     - 1 \le x \le + 1\;.
     \label{Variable_x_Chebyshev_B}
\end{eqnarray}

\noindent 
The Chebyshev polynomials of first and second kind are related by 
\begin{eqnarray}
	T_l \left(x\right) &=& U_l\left(x\right) - x\,U_{l-1}\left(x\right),
        \label{Relation_Chebyshev_Polynomials_1}
	\\
	U_l \left(x\right) &=& \frac{x \,T_{l+1}\left(x\right) - T_{l+2}\left(x\right)}{1 - x^2} \;.
        \label{Relation_Chebyshev_Polynomials_2}
\end{eqnarray}

\noindent 
The first derivative of Chebyshev polynomials of first kind (\ref{Chebyshev_Polynomials_1}) is related to
the Chebyshev polynomials of second kind as follows (see also p. $794$ in \cite{Arfken_Weber})
\begin{eqnarray}
        \frac{d\,T_l\left(x\right)}{d x} &=& l\,U_{l-1}\left(x\right). 
        \label{Chebyshev_polynomials_T_derivative}
\end{eqnarray}

\noindent
The first derivative of Chebyshev polynomials of second kind (\ref{Chebyshev_Polynomials_2}) is related to
the Chebyshev polynomials of first kind as follows
\begin{eqnarray}
	\frac{d\,U_l\left(x\right)}{d x} &=& \frac{x\,U_{l}\left(x\right) - \left(l + 1\right) T_{l+1}}{1-x^2}\;. 
        \label{Chebyshev_polynomials_U_derivative}
\end{eqnarray}

\noindent 
The Chebyshev polynomials of first kind (\ref{Chebyshev_Polynomials_1}) can be written in terms of trigonometric functions (cf. Eq.~(13.83a) in \cite{Arfken_Weber})
\begin{eqnarray} 
        T_l \left(x\right) &=& \cos \left(l\,\arccos \,x\right) 
        \label{upper_limit_M_L_20}
\end{eqnarray}

\noindent
for $l \ge 0$.
The Chebyshev polynomials of second kind (\ref{Chebyshev_Polynomials_2}) can be written in terms of trigonometric functions (cf. Eqs.~(13.83b) and (13.85a) in \cite{Arfken_Weber})
\begin{eqnarray} 
        U_{l-1} \left(x\right) &=& \frac{1}{\sqrt{1-x^2}}\,\sin \left(l\,\arccos \,x\right) .
        \label{upper_limit_S_L_20}
\end{eqnarray}

\subsection{The tangent vector of light ray at future infinity}

The tangent vector of the light trajectory at future infinity is given by Eq.~(\ref{vector_mu_M_L_S_L}), 
\begin{eqnarray}
        \ve{\nu} &=& \ve{\sigma} + \sum\limits_{l=0}^{\infty} \ve{\nu}_{\rm 1PN}^{M_L} + \sum\limits_{l=1}^{\infty} \ve{\nu}_{\rm 1.5PN}^{S_L} + {\cal O}\left(c^{-4}\right),  
        \noindent 
        \label{Total_Light_Deflection_Vector}
\end{eqnarray}
 
\noindent 
with the mass-multipole term (\ref{vector_mu_M_L_Final_Form}) and spin-multipole term (\ref{vector_mu_S_L_Final_Form}). 
In what follows it will be shown that the mass-multipole terms in (\ref{Total_Light_Deflection_Vector}) are given by Chebyshev polynomials of first kind and the spin-multipole terms
in (\ref{Total_Light_Deflection_Vector}) are given by Chebyshev polynomials of second kind.
We will consider these quantities separately.

\subsubsection{Mass-multipoles}
 
By inserting the mass-multipoles (\ref{M_L}) into the mass-multipole term (\ref{vector_mu_M_L_Final_Form}) one obtains for the spatial components of these quantities 
\begin{eqnarray}
	{\nu}_{\rm 1PN}^{i\,M_L} &=& - \frac{4 G M}{c^2}\,\frac{J_l}{l}\,P^{ij}\,\frac{\partial}{\partial d_{\sigma}^j} \left(\frac{P}{d_{\sigma}}\right)^l\,F^l_M \,, 
	\label{Tangent_nu_M}
\end{eqnarray}

\noindent 
where the auxiliary vectors $\ve{\xi}$ are replaced by the impact vectors $\ve{d}_{\sigma}$ everywhere in (\ref{Tangent_nu_M}).  
The dimensionless function $F_{M}^l$ in (\ref{Tangent_nu_M}) is given by Eq.~(\ref{Appendix_Proof1}) in Appendix~\ref{Proof1} 
and can be written in the form (details are given in Appendix~\ref{Proof1}) 
\begin{eqnarray}
        F_{M}^l &=& \frac{1}{\left(l - 1\right)!} \sum\limits_{n=0}^{[l/2]} G_n^l \,\bigg( 1 - \left(\ve{\sigma} \cdot \ve{e}_3\right)^2  \bigg)^n  
        \left(\frac{\ve{d}_{\sigma} \cdot \ve{e}_3}{d_{\sigma}}\right)^{l - 2 n}\,,    
        \label{upper_limit_M_L_15}
\end{eqnarray}

\noindent
where the auxiliary vectors $\ve{\xi}$ are replaced by the impact vectors $\ve{d}_{\sigma}$ everywhere in (\ref{upper_limit_M_L_15}) and the scalar coefficients 
are given by Eq.~(\ref{Relation_D}). 

The scalar function in (\ref{upper_limit_M_L_15}) can be expressed in terms of Chebyshev polynomials of first kind. To demonstrate this fact, we introduce the variable
\begin{eqnarray}
        x &=& \left(1 - \left(\ve{\sigma} \cdot \ve{e}_3\right)^2\right)^{-1/2}\;\left(\frac{\ve{d}_{\sigma} \cdot \ve{e}_3}{d_{\sigma}}\right),  
        \label{Variable_x_Chebyshev_A}
\end{eqnarray}

\noindent
which is a real number. Below it is shown that this variable is laying in the interval (\ref{Variable_x_Chebyshev_B}). 
By inserting the coefficients (\ref{Relation_D}) and using (\ref{Variable_x_Chebyshev_A}) one may rewrite the scalar function in (\ref{upper_limit_M_L_15}) in the following form:
\begin{eqnarray}
        F_{M}^l &=& \left[1 - \left(\ve{\sigma} \cdot \ve{e}_3\right)^2\right]^{[l/2]} 
        \, \frac{l}{2} \, \sum \limits_{n=0}^{[l/2]} \frac{\left(-1\right)^n}{n!} \frac{\left(l - n - 1\right)!}{\left(l - 2 n\right)!}
        \left(2 x\right)^{l - 2 n} . 
        \label{Function_F_M_Chebyshev}
\end{eqnarray}

\noindent
With the aid of Eq.~(\ref{Chebyshev_Polynomials_1}) one finds that the scalar function in (\ref{Function_F_M_Chebyshev}) is, up to a prefactor, just a
Chebyshev polynomial of first kind,
\begin{eqnarray}
        F_{M}^l &=& \left[1 - \left(\ve{\sigma} \cdot \ve{e}_3\right)^2\right]^{[l/2]} T_l \left(x\right).
        \label{Function_F_M_Chebyshev_5}
\end{eqnarray}

\noindent
Accordingly, by inserting (\ref{Function_F_M_Chebyshev_5}) into (\ref{Tangent_nu_M}) we obtain the mass multipole terms in (\ref{Total_Light_Deflection_Vector})  
in terms of Chebyshev polynomials of first kind: 
\begin{eqnarray}
	        {\nu}_{\rm 1PN}^{i\,M_L} &=& - \frac{4 G M}{c^2}\,\frac{J_l}{l}\,\left[1 - \left(\ve{\sigma} \cdot \ve{e}_3\right)^2\right]^{[l/2]}
        \, P^{ij}\,\frac{\partial}{\partial d_{\sigma}^j} \left(\frac{P}{d_{\sigma}}\right)^l\,T_l \left(x\right). 
        \label{Tangent_nu_M_Chebyshev}
\end{eqnarray}

\noindent
By evaluating the derivative and using (\ref{Relation_Chebyshev_Polynomials_1}) as well as (\ref{Chebyshev_polynomials_T_derivative}), 
one obtains for the mass multipole terms (\ref{Tangent_nu_M_Chebyshev}) the following expression: 
\begin{eqnarray}
	\ve{\nu}_{\rm 1PN}^{M_L} &=& + \frac{4 G M}{c^2\,d_{\sigma}}\,J_l\,\left[1 - \left(\ve{\sigma} \cdot \ve{e}_3\right)^2\right]^{[l/2]}
	\left(\frac{P}{d_{\sigma}}\right)^l\, 
	\left[U_{l}\left(x\right)\,\frac{\ve{d}_{\sigma}}{d_{\sigma}} 
	- U_{l-1}\left(x\right)\,\frac{\ve{\sigma} \times \left(\ve{e}_3 \times \ve{\sigma}\right)}{\sqrt{1 - \left(\ve{\sigma} \cdot \ve{e}_3\right)^2}}\right],  
        \label{Tangent_nu_M_Chebyshev_Final}
\end{eqnarray}

\noindent
which is valid for $l \ge 1$. The Chebyshev polynomials of second kind $U_l$ are given by Eqs.~(\ref{Relation_Chebyshev_Polynomials_1}) and (\ref{Relation_Chebyshev_Polynomials_2}) 
and variable $x$ has been defined by Eq.~(\ref{Variable_x_Chebyshev_A}).  
Some examples of (\ref{Tangent_nu_M_Chebyshev_Final}) are presented by Eqs.~(\ref{Example_Quadrupole_7}) and (\ref{Example_Octupole_7}) in Appendix~\ref{Examples}.  

The representation of the mass multipole terms (\ref{Tangent_nu_M_Chebyshev_Final}) recovers the intrinsic relation of the total light deflection 
with the Chebyshev polynomials and it allows to determine the upper limits of total light deflection, as will be demonstrated below. It has been checked that our result 
for the unit tangent vector (\ref{Tangent_nu_M_Chebyshev_Final}) agrees exactly with Eqs.~(43) and (44) in \cite{Poncin_Lafitte_Teyssandier}. This fact demonstrates that the results 
obtained with the advanced integration method developed in \cite{Kopeikin1997} coincide with the time-transfer function approach used in \cite{Poncin_Lafitte_Teyssandier}. 

The mass multipole terms (\ref{Tangent_nu_M_Chebyshev_Final}) seem apparently to be composed of Chebyshev polynomials of second kind. However, one may also rewrite 
(\ref{Tangent_nu_M_Chebyshev_Final}) fully in terms of Chebyshev polynomials of first kind by using relation (\ref{Relation_Chebyshev_Polynomials_2}). 

Finally, one still has to show that the possible values of variable $x$ in (\ref{Variable_x_Chebyshev_A}) are given by relation (\ref{Variable_x_Chebyshev_B}).
This can be seen by the following consideration.
The angles $\alpha = \delta\left(\ve{\sigma}, \ve{e}_3\right)$ and $\beta = \delta\left(\ve{d}_{\sigma}, \ve{e}_3\right)$ are not independent of each other, because the three-vectors
$\ve{\sigma}$ and $\ve{d}_{\sigma}$ are perpendicular to each other. Furthermore, the function in (\ref{Total_Light_Deflection_Mass_Chebyshev}) with variable
in (\ref{Variable_x_Chebyshev_A}) depends on scalar products of three-vectors,
$\ve{\sigma}\cdot\ve{e}_3$ and $(\ve{d}_{\sigma}\cdot\ve{e}_3)/d_{\sigma}$. Hence, that scalar function is independent of the orientation of the spatial coordinate
system. For these both reason, one may rotate the spatial axes of the coordinate system in such a way, that $\ve{\sigma}$ is aligned along the $x^1$-axis, and $\ve{d}_{\sigma}$
is aligned along the $x^2$-axis, while the symmetry axis of the massive body has three spatial components now: $\ve{e}_3 = \left(e_3^1, e_3^2. e_3^3\right)$.
Accordingly, we get $\ve{\sigma} \cdot \ve{e}_3 = e_3^1$ and $\displaystyle (\ve{d}_{\sigma} \cdot \ve{e}_3)/d_{\sigma} = e_3^2$. The numerical values of the components
of the three-vector $\ve{e}_3$ are restricted, because it is a unit-vector, that means: $\left(e_3^1\right) + \left(e_3^2\right) + \left(e_3^3\right) = 1$; for similar considerations
we refer to the endnote~$\left[99\right]$ in \cite{Zschocke_Quadrupole_1}. Taking all these aspects into account, one obtains for the variable in (\ref{Variable_x_Chebyshev_A})
$x = \pm \left(1 + y^2\right)^{-1/2}$ where $y = e_3^3/e_3^2$ is the ratio of the $x^3$-component over the $x^2$-component of unit-vector $\ve{e}_3$, which is aligned along
the symmetry axis of the body. From this consideration follows that $-1 \le x \le +1$ as stated by Eq.~(\ref{Variable_x_Chebyshev_B}).

\subsubsection{Spin-multipoles}

By inserting the spin-dipole term (\ref{S}) into (\ref{vector_mu_S_L_Final_Form}), one obtains for the spin dipole term in (\ref{Total_Light_Deflection_Vector}) the following expression,
\begin{eqnarray}
	\ve{\nu}_{\rm 1.5PN}^{S_1} &=& + \frac{4 G M}{c^3}\,\Omega\,\kappa^2\,J_0 \left(\frac{P}{d_{\sigma}}\right)^2 
	\, \bigg[2\,\frac{\left(\ve{\sigma}\times\ve{d}_{\sigma}\right) \cdot \ve{e}_3}{d_{\sigma}}\,\frac{\ve{d}_{\sigma}}{d_{\sigma}} + 
	\left(\ve{\sigma} \times \ve{e}_3\right)\bigg], 
        \label{Tangent_nu_S_Dipole}
\end{eqnarray}

\noindent
which, in view of $J_0 = -1$ and (\ref{Introduction_Spin_Dipole}), agrees with Eq.~(60) in \cite{Klioner1991}.
By inserting the spin-multipoles (\ref{S_L}) into the spin-multipole term (\ref{vector_mu_S_L_Final_Form}), 
one obtains for the spatial components of these terms the following expression, 
\begin{eqnarray}
	{\nu}_{\rm 1.5PN}^{i\,S_L} &=& - \frac{8 G M}{c^3}\,\Omega\,P\,\frac{J_{l-1}}{l+4}\,P^{ij}\,\frac{\partial}{\partial d_{\sigma}^j}
	\left(\frac{P}{d_{\sigma}}\right)^l\,F^l_S\;, 
        \label{Tangent_nu_S}
\end{eqnarray}

\noindent
which is valid for $l > 1$. 
The dimensionless function $F_{S}^l$ in (\ref{Tangent_nu_S}) is given by Eq.~(\ref{Appendix_Proof2}) in Appendix~\ref{Proof2} 
and can be written in the form (details are given in Appendix~\ref{Proof2})
\begin{eqnarray}
        F_{S}^l &=& \frac{\left(\ve{\sigma} \times \ve{d}_{\sigma}\right) \cdot \ve{e}_3}{d_{\sigma}}\,\frac{1}{\left(l-1\right)!}  
        \sum\limits_{n=0}^{[l/2]} G_n^l\,\frac{l - 2n}{l} \bigg(1 - \left(\ve{\sigma} \cdot \ve{e}_3\right)^2  \bigg)^{n}  
        \left(\frac{\ve{d}_{\sigma} \cdot \ve{e}_3}{d_{\sigma}}\right)^{l - 2 n -1} ,  
        \label{upper_limit_S_L_15}
\end{eqnarray}

\noindent
where the auxiliary vectors $\ve{\xi}$ are replaced by the impact vectors $\ve{d}_{\sigma}$ everywhere in (\ref{upper_limit_S_L_15}) and the scalar coefficients 
are given by Eq.~(\ref{Relation_D}). 

The pseudo-scalar function in (\ref{upper_limit_S_L_15}) can be expressed in terms of Chebyshev polynomials of second kind. To show that we insert the coefficients (\ref{Relation_D})
and use the variable in (\ref{Variable_x_Chebyshev_A}) and obtain the pseudo-scalar function in (\ref{upper_limit_S_L_15}) in the following form:
\begin{eqnarray}
        && \hspace{-0.75cm} F_{S}^l = \frac{\left(\ve{\sigma} \times \ve{d}_{\sigma}\right) \cdot \ve{e}_3}{d_{\sigma}}  
        \left[1 - \left(\ve{\sigma} \cdot \ve{e}_3\right)^2\right]^{[l/2]}
        \sum \limits_{n=0}^{[l/2]} \left(l - 2n\right)\frac{\left(-1\right)^n}{n!} \,\frac{\left(l - n - 1\right)!}{\left(l - 2 n\right)!} \left(2 x \right)^{l-2n-1}. 
        \label{Function_F_S_Chebyshev_1}
\end{eqnarray}

\noindent
Using
\begin{eqnarray}
        \left(l - 2n\right) \left(2 x \right)^{l-2n-1} &=& \frac{1}{2}\,\frac{d}{dx} \left(2 x\right)^{l - 2n}\;,
        \label{Derivative_Function_F_S_Chebyshev}
\end{eqnarray}

\noindent
one may write (\ref{Function_F_S_Chebyshev_1}) in the form
\begin{eqnarray}
        && \hspace{-0.75cm} F_{S}^l = \frac{\left(\ve{\sigma} \times \ve{d}_{\sigma}\right) \cdot \ve{e}_3}{d_{\sigma}}\,\frac{1}{l}  
        \left[1 - \left(\ve{\sigma} \cdot \ve{e}_3\right)^2\right]^{[l/2]}\, 
        \frac{d}{dx}\,\frac{l}{2} \sum \limits_{n=0}^{[l/2]} \frac{\left(-1\right)^n}{n!} \,\frac{\left(l - n - 1\right)!}{\left(l - 2 n\right)!} \left(2 x \right)^{l-2n} . 
        \label{Function_F_S_Chebyshev_2}
\end{eqnarray}

\noindent
In view of the power representation of Chebyshev polynomials of first kind in (\ref{Chebyshev_Polynomials_1}), one finds that the pseudo-scalar function in (\ref{Function_F_S_Chebyshev_2})
is, up to a prefactor, just the derivative of a Chebyshev polynomial of first kind,
\begin{eqnarray} 
        F_{S}^l &=&  \frac{\left(\ve{\sigma} \times \ve{d}_{\sigma}\right) \cdot \ve{e}_3}{d_{\sigma}}\,\frac{1}{l}
        \left[1 - \left(\ve{\sigma} \cdot \ve{e}_3\right)^2\right]^{[l/2]} \frac{d\,T_l\left(x\right)}{d x}\;, 
        \label{Function_F_S_Chebyshev_3}
\end{eqnarray}

\noindent
which is valid for $l \ge 3$. By means of relation (\ref{Chebyshev_polynomials_T_derivative}) one obtains for the pseudo-scalar function in (\ref{Function_F_S_Chebyshev_3})
\begin{eqnarray}
        F_{S}^l &=&  \frac{\left(\ve{\sigma} \times \ve{d}_{\sigma}\right) \cdot \ve{e}_3}{d_{\sigma}} 
        \left[1 - \left(\ve{\sigma} \cdot \ve{e}_3\right)^2\right]^{[l/2]} U_{l-1} \left(x\right).  
        \label{Function_F_S_Chebyshev_4}
\end{eqnarray}
 
\noindent 
Accordingly, by inserting (\ref{Function_F_S_Chebyshev_4}) into (\ref{Tangent_nu_S}) we obtain the spin-multipole terms in (\ref{Total_Light_Deflection_Vector}) 
in terms of Chebyshev polynomials of second kind: 
\begin{eqnarray}
	&& \hspace{-1.0cm} {\nu}_{\rm 1.5PN}^{i\,S_L} = - \frac{8 G M}{c^3}\,\Omega\,P\,\frac{J_{l-1}}{l+4}\,
	\left[1 - \left(\ve{\sigma} \cdot \ve{e}_3\right)^2\right]^{[l/2]} 
	\, P^{ij}\,\frac{\partial}{\partial d_{\sigma}^j} \frac{\left(\ve{\sigma} \times \ve{d}_{\sigma}\right) \cdot \ve{e}_3}{d_{\sigma}} 
	\left(\frac{P}{d_{\sigma}}\right)^l\,U_{l-1} \left(x\right).
        \label{Tangent_nu_S_Chebyshev}
\end{eqnarray}

\noindent 
By evaluating the derivative and using (\ref{Chebyshev_polynomials_U_derivative}) one obtains for the spin multipole terms (\ref{Tangent_nu_S_Chebyshev}) the following expression:
\begin{eqnarray}
	\ve{\nu}_{\rm 1.5PN}^{S_L} &=& + \frac{8 G M}{c^3}\,\Omega\,\frac{J_{l-1}}{l+4}\,\left[1 - \left(\ve{\sigma} \cdot \ve{e}_3\right)^2\right]^{[l/2]}
	\left(\frac{P}{d_{\sigma}}\right)^{l+1}
        \nonumber\\
        && \hspace{-1.4cm} \times \left[\!F_1\left(x\right) \frac{\ve{d}_{\sigma}}{d_{\sigma}} 
        + F_2\left(x\right) \frac{\ve{\sigma} \times \left(\ve{e}_3 \times \ve{\sigma}\right)}{\sqrt{1 - \left(\ve{\sigma} \cdot \ve{e}_3\right)^2}}
	+ F_3\left(x\right) \ve{\sigma} \times \ve{e}_3\!\right], 
	\label{Tangent_nu_S_Chebyshev_Final}
\end{eqnarray}

\noindent 
with 
\begin{eqnarray}
	F_1\left(x\right) &=& \frac{\left(\ve{\sigma} \times \ve{d}_{\sigma}\right) \cdot \ve{e}_3}{d_{\sigma}}\,\frac{U_{l-1}\left(x\right) - l\,T_{l+1}\left(x\right)}{1 - x^2}, 
	\label{F_1}
	\\
	F_2\left(x\right) &=& \frac{\left(\ve{\sigma} \times \ve{d}_{\sigma}\right) \cdot \ve{e}_3}{d_{\sigma}}\,\frac{\left(l + 1\right)T_l\left(x\right) - U_l\left(x\right)}{1 - x^2},
	\label{F_2}
	\\
	F_3\left(x\right) &=& U_{l-1}\left(x\right), 
	\label{F_3}
\end{eqnarray}

\noindent
which is valid for $l \ge 3$. The Chebyshev polynomials $T_l$ and $U_l$ are given by Eqs.~(\ref{Relation_Chebyshev_Polynomials_1}) and (\ref{Relation_Chebyshev_Polynomials_2}) 
and variable $x$ has been defined by Eq.~(\ref{Variable_x_Chebyshev_A}). Some examples of (\ref{Tangent_nu_S_Chebyshev_Final}) are presented by 
Eqs.~(\ref{Example_Spin_Dipole_nu}) and (\ref{Example_Spin_Hexapole_nu}) in Appendix~\ref{Examples}.

\subsection{The angle of total light deflection}

According to Eq.~(\ref{total_light_deflection}) the total light deflection in 1.5PN approximation can be separated into mass-multipole and spin-multipole terms,  
\begin{eqnarray}
        \delta\left(\ve{\sigma},\ve{\nu}\right) &=& 
        \sum\limits_{l=0}^{\infty} \delta\!\left(\ve{\sigma},\ve{\nu}_{\rm 1PN}^{M_L}\right) 
        + \sum\limits_{l=1}^{\infty} \delta\!\left(\ve{\sigma},\ve{\nu}_{\rm 1.5PN}^{S_L}\right) + {\cal O}\left(c^{-4}\right),
        \label{Total_Light_Deflection_Angle}
\end{eqnarray}

\noindent 
where the mass-multipole and spin-multipole terms are given by Eqs.~(\ref{light_deflection_mass_final}) and (\ref{light_deflection_spin_final}), respectively. 
We will consider these expressions for the mass-multipoles and spin-multipoles in case of an axisymmetric body in uniform rotational motion. 
It will be found that these terms of the total light deflection (\ref{Total_Light_Deflection_Angle}) are just Chebyshev polynomials.  
In particular, we will find that the mass-multipole terms in (\ref{Total_Light_Deflection_Angle}) are given by Chebyshev polynomials of first kind and the spin-multipole terms 
in (\ref{Total_Light_Deflection_Angle}) are given by Chebyshev polynomials of second kind. 

\subsubsection{Mass-multipoles}

The total light deflection for the mass-multipoles of the body is given by Eq.~(\ref{light_deflection_mass_final}). By inserting the
mass-monopole (\ref{M}) and the higher mass-multipoles (\ref{M_L}) as well as the relation for the spatial derivatives (\ref{Relation_C}) into this equation, 
one obtains the following expression for the total light deflection terms: 
\begin{eqnarray}
	\delta\left(\ve{\sigma}, \ve{\nu}_{\rm 1PN}^{M_0}\right) &=& - \,\frac{4 G M}{c^2}\,\frac{J_0}{d_{\sigma}} \;, 
        \label{upper_limit_M_0_5}
	\\
	\delta\left(\ve{\sigma}, \ve{\nu}_{\rm 1PN}^{M_L}\right) &=&  
	- \,\frac{4 G M}{c^2}\,\frac{J_l}{d_{\sigma}} \left(\frac{P}{d_{\sigma}}\right)^l\,F_{M}^l \;,  
        \label{upper_limit_M_L_5}
\end{eqnarray}

\noindent
where $l=0$ in (\ref{upper_limit_M_0_5}), while (\ref{upper_limit_M_L_5}) is valid for any natural number of $l \ge 2$. 
The total light deflection for higher mass-multipoles in (\ref{upper_limit_M_L_5}) vanishes in case of a spherically symmetric body,
because in that case the zonal harmonic coefficients, $J_l$, would be zero. The scalar function $F_{M}^l$ in (\ref{upper_limit_M_L_5}) is given by 
Eq.~(\ref{upper_limit_M_L_15}).  

By inserting (\ref{Function_F_M_Chebyshev_5}) into (\ref{upper_limit_M_L_5}) one finds that the total light deflection terms  
of higher mass-multipoles (\ref{upper_limit_M_L_5}) can be expressed by Chebyshev polynomials of first kind:
\begin{eqnarray}
        \delta\left(\ve{\sigma}, \ve{\nu}_{\rm 1PN}^{M_L}\right) &=&  
	- \frac{4 G M}{c^2 d_{\sigma}} J_l \left(\frac{P}{d_{\sigma}}\right)^l 
	\left[1 - \left(\ve{\sigma} \cdot \ve{e}_3\right)^2\right]^{[l/2]} T_l\left(x\right) 
        \label{Total_Light_Deflection_Mass_Chebyshev}
\end{eqnarray}

\noindent
which is valid for $l \ge 0$, since the mass-monopole in (\ref{upper_limit_M_0_5}) has been included. 
Some examples of (\ref{Total_Light_Deflection_Mass_Chebyshev}) are given by Eqs.~(\ref{Example_Monopole_10}),  
(\ref{Example_Quadrupole_10}) and (\ref{Example_Octupole_10}) in Appendix~\ref{Examples}. 

The expression in (\ref{Total_Light_Deflection_Mass_Chebyshev}) represents the total light deflection caused by the mass-multipole structure of an axisymmetric body at rest. 
The spatial axes of the coordinate system are oriented such that the symmetry axis $\ve{e}_3$ is aligned along the $x^3$-axis of the coordinate system.  
Of course, one may rotate the spatial axes of the coordinate system by using (\ref{Rotation_Coordinate_System}). Then, the components of the mass-multipoles are 
changed according to Eq.~(\ref{Rotation_Mass_Multipoles}) as well as the components of the three-vectors $\ve{e}_3$, $\ve{\sigma}$ and $\ve{d}_{\sigma}$, but 
the total light deflection in (\ref{Total_Light_Deflection_Mass_Chebyshev}) is invariant, because it is a scalar under rotations of the spatial axes. 

From (\ref{upper_limit_M_L_20}) follows that  
\begin{eqnarray} 
\left|T_l\left(x\right)\right| \le 1 \;.
        \label{upper_limit_M_L_21}
\end{eqnarray}

\noindent 
Hence, we get from (\ref{Total_Light_Deflection_Mass_Chebyshev}) the following upper limit for the absolute value of the total light deflection 
at some massive body with full mass-multipole structure: 
\begin{eqnarray}
	\left|\delta\left(\ve{\sigma}, \ve{\nu}_{\rm 1PN}^{M_L}\right)\right| &\le& 
	\frac{4 G M}{c^2}\,\frac{\left|J_l\right|}{d_{\sigma}} \left( 1 - \left(\ve{\sigma} \cdot \ve{e}_3\right)^2  \right)^{[l/2]}\,\left(\frac{P}{d_{\sigma}}\right)^l
        \label{upper_limit_M_L_25}
\end{eqnarray}

\noindent 
for $l \ge 0$. 
The total light deflection for higher mass multipoles in (\ref{upper_limit_M_L_25}) takes its maximal possible value in case of $\ve{\sigma} \cdot \ve{e}_3 = 0$,
that means when the unperturbed light ray propagates parallel to the equatorial plane. The light deflection for higher mass-multipoles (\ref{upper_limit_M_L_25}) vanishes in
case of $\ve{\sigma} \cdot \ve{e}_3 = 1$, that means when the light ray propagates parallel to the symmetry axis, $\ve{e}_3$, of the massive body.
Furthermore, from (\ref{upper_limit_M_0_5}) and (\ref{upper_limit_M_L_25}) one obtains the following upper limit for the total light deflection at some massive body 
with full mass-multipole structure:
\begin{eqnarray}
        \left|\delta\left(\ve{\sigma}, \ve{\nu}_{\rm 1PN}^{M_L}\right)\right| &\le&
        \frac{4 G M}{c^2}\,\frac{\left|J_l\right|}{d_{\sigma}} \left(\frac{P}{d_{\sigma}}\right)^l
        \label{upper_limit_M_L_30}
\end{eqnarray}

\noindent
for $l \ge 0$. 
This upper limit is strictly valid for any configuration between light source and massive body and observer.

The relations (\ref{upper_limit_M_L_25}) and (\ref{upper_limit_M_L_30}) agree with Eq.~(58) and (59) in \cite{Poncin_Lafitte_Teyssandier}, respectively, 
where the validity of these relations has been shown for the values $l \le 4$. In the endnote [20] the conjecture was formulated, that these relations might be valid 
for any natural number of $l \le 0$. It is also mentioned that in \cite{Zschocke_15PN} relation (\ref{upper_limit_M_L_30}) has been adopted as some kind of educated guess 
(cf. Eq.~(187) ibid.). Here, we have demonstrated that these relations (\ref{upper_limit_M_L_25}) and (\ref{upper_limit_M_L_30}) are, in fact, strictly valid 
for any natural number of $l \ge 0$.

From (\ref{upper_limit_M_L_30}) one obtains the following upper limit for the total light deflection
of grazing light rays (impact parameter $d_{\sigma}$ equals equatorial radius $P$) at some massive body with full mass-multipole structure:
\begin{eqnarray}
        && \hspace{-0.5cm} \left|\delta\left(\ve{\sigma}, \ve{\nu}_{\rm 1PN}^{M_L}\right)\right| \le
        \frac{4 G M}{c^2}\,\frac{\left|J_l\right|}{P}\;,  
        \label{upper_limit_M_L_35}
\end{eqnarray}

\noindent 
which is valid for $l \ge 0$. 
The upper limits, presented by Eqs.~(\ref{upper_limit_M_L_25}) - (\ref{upper_limit_M_L_35}),
are strictly valid for any configuration between light source and massive body and observer.

\subsubsection{Spin-multipoles}

The total light deflection caused by the spin-multipole structure of the body is given by Eq.~(\ref{light_deflection_spin_final}). By inserting the 
spin-dipole (\ref{S}) and the higher spin-multipoles (\ref{S_L}) as well as the relation for the spatial derivatives (\ref{Relation_C}) into this equation, 
one obtains the following expression for the total light deflection terms: 
\begin{eqnarray}
	\delta\left(\ve{\sigma}, \ve{\nu}_{\rm 1.5PN}^{S_1}\right) &=& - \,\frac{4 G M}{c^3}\,\Omega\,\kappa^2\,J_0\,\left(\frac{P}{d_{\sigma}}\right)^2\,
	\frac{\left(\ve{\sigma} \times \ve{d}_{\sigma}\right) \cdot \ve{e}_3}{d_{\sigma}}\;, 
        \label{upper_limit_S_1_5}
        \\
	\delta\left(\ve{\sigma}, \ve{\nu}_{\rm 1.5PN}^{S_L}\right) &=&
	-\,\frac{8 G M}{c^3}\,\Omega\,J_{l-1}\,\left(\frac{P}{d_{\sigma}}\right)^{l+1}\,F_{S}^l \,, 
        \label{upper_limit_S_L_5}
\end{eqnarray}

\noindent
where the light deflection caused by the spin-dipole (\ref{upper_limit_S_1_5}) agrees with Eq.~(61) in \cite{Klioner1991} 
\footnote{This can be seen if the spin-dipole, as given by Eq.~(\ref{Introduction_Spin_Dipole}) as well as by Eq.~(\ref{S}), is inserted into (\ref{upper_limit_S_1_5}) 
and using $J_0 = -1$ which follows from the definition of the zonal harmonic coefficients (\ref{zonal_harmonic_coefficients}).},  
while the light deflection caused by higher spin-multipoles (\ref{upper_limit_S_L_5}) is valid for any natural number of $l \ge 3$. 
The total light deflection for higher spin-multipoles in (\ref{upper_limit_S_L_5}) vanishes in case of a spherically symmetric body,
because in that case the zonal harmonic coefficients, $J_l$, would be zero. The pseudo-scalar function $F_{S}^l$ in (\ref{upper_limit_S_L_5}) is given by Eq.~ (\ref{upper_limit_S_L_15}). 

By inserting (\ref{Function_F_S_Chebyshev_4}) into (\ref{upper_limit_S_L_5}) one finds that the total light deflection terms of higher spin-multipoles 
can be expressed by Chebyshev polynomials of second kind:
\begin{eqnarray}
        && \hspace{-0.25cm} \delta\left(\ve{\sigma}, \ve{\nu}_{\rm 1.5PN}^{S_L}\right) =
        - \frac{8 G M}{c^3}\,\Omega\,J_{l-1} \left(\frac{P}{d_{\sigma}}\right)^{l+1} \, 
	\frac{\left(\ve{\sigma} \times \ve{d}_{\sigma}\right) \cdot \ve{e}_3}{d_{\sigma}} 
        \frac{l}{l+4} \, 
	\left[1 - \left(\ve{\sigma} \cdot \ve{e}_3\right)^2\right]^{[l/2]}\;U_{l-1}\left(x\right), 
	\nonumber\\ 
        \label{Total_Light_Deflection_Spin_Chebyshev_B}
\end{eqnarray}

\noindent 
which is valid for of $l \ge 3$. 
The expression in (\ref{Total_Light_Deflection_Spin_Chebyshev_B}) represents the total light deflection caused by the spin-multipole structure of an axisymmetric body at rest, 
which is in uniform rotational motion. The spatial axes of the coordinate system are oriented such that the symmetry axis $\ve{e}_3$ is aligned along the $x^3$-axis of the 
coordinate system. Some examples of the total light deflection caused by spin-multipoles (\ref{Total_Light_Deflection_Spin_Chebyshev_B}) are given in Appendix~\ref{Examples}.  
Of course, one may rotate the spatial axes of the coordinate system by using (\ref{Rotation_Coordinate_System}). Then, the components of the spin-multipoles are
changed according to Eq.~(\ref{Rotation_Spin_Multipoles}) as well as the components of the three-vectors $\ve{e}_3$, $\ve{\sigma}$ and $\ve{d}_{\sigma}$, but 
the total light deflection in (\ref{Total_Light_Deflection_Spin_Chebyshev_B}) does not change, because it is a pseudo-scalar, hence is invariant under rotations of the spatial axes.    

Taking account of the fact that (cf. Eq.~(22.14.6) in \cite{Abramowitz_Stegun}) 
\begin{eqnarray} 
	\left|U_{l-1}\left(x\right)\right| \le l \;,
        \label{upper_limit_S_L_21}
\end{eqnarray}

\noindent
we get from (\ref{Total_Light_Deflection_Spin_Chebyshev_B}) the following upper limit for the absolute value of the total light deflection at some massive body 
with full spin-multipole structure: 
\begin{eqnarray}
	\left|\delta\left(\ve{\sigma}, \ve{\nu}_{\rm 1.5PN}^{S_L}\right)\right| &\le& 
	\frac{8 G M}{c^3}\,\Omega\,\frac{l^2}{l+4}\,\left|J_{l-1}\right| \left(\frac{P}{d_{\sigma}}\right)^{l+1} 
	\left(1 - \left(\ve{\sigma} \cdot \ve{e}_3\right)^2\right)^{[l/2]} 
        \label{upper_limit_S_L_25}
\end{eqnarray}

\noindent
for $l \ge 3$. The total light deflection for higher spin-multipoles in (\ref{upper_limit_S_L_25}) takes its maximal possible value in case of 
$\ve{\sigma} \cdot \ve{e}_3 = 0$, that means when the unperturbed light ray propagates parallel to the equatorial plane. The light deflection for higher 
spin-multipoles (\ref{upper_limit_S_L_25}) vanishes in case of $\ve{\sigma} \cdot \ve{e}_3 = 1$, that means when the unperturbed light ray propagates parallel to the symmetry axis, 
$\ve{e}_3$, of the massive body. Furthermore, from (\ref{upper_limit_M_L_25}) one obtains the following upper limit for the total light deflection at some massive body
with full spin-multipole structure:
\begin{eqnarray}
        \left|\delta\left(\ve{\sigma}, \ve{\nu}_{\rm 1.5PN}^{S_L}\right)\right| &\le&
        \frac{8 G M}{c^3}\,\Omega\,\frac{l^2}{l+4}\,\left|J_{l-1}\right|\,\left(\frac{P}{d_{\sigma}}\right)^{l+1}
        \label{upper_limit_S_L_30}
\end{eqnarray}

\noindent  
for any natural number of $l \ge 3$. 

From (\ref{upper_limit_S_1_5}) and (\ref{upper_limit_S_L_30}) one obtains the following upper limit for the total light deflection of grazing light rays 
(impact parameter $d_{\sigma}$ equals equatorial radius $P$) at some massive body with full spin-multipole structure:
\begin{eqnarray}
        \left|\delta\left(\ve{\sigma}, \ve{\nu}_{\rm 1.5PN}^{S_1}\right)\right| &\le&
        \frac{4 G M}{c^3}\,\Omega\,\kappa^2\;,
        \label{upper_limit_S_1_35}
        \\
        \left|\delta\left(\ve{\sigma}, \ve{\nu}_{\rm 1.5PN}^{S_L}\right)\right| &\le&
        \frac{8 G M}{c^3}\,\Omega\,\frac{l^2}{l+4}\,\left|J_{l-1}\right| ,
        \label{upper_limit_S_L_35}
\end{eqnarray}

\noindent 
where $l=1$ in (\ref{upper_limit_S_1_35}), while (\ref{upper_limit_S_L_35}) is valid for any natural number of $l \ge 3$. 
The upper limits, presented by Eqs.~(\ref{upper_limit_S_L_25}) - (\ref{upper_limit_S_L_35}), 
are strictly valid for any configuration between light source and massive body and observer.

\section{Total light deflection at solar system bodies}\label{Section_Light_Deflection} 

In this Section the magnitude of the upper limits of total light deflection (\ref{total_light_deflection}) are calculated for the most massive solar system bodies, 
that is the Sun and the giant planets. The light deflection takes its maximal value in case of grazing light rays, where the upper limits for mass-multipoles 
are given by Eq.~(\ref{upper_limit_M_L_35}), and the upper limits for spin-multipoles are given by Eqs.~(\ref{upper_limit_S_1_35}) - (\ref{upper_limit_S_L_35}). 
The results of this Section allow to clarify, which mass-multipoles and spin-multipoles are required for an astrometric accuracy on the nano-arcsecond level in 
astrometric angular measurements.

\subsection{Numerical values of the parameter of solar system bodies}

The numerical parameter for determining the total light deflection in the gravitational fields of the most massive solar system bodies are given in Table~\ref{Table1}.

\begin{widetext}
\begin{table*}[h!]
\caption{Numerical parameter for mass $M$, equatorial radius $P$, actual zonal harmonic coefficients $J_l$, angular velocity $\Omega = 2\pi/T$ (with rotational period $T$), 
	dimensionless moment of inertia $\kappa^2$ of the Sun and the giant planets of the solar system. The values for $G M/c^2$ and $P$ are taken from \cite{Ellipticity}. 
	The value for $J_l$ of the Sun are taken from \cite{J_n_Sun} and references therein. The values $J_l$ with $l=2,4,6$ of Jupiter and Saturn are taken 
	from \cite{Book_Zonal_Harmonics}, while $J_l$ with $l=8,10$ of Jupiter and Saturn are taken from \cite{Zonal_Harmonics_Jupiter} and \cite{Zonal_Harmonics_Saturn}, respectively.
The values $J_l$ with $l=2,4,6$ of Uranus and Neptune are taken from \cite{Zonal_Harmonics_Uranus_Neptune}, while $J_8$ of Uranus and 
Neptune is taken from \cite{Zonal_Harmonics_Uranus_Neptune_J8}. 
The angular velocities $\Omega$ are taken from NASA planetary fact sheets.
The values for the dimensionless moment of inertia $\kappa^2$ are taken from \cite{Ellipticity}. 
A blank entry means the values are not known.} 
\begin{tabular}{| c | c | c | c | c | c|}
\hline
&&&&&\\[-12pt]
Parameter
&\hbox to 20mm{\hfill Sun \hfill}
&\hbox to 20mm{\hfill Jupiter \hfill}
&\hbox to 20mm{\hfill Saturn \hfill}
&\hbox to 20mm{\hfill Uranus \hfill}
&\hbox to 20mm{\hfill Neptune \hfill}\\[3pt] 
\hline
&&&&&\\[-12pt]
$GM/c^2\,[{\rm m}]$ & $1476.8$ & $1.410$ & $0.422$ & $0.064$  & $0.076$ \\[3pt]
$P\,[{\rm m}]$ & $696 \times 10^6$ & $71.49 \times 10^6$ & $60.27 \times 10^6$ & $25.56 \times 10^6$  & $24.76 \times 10^6$ \\[3pt]
$J_2$ & $+ 1.7 \times 10^{-7}$ & $+ 14.696 \times 10^{-3}$ & $+ 16.291 \times 10^{-3}$ & $+ 3.341 \times 10^{-3}$  & $+ 3.408 \times 10^{-3}$ \\[3pt]
$J_4$ & $+ 9.8 \times 10^{-7} $ & $ - 0.587 \times 10^{-3}$ & $ - 0.936 \times 10^{-3}$ & $ - 0.031 \times 10^{-3}$  & $ - 0.031 \times 10^{-3}$ \\[3pt]
$J_6$ & $+ 4 \times 10^{-8} $ & $+ 0.034 \times 10^{-3}$ & $+ 0.086 \times 10^{-3}$ & $+ 0.444 \times 10^{-6}$  & $+ 0.433 \times 10^{-6}$ \\[3pt]
$J_8$ & $ - 4 \times 10^{-9} $ & $ - 2.5 \times 10^{-6}$ & $ - 10.0 \times 10^{-6}$ & $ - 0.008 \times 10^{-6}$  & $ - 0.007 \times 10^{-6}$ \\[3pt]
$J_{10}$ & $ - 2 \times 10^{-10} $ & $+ 0.21 \times 10^{-6}$ & $+ 2.0 \times 10^{-6}$ & $ - $  & $ - $ \\[3pt]
$\Omega\,[{\rm sec}^{-1}]$ & $2.865 \times 10^{-6}$ & $1.758 \times 10^{-4}$ & $1.638 \times 10^{-4}$ & $1.012 \times 10^{-4}$  & $1.083 \times 10^{-4}$ \\[3pt]
$\kappa^2$ & $0.059$ & $0.254$ & $0.210$ & $0.225$  & $0.240$ \\[3pt]
\hline
\end{tabular}
\label{Table1}
\end{table*}
\end{widetext}
\begin{widetext}
\begin{table*}[h!]
\caption{The upper limit of total light deflection at the Sun and the giant planets of the solar system caused by their mass-multipole structure according to 
	Eq.~(\ref{upper_limit_M_L_35}). All values are given in micro-arcsecond ($\muas$). A blank entry indicates the light deflection is smaller than a nano-arcsecond (nas).}
\begin{tabular}{| c | c | c | c | c | c|}
\hline
&&&&&\\[-12pt]
Light deflection
&\hbox to 20mm{\hfill Sun \hfill}
&\hbox to 20mm{\hfill Jupiter \hfill}
&\hbox to 20mm{\hfill Saturn \hfill}
&\hbox to 20mm{\hfill Uranus \hfill}
&\hbox to 20mm{\hfill Neptune \hfill}\\[3pt]
\hline
&&&&&\\[-12pt]
$|\delta(\ve{\sigma}, \ve{\nu}_{\rm 1PN}^{M_0})|$ & $1.75 \times 10^{6}$ & $16.3 \times 10^{3}$ & $5.8 \times 10^{3}$ & $ 2.1 \times 10^{3} $  & $ 2.5 \times 10^{3} $ \\[3pt]
$|\delta(\ve{\sigma}, \ve{\nu}_{\rm 1PN}^{M_2})|$ & $ 0.35 $ & $ 239 $ & $ 94$ & $ 6.9 $  & $ 8.6$ \\[3pt]
$|\delta(\ve{\sigma}, \ve{\nu}_{\rm 1PN}^{M_4})|$ & $ 1.72 $ & $ 9.6 $ & $ 5.41 $ & $ 0.06 $  & $ 0.08 $ \\[3pt]
$|\delta(\ve{\sigma}, \ve{\nu}_{\rm 1PN}^{M_6})|$ & $ 0.07 $ & $ 0.55 $ & $ 0.50 $ & $ 0.001 $  & $ 0.001 $ \\[3pt]
$|\delta(\ve{\sigma}, \ve{\nu}_{\rm 1PN}^{M_8})|$ & $ 0.007 $ & $0.04 $ & $ 0.06 $ & $ - $  & $ - $ \\[3pt]
$|\delta(\ve{\sigma}, \ve{\nu}_{\rm 1PN}^{M_{10}})|$ & $ - $ & $0.003 $ & $ 0.01 $ & $ - $  & $ - $ \\[3pt]
\hline
\end{tabular}
\label{Table2}
\end{table*}
\end{widetext}
\begin{widetext}
\begin{table*}[h!]
\caption{The upper limit of total light deflection at the Sun and the giant planets of the solar system caused by their spin-multipole structure according to 
	Eqs.~(\ref{upper_limit_S_1_35}) - (\ref{upper_limit_S_L_35}). 
	All values are given in micro-arcsecond ($\muas$). A blank entry indicates the light deflection is smaller than a nano-arcsecond (nas).}
\begin{tabular}{| c | c | c | c | c | c|}
\hline
&&&&&\\[-12pt]
Light deflection
&\hbox to 20mm{\hfill Sun \hfill}
&\hbox to 20mm{\hfill Jupiter \hfill}
&\hbox to 20mm{\hfill Saturn \hfill}
&\hbox to 20mm{\hfill Uranus \hfill}
&\hbox to 20mm{\hfill Neptune \hfill}\\[3pt]
\hline
&&&&&\\[-12pt]
$|\delta(\ve{\sigma}, \ve{\nu}_{\rm 1.5PN}^{S_1})|$ & $ 0.7 $ & $ 0.17 $ & $ 0.04 $ & $ 0.004 $  & $ 0.005 $ \\[3pt]
$|\delta(\ve{\sigma}, \ve{\nu}_{\rm 1.5PN}^{S_3})|$ & $ - $ & $ 0.026 $ & $ 0.008 $ & $ - $  & $ - $ \\[3pt]
$|\delta(\ve{\sigma}, \ve{\nu}_{\rm 1.5PN}^{S_5})|$ & $ - $ & $ 0.001 $ & $ - $ & $ - $  & $ - $ \\[3pt]
\hline
\end{tabular}
\label{Table3}
\end{table*}
\end{widetext}


\subsection{Total light deflection of the mass-multipole structure of solar system bodies}

Numerical values for the light deflection of grazing rays (\ref{upper_limit_M_L_35})
at the most massive solar system bodies caused by their mass-multipole structure are presented in Table~\ref{Table2}.
These results show that only the very first few mass-multipoles with $l \le 10$ are required for an astrometric accuracy on the nano-arcsecond level.
That means, a light propagation model needs to implement only the first mass-multipoles up to at most $l = 10$ in order to determine the light trajectory 
on the nano-arcsecond level of accuracy. From the scaling behavior one may estimate that the mass-multipoles of the Sun and the giant planets with $l \ge 12$ 
contribute less that $0.001\,\muas$ to the total light deflection. Thus, the effect of light deflection caused by mass-multipoles with $l \ge 12$ can 
be neglected in astrometric measurements, even on the nano-arcsecond scale of accuracy.

\subsection{Total light deflection of the spin-multipole structure of solar system bodies}

Numerical values for the light deflection of grazing rays (\ref{upper_limit_S_1_35}) - (\ref{upper_limit_S_L_35}) 
at the most massive solar system bodies caused by their spin-multipole structure are presented in Table~\ref{Table3}. 
These results show that only the very first few spin-multipoles with $l \le 3$ are required for an astrometric accuracy on the 
nano-arcsecond level. That means, a light propagation model needs to implement only the first spin-multipoles up to at most $l = 3$ in order to determine the light trajectory
on the nano-arcsecond level of accuracy. Only in case of light deflection at Jupiter, the spin-multipole with $l=5$ contributes about $0.001\,\muas$ to the total light 
deflection. Thus, the effect of light deflection caused by spin-multipoles with $l \ge 5$ is at the outer nas-limit and 
can, most probably, not be detected in realistic astrometric measurements, even on the nano-arcsecond scale of accuracy.

\clearpage 

\section{Summary}\label{Summary}

The total light deflection in weak gravitational fields is defined as angle $\delta\left(\ve{\sigma}, \ve{\nu}\right)$ between the unit tangent vectors $\ve{\sigma}$ and $\ve{\nu}$ 
along the light ray at past infinity and future infinity, respectively, and represents an upper limit for the effect of bending of light by some massive body. In the general case, 
massive bodies can be of arbitrary shape, inner structure and oscillations, described by their mass-multipoles $M_L$, and can also be in arbitrary rotational motions and inner 
currents, described by their spin-multipoles $S_L$. Accordingly, the unit tangent vector of the light trajectory at future infinity 
is given by an infinite multipole series, which in the post-Newtonian scheme reads as follows 
\begin{eqnarray}
        \ve{\nu} &=& \ve{\sigma} + \sum\limits_{l=0}^{\infty} \ve{\nu}_{\rm 1PN}^{M_L} + \sum\limits_{l=1}^{\infty} \ve{\nu}_{\rm 1.5PN}^{S_L} + {\cal O}\left(c^{-4}\right),
        \label{Summary_nu}
\end{eqnarray}

\noindent 
where the mass-multipole terms are given by Eq.~(\ref{vector_mu_M_L}) and the spin-multipole terms are given by Eq.~ (\ref{vector_mu_S_L}). It has been shown that 
these terms take a considerably simpler form as given by Eqs.~(\ref{vector_mu_M_L_Final_Form}) and (\ref{vector_mu_S_L_Final_Form}). 

The multipole decomposition of the unit tangent vector in (\ref{Summary_nu}) implies a corresponding infinite multipole series of the angle of total light deflection 
in the gravitational field of such an arbitrarily shaped body of the solar system 
\begin{eqnarray}
        \delta\left(\ve{\sigma},\ve{\nu}\right) &=&
        \sum\limits_{l=0}^{\infty} \delta\!\left(\ve{\sigma},\ve{\nu}_{\rm 1PN}^{M_L}\right)
        + \sum\limits_{l=1}^{\infty} \delta\!\left(\ve{\sigma},\ve{\nu}_{\rm 1.5PN}^{S_L}\right) + {\cal O}\left(c^{-4}\right),
        \label{Summary_Total_Light_deflection}
\end{eqnarray}

\noindent
where the mass-multipole terms are given by Eq.~(\ref{light_deflection_mass}) and the spin-multipole terms are given by Eq.~(\ref{light_deflection_spin}). In this work it has been 
shown that these mass-multipole terms and spin-multipole terms can be written in a considerably simpler form as given by Eqs.~(\ref{light_deflection_mass_final}) and 
(\ref{light_deflection_spin_final}), respectively.  

The expressions in Eqs.~(\ref{Summary_nu}) and Eqs.~(\ref{Summary_Total_Light_deflection}) have been calculated for light signals which propagate in the gravitational field 
of an isolated axisymmetric body at rest being in uniform rotational motion around its axis of symmetry, $\ve{e}_3$, and having full mass-multipole and spin-multipole structure. 
The mass multipole and spin multipole terms are given by Eqs.~(\ref{Tangent_nu_M_Chebyshev_Final}) and (\ref{Tangent_nu_S_Chebyshev_Final}), respectively. 
It has been found that the evaluation of the unit tangent vector in (\ref{Summary_nu}) and of the angle of total light deflection in (\ref{Summary_Total_Light_deflection}) 
leads directly and in a compelling way to Chebyshev polynomials of first and second kind, respectively. 

This remarkable fact allows to obtain the expressions for the total light deflection for multipoles in arbitrary order; some examples are presented in Appendix~\ref{Examples}. 
Furthermore, that fact allows to determine their upper limits in a straightforward manner, because the upper limits of their absolute values are given by 
Eqs.~(\ref{upper_limit_M_L_21}) and (\ref{upper_limit_S_L_21}) and read 
\begin{eqnarray}
	\left|T_l\left(x\right)\right| &\le& 1 \;,
        \label{Summary_upper_limit_M_L_21}
\\ 
	\left|U_{l-1}\left(x\right)\right| &\le& l \;. 
        \label{Summary_upper_limit_S_L_21}
\end{eqnarray}

\noindent 
In this way, expressions for the upper limit of the total light deflection terms in (\ref{Summary_Total_Light_deflection}) 
have been obtained which are strictly valid in the 1PN and 1.5PN approximation 
for any astrometric configuration between light source, massive body and observer. For mass-multipoles they are given by Eq.~(\ref{upper_limit_M_L_25}) 
and for the spin-multipoles they are given by Eq.~(\ref{upper_limit_S_L_25}). 

The upper limits take their simplest form for grazing light rays, where they read 
(cf. Eqs.~(\ref{upper_limit_M_L_35}) as well as (\ref{upper_limit_S_1_35}) and (\ref{upper_limit_S_L_35}))
\begin{eqnarray}
\left|\delta\left(\ve{\sigma}, \ve{\nu}_{\rm 1PN}^{M_L}\right)\right| &\le&
        \frac{4 G M}{c^2}\,\frac{\left|J_l\right|}{P} \quad {\rm for} \; l \ge 0\,,
\label{Summary_M_L_Grazing}
\\
\left|\delta\left(\ve{\sigma}, \ve{\nu}_{\rm 1.5PN}^{S_1}\right)\right| &\le& \frac{4 G M}{c^3}\,\Omega\,\kappa^2 \quad {\rm for} \; l=1\,,
\label{Summary_S_Grazing}
\\
        \left|\delta\left(\ve{\sigma}, \ve{\nu}_{\rm 1.5PN}^{S_L}\right) \right| &\le&
        \frac{8 G M}{c^3}\,\Omega\,\frac{l^2}{l+4}\,\left|J_{l-1}\right| \;\; {\rm for} \; l \ge 3\,, 
        \label{Summary_S_L_Grazing}
\end{eqnarray}

\noindent
where $M$, $P$, $J_l$, and $\Omega$ are the mass, equatorial radius, zonal harmonic coefficient, and angular velocity of the massive body, while 
the real integer $l$ is the order of the multipole and $\kappa^2$ is the dimensionless moment of inertia. 
Each of these upper limits are realistic in the sense that there exist real astrometric configurations, where these upper bounds can be reached. 
In this sense they are just {\it the upper limits}. These formulae (\ref{Summary_M_L_Grazing}) - (\ref{Summary_S_L_Grazing}) are strictly valid 
in the 1PN and 1.5PN approximation and represent a criterion about the maximal possible deflection of 
light in the gravitational field of an axisymmetric body in uniform rotational motion. They can be used to decide whether these multipole terms need to be taken into account if 
ultra-high-accuracy astrometric measurements are to be modeled in the framework of future astrometry missions aiming at the sub-micro-arcsecond level. 

Numerical results of the upper limits of total light deflection (\ref{Summary_M_L_Grazing}) - (\ref{Summary_S_L_Grazing}) 
for some solar system bodies were presented in Table~\ref{Table2} and in Table~\ref{Table3}. 
According to these numerical values, it is found that the first few mass-multipoles with $l \le 10$ and the first few spin-multipoles with $l \le 3$ 
are sufficient in order to achieve an accuracy on the nano-arcsecond level in astrometric measurements. 

Finally, a comment is made about the 2PN terms which have been neglected in the total light deflection (\ref{Summary_Total_Light_deflection}). The 2PN mass-monopole terms 
have been determined in several investigations for the case of light propagation in the gravitational field of a spherically symmetric body 
\cite{Epstein_Shapiro1,Fischbach_Freeman,Richter_Matzner,Cowling,Article_Zschocke1,Teyssandier2,KlionerKopeikin1992,Ashby_Bertotti,Deng_Xie,Deng,Hees_Bertone_Poncin_Lafitte} 
(e.g. Table~$1$ in \cite{Article_Zschocke1}). 
Also the 2PN terms of total light deflection (\ref{Summary_Total_Light_deflection}) caused by the mass-quadrupole structure of solar system bodies have recently been determined 
in \cite{Zschocke_Quadrupole_1} (cf. Table IV in \cite{Zschocke_Quadrupole_1}). 
These investigations show that the 2PN corrections contribute less than $1\,{\rm percent}$ to the corresponding 
post-Newtonian terms in (\ref{Summary_M_L_Grazing}). Therefore, the upper limits in (\ref{Summary_M_L_Grazing}) - (\ref{Summary_S_L_Grazing})
constitute by far the most essential component to the angle of total light deflection. 

In summary, the primary results of this investigation for the case of light propagation in the gravitational field of an axisymmetric body in uniform rotation are: 
\begin{enumerate}
	\item[$\bullet$] mass-multipoles terms of the unit tangent vector of a light ray given by derivative of Chebyshev polynomials of first kind 
		(\ref{Tangent_nu_M_Chebyshev}),  
	\item[$\bullet$] mass-multipoles terms of the total light deflection are Chebyshev polynomials of first kind (\ref{Total_Light_Deflection_Mass_Chebyshev}), 
	\item[$\bullet$] spin-multipole terms of the unit tangent vector of a light ray given by derivative of Chebyshev polynomials of second kind 
		(\ref{Tangent_nu_S_Chebyshev}), 
        \item[$\bullet$] spin-multipole terms of the total light deflection are Chebyshev polynomials of second kind (\ref{Total_Light_Deflection_Spin_Chebyshev_B}), 
        \item[$\bullet$] strict upper limits for mass-multipoles terms of the total light deflection (\ref{upper_limit_M_L_25}),
        \item[$\bullet$] strict upper limits for spin-multipole terms of the total light deflection (\ref{upper_limit_S_L_25}). 
\end{enumerate}

\noindent
It has also been shown that in case of mass-multipoles of an axisymmetric body the advanced integration method developed in \cite{Kopeikin1997} yield the same unit tangent vector 
(\ref{Tangent_nu_M_Chebyshev_Final}) and the same total light deflection angle (\ref{upper_limit_M_L_15}) as the corresponding results obtained by means of the time-transfer function 
approach in \cite{Poncin_Lafitte_Teyssandier} (cf. text below Eq.~(\ref{Tangent_nu_M_Chebyshev_Final})). 

For the moment being, it remains an open question whether the unit tangent vector in (\ref{vector_mu_M_L_S_L}) and the angle of total light deflection 
in (\ref{total_light_deflection}), when applied to the case of bodies of arbitrary shape and inner structure and in arbitrary rotational motions, 
can also be expressed in terms of Chebyshev polynomials.

\section{Acknowledgment}

This work was funded by the German Research Foundation (Deutsche Forschungsgemeinschaft DFG) under grant number 447922800. Sincere gratitude is expressed to 
Prof. Sergei A. Klioner for kind encouragement and enduring support. Prof. Michael H. Soffel, Prof. Ralf Sch\"utzhold, Prof. William G. Unruh, Dr. Alexey Butkevich, 
Priv.-Doz. G\"unter Plunien, Dipl-Inf. Robin Geyer, Prof. Burkhard K\"ampfer, and Prof. Laszlo Csernai, are greatly acknowledged for inspiring discussions about astrometry 
and general theory of relativity.

\appendix

\section{Notation}\label{Appendix0}

The following notation is in use: 
\begin{itemize}
\item Newtonian constant of gravitation: $G$. 
\item vacuum speed of light in flat space-time: $c$. 
\item Newtonian mass of the body: $M$.
\item Equatorial radius of the body: $P$. 
\item Angular velocity of the body: $\Omega$. 
\item Zonal harmonic coefficients of the body: $J_l$. 
\item $\eta_{\alpha\beta} = {\rm diag}\left(-1,+1,+1,+1\right)$ is the metric tensor of flat space-time.
\item $g^{\alpha\beta}$ and $g_{\alpha\beta}$ are the contravariant and covariant components of the metric tensor with signature $\left(-,+,+,+\right)$.
\item $\displaystyle 1\,{\rm mas}\; ({\rm milli-arcsecond}) \simeq 4.85 \times 10^{-9}\,{\rm rad}$.
\item $\displaystyle 1\,\muas\; ({\rm micro-arcsecond}) \simeq 4.85 \times 10^{-12}\,{\rm rad}$.
\item $\displaystyle 1\,{\rm nas}\; ({\rm nano-arcsecond}) \simeq 4.85 \times 10^{-15}\,{\rm rad}$. 
\item $n! = n \left(n-1\right)\left(n-2\right)\cdot\cdot\cdot 2 \cdot 1$ is the factorial; by definition: $0! = 1$ \cite{Arfken_Weber}.
\item $n!! = n \left(n-2\right) \left(n-4\right)\cdot\cdot\cdot \left(2\;{\rm or}\;1\right)$ is the double factorial;  
	by definition: $0 !! = 1$ and $(-1)!! = 1$ \cite{Arfken_Weber}.
\item Lower case Greek indices take values 0,1,2,3.
\item The contravariant components of four-vectors: $a^{\mu} = \left(a^0,a^1,a^2,a^3\right)$.
\item Lower case Latin indices take values 1,2,3.
\item The three-dimensional coordinate quantities (three-vectors) referred to
the spatial axes of the reference system are in boldface: $\ve{a}$.
\item The contravariant components of three-vectors: $a^{i} = \left(a^1,a^2,a^3\right)$.
\item The absolute value of a three-vector:
$a = |\ve{a}| = \sqrt{a^1\,a^1+a^2\,a^2+a^3\,a^3}$.
\item The scalar product of two three-vectors:
$\ve{a}\,\cdot\,\ve{b}=\delta_{ij}\,a^i\,b^j=a^i\,b^i$ with Kronecker delta $\delta_{ij}$.
\item The vector product of two three-vectors reads
$\left(\ve{a}\times\ve{b}\right)^i=\varepsilon_{ijk}\,a^j\,b^k$
with Levi-Civita symbol $\varepsilon_{ijk}$.
\item The angle between two three-vectors $\ve{a}$ and $\ve{b}$ is designated as $\delta\left(\ve{a},\ve{b}\right)$ 
which can be computed by $\displaystyle \delta\left(\ve{a},\ve{b}\right) = \arccos \frac{\ve{a} \cdot \ve{b}}{|\ve{a}|\,|\ve{b}|}$ $\quad$ or by $\quad$ 
		$\displaystyle \delta\left(\ve{a},\ve{b}\right) = \arcsin \frac{\left| \ve{a} \times \ve{b} \right|}{|\ve{a}|\,|\ve{b}|}$. 
\end{itemize}

\section{STF tensors}\label{Appendix_STF}

Here we will present only those few standard notations about symmetric tracefree (STF) tensors,
which are necessary four our considerations, while further STF relations can be found
in \cite{Poisson_Will,Thorne,Blanchet_Damour1,Multipole_Damour_2,Hartmann_Soffel}.
\begin{itemize}
\item $L=i_1 i_2 ...i_l$ is a Cartesian multi-index of a given tensor $T$, that means
$T_L \equiv T_{i_1 i_2 \,.\,.\,.\,i_l}$.
\item two identical multi-indices imply summation:
\begin{eqnarray}
	A_L\,B_L \equiv \sum\limits_{i_1\,.\,.\,.\,i_l}\,A_{i_1\,.\,.\,.\,i_l}\,B_{i_1\,.\,.\,.\,i_l}\;. 
	\end{eqnarray}

\noindent 
\item The symmetric part of a Cartesian tensor $T_L$ is (cf. Eq.~(2.1) in \cite{Thorne}):
\begin{eqnarray}
T_{\left(L\right)} &=& T_{\left(i_1 ... i_l \right)} = \frac{1}{l!} \sum\limits_{\sigma}
T_{i_{\sigma\left(1\right)} ... i_{\sigma\left(l\right)}}
\end{eqnarray}

\noindent
where $\sigma$ is running over {\it all permutations} of $\left(1,2,...,l\right)$. 

For instance: let $T_{i_1 i_2 i_3}$ be a Cartesian tensor which is already symmetric in all of its Cartesian indices. Then
the symmetric part reads:
\begin{eqnarray}
        && \hspace{0.75cm} T_{\left(i_1 i_2 i_3\right)} = T_{i_1 i_2 i_3}\;. 
\end{eqnarray}

\noindent 
For instance: let $T_{i_1 i_2 i_3}$ be a Cartesian tensor which is not symmetric in any of its Cartesian indices. Then 
the symmetric part reads: 
\begin{eqnarray}
	T_{\left(i_1 i_2 i_3\right)} &=& \frac{1}{3!} 
	\left(T_{i_1 i_2 i_3}\!+\!T_{i_1 i_3 i_2}\!+\!T_{i_2 i_1 i_3}\!+\!T_{i_2 i_3 i_1}\!+\!T_{i_3 i_1 i_2}\!+\!T_{i_3 i_2 i_1}\right). 
\end{eqnarray}

\noindent
\item The un-normalized symmetric part of a Cartesian tensor $T_L$ is (cf. text above Eq.~(A19) in \cite{Blanchet_Damour1}):
\begin{eqnarray}
	T_{\{L\}} &=& T_{\{i_1 ... i_l \}} = \sum\limits_{\sigma \in S}
T_{i_{\sigma\left(1\right)} ... i_{\sigma\left(l\right)}}
\end{eqnarray}

\noindent
where $S$ is the {\it smallest set of permutations} of $\left(1,2,...,l\right)$ which makes $T_{i_1} ... i_{l}$ 
fully symmetric in its indices. Let us give some examples:  
\begin{eqnarray}
\delta_{\{i_1 i_2\}} &=& \delta_{i_1 i_2}\;,  
        \\ 
\delta_{\{i_1 i_2}\,n_{i_3\}} &=& \delta_{i_1 i_2}\,n_{i_3} + \delta_{i_1 i_3}\,n_{i_2} + \delta_{i_2 i_3}\,n_{i_1}\;,  
	\\
\delta_{\{i_1 i_2} \delta_{i_3 i_4\}} &=& \delta_{i_1 i_2} \delta_{i_3 i_4} + \delta_{i_1 i_3} \delta_{i_2 i_4} + \delta_{i_1 i_4} \delta_{i_2 i_3}\;.
\end{eqnarray}

\noindent
In case $T_{i_1 i_2 i_3}$ is a Cartesian tensor which is not symmetric in any of its indices, then the un-normalized symmetric part reads: 
\begin{eqnarray}
	T_{\{i_1 i_2 i_3\}} &=& T_{i_1 i_2 i_3} + T_{i_1 i_3 i_2} + T_{i_2 i_1 i_3} 
        + T_{i_2 i_3 i_1} + T_{i_3 i_1 i_2} + T_{i_3 i_2 i_1}\,. 
\end{eqnarray}

\noindent
\item The symmetric tracefree part of a Cartesian tensor $T_L$ (notation: $\hat{T}_L \equiv {\rm STF}_L\,T_L = T_{<i_1 \dots i_l>}$) is (cf. Eq.~(2.2) in \cite{Thorne}):
\begin{eqnarray}
	&& \hspace{0.75cm} \hat{T}_L = \sum_{k=0}^{\left[l/2\right]} a_{l k}\,\delta_{(i_1 i_2 ...} \delta_{i_{2k-1} i_{2k}}\,
	S_{i_{2k+1 ... i_l)} \,a_1 a_1 ... a_k a_k}
\label{anti_symmetric_1}
\end{eqnarray}

\noindent
where $\left[l/2\right]$ means the largest integer less than or equal to $l/2$, 
\begin{eqnarray}
	\left[ l/2 \right] &=& l/2 \quad {\rm for} \quad {\rm even}\;{\rm values}\;{\rm of}\; l\;,
\label{even_l}
\\
	\left[ l/2 \right] &=& (l-1)/2 \quad {\rm for} \quad {\rm odd}\;{\rm values}\;{\rm of}\; l\;,
\label{odd_l}
\end{eqnarray}

\noindent 
and $S_L \equiv T_{\left(L\right)}$ abbreviates the symmetric part of tensor $T_L$. The coefficient in (\ref{anti_symmetric_1}) is given by
\begin{eqnarray}
a_{l k} &=& \left(-1\right)^k \frac{l!}{\left(l - 2 k\right)!}\,
\frac{\left(2 l - 2 k - 1\right)!!}{\left(2 l - 1\right)!! \left(2k\right)!!}\,.
\label{coefficient_anti_symmetric}
\end{eqnarray}
\end{itemize}

\noindent 
For instance: 
\begin{eqnarray}
	\hat{T}_{i_1 i_2 i_3} &=& T_{\left(i_1 i_2 i_3\right)}
- \frac{1}{5} \left(\delta_{i_1 i_2}\,T_{\left(i_3 kk\right)} + \delta_{i_2 i_3}\,T_{\left(i_1 kk\right)} + \delta_{i_3 i_1}\,T_{\left(i_2 kk\right)} \right)\,. 
\end{eqnarray}

\noindent 
Three comments are in order about STF. First of all, the Kronecker delta has no symmetric tracefree part,
\begin{eqnarray}
        {\rm STF}_{ab} \,\delta^{ab} &=& 0\;.
        \label{STF_comment_1}
\end{eqnarray}

\noindent
Second, the symmetric tracefree part of any tensor which contains Kronecker delta is zero, if the Kronecker delta
has not any summation (dummy) index, for instance,
\begin{eqnarray}
        {\rm STF}_{abc} \,\delta^{ab}\,d_{\sigma}^c &=& 0\;,
        \label{STF_comment_2a}
        \\
        {\rm STF}_{abc} \,\delta^{ab}\,\sigma^c &=& 0\;.
        \label{STF_comment_2b}
\end{eqnarray}

\noindent
And third, the following relation is very useful (cf. Eq.~(1.158) in \cite{Poisson_Will} or Eq.~(A1) in \cite{Hartmann_Soffel}),
\begin{eqnarray}
A_{<L>}\,B_{<L>} &=& A_{L}\,B_{<L>}  
\label{STF_comment_3}
\end{eqnarray}

\noindent
which often considerably simplifies analytical evaluations.
In particular, we need the following Cartesian STF tensor,
\begin{eqnarray}
        \hat{n}_L = \frac{x_{<\,i_1}}{r}\,\dots\,\frac{x_{i_l\,>}}{r} \;, 
\label{Appendix_Cartesian_Tensor}
\end{eqnarray}

\noindent
where $x_i$ are the spatial coordinates of some arbitrary field point and $r = \left|\ve{x}\right|$; we note that $x_i = x^i$ and $\hat{n}_L = \hat{n}^L$. A very useful 
relation for later purposes is the expansion of the Cartesian STF tensor $\hat{n}_L$ in terms of Cartesian tensor $n_L$ (cf. Eq.~(A20a) in \cite{Blanchet_Damour1}), 
\begin{eqnarray}
	\hat{n}_L &=& \sum\limits_{k=0}^{[l/2]} \left(-1\right)^k \frac{\left(2 l - 2 k - 1\right)!!}{\left(2 l - 1\right)!!}  
	\delta_{\{ i_1 i_2 \,\dots} \delta_{i_{2 k - 1} i _{2 k}}  n_{i_{2 k + 1} \dots i_l \}} \,.  
\label{STF_n_L}
\end{eqnarray}


\section{Proof of Eq.~(\ref{Relation_C})}\label{Appendix2}

In this Appendix we will show that (\ref{Relation_C}) follows from (\ref{term_partial_ln}), given by 
\begin{eqnarray} 
\hat{\partial}_{L}\,\ln \left|\ve{\xi}\right| &=&
        {\rm STF}_{i_1 \dots i_l}\,P_{i_1}^{j_1}\,\dots\,P_{i_l}^{j_l}\,\partial_{j_1 \dots j_l}\,\ln \left|\ve{\xi}\right|.
\label{Appendix_Relation_C_1}
\end{eqnarray}

\noindent 
First of all, we consider the following term in Eq.~(\ref{Appendix_Relation_C_1}): 
\begin{eqnarray}
	\partial_{j_1 \dots j_l} \, \ln \left|\ve{\xi}\right| &=& 
	\frac{\partial}{\partial \xi^{j_1}} \, \dots \, \frac{\partial}{\partial \xi^{j_l}} \ln \left|\ve{\xi}\right|.
        \label{Appendix_Relation_C_5}
\end{eqnarray}

\noindent
By means of relation (A21b) in \cite{Blanchet_Damour1} one obtains 
\begin{eqnarray}
	\partial_{j_1 \dots j_l} \, \ln \left|\ve{\xi}\right| &=& \sum\limits_{k=0}^{[l/2]} \frac{\left(2 l - 4 k + 1\right)!!}{\left(2 l - 2 k + 1\right)!!} 
	 \, \delta_{\{j_1 j_2 \dots}\,\delta_{j_{2k-1} j_{2k}}\,\hat{\partial}_{j_{2k+1} \dots j_l\}} 
	\;\left(\Delta\right)^k \,\ln \left|\ve{\xi}\right| ,
        \label{Appendix_Relation_C_10}
\end{eqnarray}

\noindent
where $\Delta$ is the Laplace operator in terms of three-vector $\ve{\xi} = \left(\xi^1, \xi^2, \xi^3\right)$.  
One obtains 
\begin{eqnarray}
	\left(\Delta\right)^0\,\ln \left|\ve{\xi}\right| &=& \ln \left|\ve{\xi}\right|\,,
	\label{Appendix_Relation_C_15}
	\\
	\left(\Delta\right)^k \,\ln \left|\ve{\xi}\right| &=& \frac{\left(2 k - 2 \right)!}{\left|\ve{\xi}\right|^{2 k}} \quad {\rm for} \quad k = 1,2,3,\,\dots\,.  
	\label{Appendix_Relation_C_20}
\end{eqnarray}

\noindent 
Inserting (\ref{Appendix_Relation_C_15}) and (\ref{Appendix_Relation_C_20}) into (\ref{Appendix_Relation_C_10}) yields 
\begin{eqnarray}
	\partial_{j_1 \dots j_l}  \, \ln \left|\ve{\xi}\right| &=& \hat{\partial}_{\{j_1 \dots j_l\}}\,\ln \left|\ve{\xi}\right|
	+ \sum\limits_{k=1}^{[l/2]} \frac{\left(2 l - 4 k + 1\right)!!}{\left(2 l - 2 k + 1\right)!!} 
	\,\delta_{\{j_1 j_2 \dots}\,\delta_{j_{2k-1} j_{2k}}\,\hat{\partial}_{j_{2k+1} \dots j_l\}}
        \frac{\left(2 k - 2 \right)!}{\left|\ve{\xi}\right|^{2 k}}\,. 
        \label{Appendix_Relation_C_25}
\end{eqnarray}

\noindent 
Using relation (A30) in \cite{Blanchet_Damour1} one obtains for the partial derivatives 
\begin{eqnarray}
	&& \hspace{-0.5cm} 
	\hat{\partial}_{j_1 \dots j_l}\,\ln \left|\ve{\xi}\right| = \left(-1\right)^{l+1} \left(2 l - 2\right)!!\,
	\frac{\hat{n}_{j_1 \dots j_l}}{\left|\xi\right|^l}\;,
	\label{Appendix_Relation_C_30}
	\\
	&& \hspace{-0.5cm} 
	\hat{\partial}_{j_{2k+1} \dots j_l}\,\frac{1}{\left|\ve{\xi}\right|^{2k}} = \left(-1\right)^l \frac{\left(2 l - 2 k - 2\right)!!}{\left(2 k - 2\right)!!}
	\;\frac{\hat{n}_{j_{2k+1} \dots j_l}}{\left|\xi\right|^l}\,,
	\label{Appendix_Relation_C_35}
\end{eqnarray}

\noindent
where the STF tensor $\hat{n}_{j_1 \dots j_l} = {\rm STF}_{j_1 \dots j_l}\,\xi^{j_1} \dots \xi^{j_l}/\left|\xi\right|^l$; 
we notice that the allowed values $l \ge 2$ with $1 \le k \le l/2$ for even $l$ and $1 \le k \le (l-1)/2$ for odd $l$ in (\ref{Appendix_Relation_C_35}). 
By inserting (\ref{Appendix_Relation_C_30}) and (\ref{Appendix_Relation_C_35}) into (\ref{Appendix_Relation_C_25}) one obtains 
\begin{eqnarray}
	\partial_{j_1 \dots j_l}  \, \ln \left|\ve{\xi}\right| = 
	\frac{\left(-1\right)^l}{\left|\xi\right|^l} \sum\limits_{k=0}^{[l/2]} B^l_k\; \delta_{\{j_1 j_2 \dots}\,\delta_{j_{2k-1} j_{2k}}\,\hat{n}_{j_{2k+1} \dots j_l\}}    
        \label{Appendix_Relation_C_40}
\end{eqnarray}

\noindent 
where the coefficients are given by 
\begin{eqnarray}
	B^l_k &=& - \left(2 l - 2\right)!! \quad\quad {\rm for} \quad k = 0\;,  
	\label{Appendix_Relation_C_B0}
	\\
	B^l_k &=& + \frac{\left(2 l - 4 k + 1\right)!!}{\left(2 l - 2 k + 1\right)!!}\, 
	\frac{\left(2 k - 2 \right)!}{\left(2 k - 2\right)!!} \left(2 l - 2 k - 2\right)!! \quad\quad {\rm for} \quad k \ge 1\,. 
        \label{Appendix_Relation_C_Bk}
\end{eqnarray}

\noindent
From relation (A19) in \cite{Blanchet_Damour1} we get 
\begin{eqnarray}
	&& \hspace{-1.25cm} \delta_{\{j_1 j_2 \dots}\,\delta_{j_{2k-1} j_{2k}}\,\hat{n}_{j_{2k+1} \dots j_l\}} =  
	\frac{l!\,\left(2 k - 1\right)!!}{\left(2 k \right)!\,\left(l - 2 k \right)!}\;
	\,\delta_{(j_1 j_2 \dots}\,\delta_{j_{2k-1} j_{2k}}\,\hat{n}_{j_{2k+1} \dots j_l)}\;. 
	\label{Appendix_Relation_C_50}
\end{eqnarray}

\noindent 
Accordingly, by means of this relation one obtains for the expression in (\ref{Appendix_Relation_C_1}) 
\begin{eqnarray} 
	&& \hat{\partial}_{L}\,\ln \left|\ve{\xi}\right| = 
	{\rm STF}_{i_1 \dots i_l}\, \frac{\left(-1\right)^l}{\left|\xi\right|^l} \sum\limits_{k=0}^{[l/2]} C^l_k\;
        P_{i_1}^{j_1}\,\dots\,P_{i_l}^{j_l}\,
	\delta_{j_1 j_2 \dots}\,\delta_{j_{2k-1} j_{2k}}\,\hat{n}_{j_{2k+1} \dots j_l}
       \label{Appendix_Relation_C_55_A}
\end{eqnarray}

\noindent
where the symmetrization from the STF operation allows to remove the round brackets. The new coefficients in (\ref{Appendix_Relation_C_55_A}) are given by 
\begin{eqnarray}
	C^l_k &=& - \left(2 l - 2\right)!!  \quad\quad {\rm for} \quad k = 0\;,
        \label{Appendix_Relation_C_C0}
        \\ 
	C^l_k &=& + \frac{\left(2 l - 4 k + 1\right)!!}{\left(2 l - 2 k + 1\right)!!}\; 
        \frac{\left(2 k - 2 \right)!}{\left(2 k - 2\right)!!}\; \left(2 l - 2 k - 2\right)!! \;
	\frac{l!\,\left(2 k - 1\right)!!}{\left(2 k \right)!\,\left(l - 2 k \right)!}   
	\quad\quad {\rm for} \quad k \ge 1\;, 
        \label{Appendix_Relation_C_Ck}
\end{eqnarray}

\noindent
which are a combination of the coefficients in (\ref{Appendix_Relation_C_B0}) - (\ref{Appendix_Relation_C_Bk}) and the factorial coefficients on the r.h.s. 
of relation (\ref{Appendix_Relation_C_50}). The contraction of the product of two projectors by a Kronecker symbol yields a projector, e.g.
$P_{i_1}^{j_1}\,P_{i_2}^{j_2}\,\delta_{j_1 j_2} = P_{i_1 i_2}$. Hence, one may write (\ref{Appendix_Relation_C_55_A}) in the form 
\begin{eqnarray} 
	\hat{\partial}_{L}\,\ln \left|\ve{\xi}\right| &=& 
        {\rm STF}_{i_1 \dots i_l}\, \frac{\left(-1\right)^l}{\left|\xi\right|^l} \sum\limits_{k=0}^{[l/2]} C^l_k\;
	P_{i_1 i_2}\,\dots\,P_{i_{2k-1} i_{2k}}\,
	P^{j_{2k+1}}_{i_{2k+1}}\,\dots\,P_{i_l}^{j_l}\,\hat{n}_{j_{2k+1} \dots j_l}\;.
       \label{Appendix_Relation_C_55_B}
\end{eqnarray}

\noindent
Because the three-vector $\ve{\xi}$ is laying in the plane orthogonal to three-vector $\ve{\sigma}$, the projectors acting on three-vector $\ve{\xi}$ yield the same three-vector: 
$P^{j_1}_{i_1}\,\xi_{j_1} = \xi_{i_1}$. Accordingly, if the projectors are acting on the tensor $n_L$ then the result will  be the same tensor $n_L$. For instance:  
$P^{j_1}_{i_1}\,P^{j_2}_{i_2}\,n_{j_1 j_2} = {n}_{i_1 i_2}$. However, it is emphasized that if the projectors are acting on the STF version of this tensor, $\hat{n}_L$, then the result 
will in general not be the same STF tensor $\hat{n}_L$. For instance: $P^{j_1}_{i_1}\,P^{j_2}_{i_2}\,\hat{n}_{j_1 j_2} \neq \hat{n}_{i_1 i_2}$. Therefore, in order to evaluate the action 
of the projectors on STF tensor $\hat{n}_L$ in (\ref{Appendix_Relation_C_55_B}) one has to expand the STF tensor $\hat{n}_L$ in terms of $n_L$, which allows to determine the action of 
the projectors in (\ref{Appendix_Relation_C_55_B}). By means Eq.~(\ref{STF_n_L}) (cf. Eq.~(A20a) in \cite{Blanchet_Damour1}) one obtains for this expansion the following series, 
\begin{eqnarray}
	\hat{n}_{j_{2k+1} \dots j_l} &=& \sum\limits_{p=0}^{[(l-2k)/2]} D^l_{k,p}\; 
	\delta_{\{ j_{2 k + 1} j _{2 k + 2}} \dots \delta_{j_{2 k + 2 p - 1} j _{2 k + 2 p}}\; n_{j_{2 k + 2 p + 1} \dots j_l \}} \;, 
\label{Appendix_Relation_C_60}
\end{eqnarray}

\noindent
where the coefficients
\begin{eqnarray} 
        D^l_{k,p} &=& \left(-1\right)^p \frac{\left(2 l - 4 k - 2 p - 1\right)!!}{\left(2 l - 4 k - 1\right)!!} \;. 
        \label{Appendix_Relation_C_65_A}
\end{eqnarray}

\noindent
These coefficients are well defined, because the arguments of the double faculty are zero or positive, and only in the most extreme possible case we have $\left(-1\right)!! = 1$ 
(cf. p. $301$ in \cite{Arfken_Weber}). 
However, later we will rewrite the double factorial in a more compact form of factorials and binomial coefficients, where $\left(-1\right)!$ is not well defined. 
Therefore, we have to rewrite (\ref{Appendix_Relation_C_65_A}) as follows, 
\begin{eqnarray}
	D^l_{k,p} &=& \left(-1\right)^p \frac{\left(2 l - 4 k - 2 p + 1\right)!!}{\left(2 l - 4 k + 1\right)!!}\; \frac{2 l - 4 k + 1}{2 l - 4 k - 2 p + 1}\;. 
        \label{Appendix_Relation_C_65_B}
\end{eqnarray}

\noindent
This form allows to apply the relation between double factorial and factorial in Eqs.~(\ref{relation_factorial_doblefactorial}), which will be used later. 
As stated, when the projectors are acting on the tensor $n_L$ then the result will be the same tensor. Accordingly, using (\ref{Appendix_Relation_C_60}), 
one obtains for the last term in (\ref{Appendix_Relation_C_55_B}) the following expression,  
\begin{eqnarray} 
	P^{j_{2k+1}}_{i_{2k+1}}\,\dots\,P_{i_l}^{j_l}\,\hat{n}_{j_{2k+1} \dots j_l} 
	&=& \sum\limits_{p=0}^{[(l-2k)/2]} D^l_{k,p}\;
	P_{\{ i_{2 k + 1} i _{2 k + 2}} \dots P_{i_{2 k + 2 p - 1} i _{2 k + 2 p}}\; n_{i_{2 k + 2 p + 1} \dots i_l \}}\,. 
	\label{Appendix_Relation_C_70}
\end{eqnarray}

\noindent 
Accordingly, by inserting (\ref{Appendix_Relation_C_60}) into (\ref{Appendix_Relation_C_55_B}), by means of  
the explicit form in (\ref{Appendix_Relation_C_70}), one obtains  
\begin{eqnarray} 
	\hat{\partial}_{L} \ln \left|\ve{\xi}\right| &=&  \frac{\left(-1\right)^l}{\left|\xi\right|^l} 
        {\rm STF}_{i_1 \dots i_l} \sum\limits_{k=0}^{[l/2]}   
	\sum\limits_{p=0}^{[(l-2k)/2]} E^l_{k,p}\,
	\nonumber\\ 
	&& \times\, P_{i_1 i_2 \,\dots\,} P_{i_{2k-1} i_{2k}}  
	\,P_{\{i_{2k+1} i_{2k+2}}\,\dots\,P_{i_{2k+2p-1} i_{2k+2p}}\,n_{i_{2k+2p+1 \dots} i_l \} }\;,
	\label{Appendix_Relation_C_85} 
\end{eqnarray}

\noindent 
where the coefficients $E^l_{k,p} = C^l_{k}\,D^l_{k,p}\;$ (no summation over $l$ and $k$) are given by   
\begin{eqnarray}
	E^l_{k,p} &=& - \left(-1\right)^p \frac{\left(2 l - 2\right)!!}{\left(2 l - 1\right)!!} \;\left(2 l - 2 p - 1\right)!! 
	\quad\quad {\rm for} \quad k = 0\;, 
        \label{Appendix_Relation_C_90}
	\\
	E^l_{k,p} &=& + \left(-1\right)^p \frac{2 l - 4 k + 1}{2 l - 4 k - 2 p + 1}\; 
        \frac{\left(2 k - 2 \right)!}{\left(2 k - 2\right)!!}
	\left(2 l - 2 k - 2\right)!!\; 
	\nonumber\\ 
	&& \times\, \frac{l!\,\left(2 k - 1\right)!!}{\left(2 k \right)!\,\left(l - 2 k \right)!} 
	\frac{\left(2 l - 4 k - 2 p + 1\right)!!}{\left(2 l - 2 k + 1\right)!!}
        \quad\quad {\rm for} \quad k \ge 1\;.  
        \label{Appendix_Relation_C_95}
\end{eqnarray}

\noindent
Using relation (A19) in \cite{Blanchet_Damour1} we get (cf. Eq.~(\ref{Appendix_Relation_C_50}))  
\begin{eqnarray}
	&& P_{\{i_{2k+1} i_{2k+2}}\,\dots\,P_{i_{2k+2p-1} i_{2k+2p}}\,n_{i_{2k+2p+1 \dots} i_l \} }  
	\nonumber\\ 
	&=& \frac{(l - 2 k)!\,\left(2 p - 1\right)!!}{\left(2 p \right)!\,\left(l - 2 k - 2 p \right)!}\, 
	P_{( i_{2k+1} i_{2k+2}}\,\dots\,P_{i_{2k+2p-1} i_{2k+2p}}\,n_{i_{2k+2p+1 \dots} i_l )}\;. 
        \label{Appendix_Relation_C_100}
\end{eqnarray}

\noindent
Inserting (\ref{Appendix_Relation_C_100}) into (\ref{Appendix_Relation_C_85}) yields 
\begin{eqnarray} 
        &&  
        \hat{\partial}_{L}\,\ln \left|\ve{\xi}\right| =  
	{\rm STF}_{i_1 \dots i_l}\, \frac{\left(-1\right)^{l+1}}{\left|\xi\right|^l} \sum\limits_{k=0}^{[l/2]}\;\;  
        \sum\limits_{p=0}^{[(l-2k)/2]} F^l_{k,p}\, 
	P_{i_1 i_2 \dots}\,P_{i_{2k+2p-1} i_{2k+2p}}\,n_{i_{2k+2p+1 \dots} i_l } \;,  
        \label{Appendix_Relation_C_105} 
\end{eqnarray}

\noindent
where the STF operation allows to remove the round brackets (cf. Eq.~(\ref{Appendix_Relation_C_55_A})) and the new coefficients are given by  
\begin{eqnarray} 
	F^l_{k,p} &=& + \left(-1\right)^p \frac{\left(2 l - 2\right)!!}{\left(2 l - 1\right)!!} \;\left(2 l - 2 p - 1\right)!! 
        \frac{l!\,\left(2 p - 1\right)!!}{\left(2 p \right)!\,\left(l - 2 p \right)!}
	\quad\quad {\rm for} \quad k = 0\;,
        \label{Appendix_Relation_C_110}
        \\
	\nonumber\\
	F^l_{k,p} &=& - \left(-1\right)^p \frac{2 l - 4 k + 1}{2 l - 4 k - 2 p + 1}\,  
        \frac{\left(2 k - 2 \right)!}{\left(2 k - 2\right)!!}\,
        \frac{l!\,\left(2 k - 1\right)!!}{\left(2 k \right)!} 
        \nonumber\\ 
        && \times \left(2 l - 2 k - 2\right)!!  
        \frac{\left(2 l - 4 k - 2 p + 1\right)!!}{\left(2 l - 2 k + 1\right)!!}\,
	\frac{\left(2 p - 1\right)!!}{\left(2 p \right)!\,\left(l - 2 k - 2 p \right)!}
	\quad\quad {\rm for} \quad k \ge 1\;. 
        \label{Appendix_Relation_C_115}
\end{eqnarray}

\noindent
The expression in (\ref{Appendix_Relation_C_105}) represents a double sum over a finite number of terms which are functions of 
the summation indices $k$ and $p$. All terms with $k + p = {\rm const.}$ have the same number of projectors and unit-vectors. Therefore, 
it is useful to regroup the expression in (\ref{Appendix_Relation_C_105}) into a double sum over terms with have the same structure. 
That means, it is useful to introduce the new summation indices $q = k$ and $n = p + q$. All those terms with the same number $n$ have the 
same number of projectors and unit-vectors. Then, the double summation over $k$ and $p$ is replaced by a double summation over these 
new indices $n$ and $q$, that means  
\begin{eqnarray} 
	&& \hspace{-2.0cm}  
	\hat{\partial}_{L}\,\ln \left|\ve{\xi}\right| = \frac{\left(-1\right)^{l+1}}{\left|\xi\right|^l}\, 
        {\rm STF}_{i_1 \dots i_l} \sum\limits_{n=0}^{[l/2]}\;\;  
        \sum\limits_{q=0}^{n} F^l_{q,n-q}
	\,P_{i_1 i_2 \dots}\,P_{i_{2n-1} i_{2n}}\,n_{i_{2n+1 \dots} i_l }  \;. 
        \label{Appendix_Relation_C_107} 
\end{eqnarray}

\noindent
The scalar coefficients in (\ref{Appendix_Relation_C_110}) and (\ref{Appendix_Relation_C_115}), after some algebraic manipulations 
and expressed in terms of these new summation indices $q = k$ and $n = p + q$, are given as follows: 
\begin{eqnarray} 
	F^l_{q,n-q} &=& \left(-1\right)^n 2^{l - 1} \left(l - 1\right)! {l - 1  \choose n} \! {l  \choose 2 n}\! 
        \left[{2 l - 1 \choose 2 n} \right]^{-1}
	\quad\quad {\rm for} \quad q = 0\;,
        \label{Coefficients_F_0} 
        \\
	\nonumber\\ 
        F^l_{q,n-q} &=& \left(-1\right)^{n - q + 1} 2^{l-2 q}\,\left(l - 2\right)!\,\frac{2 l - 4 q + 1}{2 n + 1}\,{l \choose 2 n} {n \choose q} 
        \nonumber\\ 
	&& \times\, {l - q \choose n}
	{2 q - 2  \choose q - 1} \left[{l - 2 \choose q - 1} \right]^{-1} \, \left[{2 l - 2 q + 1 \choose 2 n + 1} \right]^{-1} 
        \quad\quad {\rm for} \quad q \ge 1\;,  
        \label{Coefficients_F_n}
\end{eqnarray}

\noindent 
where the binomial coefficients are defined by Eq.~(\ref{binomial_coefficients}). 
In order to deduce these scalar coefficients in (\ref{Coefficients_F_0}) - (\ref{Coefficients_F_n}) 
from (\ref{Appendix_Relation_C_110}) - (\ref{Appendix_Relation_C_115}), the relations 
\begin{eqnarray}
	&& \hspace{-1.0cm} \left(2 m \right)!! = 2^m\,m! \quad {\rm and} \quad \left(2 m + 1\right)!! = \frac{\left(2 m + 1\right)!}{2^m\,m!} 
	\label{relation_factorial_doblefactorial}
\end{eqnarray}

\noindent
have been used \cite{Arfken_Weber}, which allows one to rewrite  (\ref{Appendix_Relation_C_110}) and (\ref{Appendix_Relation_C_115}) fully in terms of factorials. 
If one performs the summation over variable $q$ in (\ref{Appendix_Relation_C_107}), one obtains  
\begin{eqnarray}
        && \hspace{-2.0cm}  
	\hat{\partial}_{L}\,\ln \left|\ve{\xi}\right| = \frac{\left(-1\right)^{l+1}}{\left|\xi\right|^l}\, 
        {\rm STF}_{i_1 \dots i_l} \sum\limits_{n=0}^{[l/2]}\;G^l_n  
	\, P_{i_1 i_2 \dots}\,P_{i_{2n - 1} i_{2n}}\,n_{i_{2n + 1 \dots} i_l} \;.  
        \label{Appendix_Relation_C_120}
\end{eqnarray}

\noindent
These new coefficients in (\ref{Appendix_Relation_C_120}) are defined by 
\begin{eqnarray}
	G^l_n &=& \sum\limits_{q=0}^{n} F^{l}_{q, n - q} \;. 
        \label{Appendix_Relation_C_125}
\end{eqnarray}

\noindent 
By inserting (\ref{Coefficients_F_0}) - (\ref{Coefficients_F_n}) into (\ref{Appendix_Relation_C_125}) and performing the summation one obtains 
\footnote{Relation (\ref{Appendix_Relation_C_125}) with the result in (\ref{Appendix_Relation_G_l_n}) has been calculated by means of the computer algebra 
systems {\it Mathematica} \cite{Mathematica} as well as {\it Maple} \cite{Maple}, while an exact proof can be performed by mathematical induction.} 
\begin{eqnarray}
	G^l_n &=& \left(-1\right)^{n}\,2^{l - 2 n - 1}\,\frac{l!}{n!}\,\frac{\left(l - n - 1\right)!}{\left(l - 2 n\right)!}\;.  
        \label{Appendix_Relation_G_l_n}
\end{eqnarray}

\noindent 
These are the coefficients given by Eq.~(\ref{Relation_D}) and the expression in (\ref{Appendix_Relation_C_120}) coincides with (\ref{Relation_C}).

\section{Proof of Eq.~(\ref{upper_limit_M_L_15})}\label{Proof1} 

The scalar function $F_{M}^l$ in (\ref{Tangent_nu_M}) reads  
\begin{eqnarray}
        F_{M}^l &=& \frac{1}{\left(l-1\right)!}\;
	\delta^{\;3}_{<{i_1}} \; \dots \; \delta^{3}_{{i_l}>}\;\; \sum\limits_{n=0}^{[l/2]} G_n^l\;P_{i_1 i_2}\, \dots \, P_{i_{2 n - 1} i_{2 n}}\,  
        \frac{\xi_{i_{2 n + 1}}\,\dots\,\xi_{i_{l}}}{\left|\ve{\xi}\right|^{l-2n}} \;,
        \label{Appendix_Proof1}
\end{eqnarray}

\noindent
where the coefficients $G_n^l$ are given by Eq.~(\ref{Relation_D}). In this Appendix we will show that (\ref{upper_limit_M_L_15}) follows from (\ref{Appendix_Proof1}). 

The projectors are defined by Eq.~(\ref{Projection_Operator}). In view of the fact that the contraction of the STF tensor 
$\delta^{\;3}_{<{i_1}} \; \dots \; \delta^{3}_{{i_l}>}$ with a Kronecker symbol vanishes, one may omit the Kronecker symbol in all projectors in (\ref{Appendix_Proof1}), 
e.g.: $P_{i_1 i_2} \rightarrow - \,\sigma_{i_1} \sigma_{i_2}$. That means 
\begin{eqnarray}
	\hspace{-0.5cm} P_{i_1 i_2}\, \dots \, P_{i_{2 n - 1} i_{2 n}} \rightarrow \left(-1\right)^n \sigma_{\i_1} \sigma_{i_2} \, \dots \, \sigma_{i_{2 n - 1}} \sigma_{i_{2 n}} \,.
        \label{Appendix_Proof1_1}
\end{eqnarray}

\noindent 
Furthermore, in each term in (\ref{Appendix_Proof1}) the auxiliary three-vector $\ve{\xi}$ can be replaced by the impact vector $\ve{d}_{\sigma}$ 
(cf. text below Eq.~(\ref{Relation_C})). Then, the scalar function in (\ref{Appendix_Proof1}) takes the following form, 
\begin{eqnarray}
	F_{M}^l &=& \frac{1}{\left(l-1\right)!}\;\sum\limits_{n=0}^{[l/2]} G_n^l\;\; 
	\delta^{\;3}_{<{i_1}} \,\dots\,\delta^{3}_{{i_l}>}
	\,\left(-1\right)^n \sigma_{i_1} \, \dots \, \sigma_{i_{2 n}}
        \frac{d_{\sigma}^{i_{2 n + 1}}\,\dots\,d_{\sigma}^{i_l}}{\left(d_{\sigma}\right)^{l-2n}} \,.
        \label{Appendix_Proof1_5}
\end{eqnarray}

\noindent 
Now the expression of the STF tensor in (\ref{delta}) with the coefficients in (\ref{Coefficients_G_l_p}) is inserted into (\ref{Appendix_Proof1_5}) which yields 
\begin{eqnarray}
	F_M^l &=& \frac{1}{\left(l-1\right)!}\;\sum\limits_{n=0}^{[l/2]} G_n^l \sum\limits_{p=0}^{[l/2]} \left(-1\right)^p 
	\frac{\left(2 l - 2 p - 1\right)!!}{\left(2 l - 1 \right)!!} 
        \nonumber\\ 
	&& \times\, \delta_{\{ i_1 i_2}\,\dots\, \delta_{i_{2p - 1} i_{2p}} \,\delta^{3}_{i_{2 p + 1}}\,\dots\, \delta^{3}_{i_l \}}
	\nonumber\\ 
	&& \times \,\left(-1\right)^n  \sigma_{i_1} \, \dots \, \sigma_{i_{2 n}}
         \frac{d_{\sigma}^{i_{2 n + 1}}\,\dots\,d_{\sigma}^{i_l}}{\left(d_{\sigma}\right)^{l-2n}} \,.
        \label{Appendix_Proof1_20}
\end{eqnarray}

\noindent
It is appropriate to perform the summation over $n$ rather than $p$ in (\ref{Appendix_Proof1_20}). Using relations like  
\begin{eqnarray}
	\delta_{i_1}^{3}\,\sigma^{i_1} &=& \left(\ve{\sigma} \cdot \ve{e}_3\right),  
        \label{scalar_products_1}
	\\ 
	\delta_{i_1}^{3}\,d_{\sigma}^{i_1} &=& \left(\ve{d}_{\sigma} \cdot \ve{e}_3\right), 
        \label{scalar_products_2}
\end{eqnarray}

\noindent 
as well as 
\begin{eqnarray}
        \delta_{i_1 i_2}\,\sigma^{i_1}\,\sigma^{i_2} &=& 1 \;,
        \label{scalar_products_3}
	\\ 
	\delta_{i_1 i_2}\,\frac{d_{\sigma}^{i_1}\,d_{\sigma}^{i_2}}{\left(d_{\sigma}\right)^2} &=& 1\;,
        \label{scalar_products_4}
\end{eqnarray}

\noindent 
one obtains for the summation over variable $n$ in (\ref{Appendix_Proof1_20}) the following result: 
\begin{eqnarray}
	&& \sum\limits_{n=0}^{[l/2]} G_n^l \;  
        \delta_{\{ i_1 i_2}\,\dots\, \delta_{i_{2p - 1} i_{2p}} \,\delta^{3}_{i_{2 p + 1}}\,\dots\, \delta^{3}_{i_l \}}
        \,\left(-1\right)^n  \sigma_{i_1} \, \dots \, \sigma_{i_{2 n}}
         \frac{d_{\sigma}^{i_{2 n + 1}}\,\dots\,d_{\sigma}^{i_l}}{\left(d_{\sigma}\right)^{l-2n}} 
	\nonumber\\
	&=& \sum\limits_{m=p}^{l/2} K_m^l \left(\frac{\ve{d}_{\sigma} \cdot \ve{e}_3}{d_{\sigma}}\right)^{l - 2m} \left(\ve{\sigma} \cdot \ve{e}_3\right)^{2m-2p} \,. 
        \label{Appendix_Proof1_25}
\end{eqnarray}

\noindent 
The scalar coefficients in (\ref{Appendix_Proof1_25}) are given by 
\begin{eqnarray}
	K_m^l &=& 2^{l - 2 m - 1}\;\frac{\left(l - m - 1\right)!}{\left(l - 2 m\right)!} \; 
	{m \choose p} \frac{\left(2 l - 1\right)!!}{\left(2 l - 2 p - 1\right)!!} \;\frac{l!}{m!}\;,
        \label{Appendix_Proof1_30}
\end{eqnarray}

\noindent 
which are a combination of the coefficients in (\ref{Appendix_Relation_G_l_n}) times a binomial coefficient times the inverse of the coefficients in (\ref{Coefficients_G_l_p}). 
Now the result (\ref{Appendix_Proof1_30}) is inserted into (\ref{Appendix_Proof1_20}) and the summation index $m$ is renamed into $n$. Furthermore, the relation 
\begin{eqnarray}
	\sum\limits_{p=0}^{[l/2]} \sum\limits_{n=p}^{[l/2]} f\left(n,p\right) &=& \sum\limits_{n=0}^{[l/2]} \sum\limits_{p=0}^{n} f\left(n,p\right) 
        \label{Appendix_Proof1_35}
\end{eqnarray}

\noindent 
is used. By these steps one arrives at  
\begin{eqnarray}
	F_M^l &=&  \frac{1}{\left(l-1\right)!}\;\sum\limits_{n=0}^{[l/2]} G_n^l \left(-1\right)^n \left(\frac{\ve{d}_{\sigma} \cdot \ve{e}_3}{d_{\sigma}}\right)^{l - 2n} 
	\sum\limits_{p=0}^{n} \left(-1\right)^p {n \choose p} \left(\ve{\sigma} \cdot \ve{e}_3\right)^{2n-2p}\;. 
        \label{Appendix_Proof1_40}
\end{eqnarray}

\noindent 
From the binomial theorem (\ref{binomial_theorem_1}) we get the relation  
\begin{eqnarray}
	\sum\limits_{p=0}^n \left(-1\right)^p {n \choose p} a^{2 n - 2p} &=& \left(-1\right)^n \left(1 - a^2\right)^n \;. 
        \label{binomial_theorem_2}
\end{eqnarray}

\noindent
Using (\ref{binomial_theorem_2}) one finally obtains (with $\left(-1\right)^n \left(-1\right)^n = 1$)
\begin{eqnarray}
	F_{M}^l &=& \frac{1}{\left(l-1\right)!}\;\sum\limits_{n=0}^{[l/2]} G_n^l \left(\frac{\ve{d}_{\sigma} \cdot \ve{e}_3}{d_{\sigma}}\right)^{l - 2n}  
	\bigg(1 - \left(\ve{\sigma} \cdot \ve{e}_3\right)^2\bigg)^n .   
        \label{Appendix_Proof1_45}
\end{eqnarray}

\noindent 
The expression in Eq.~(\ref{Appendix_Proof1_45}) coincides with the expression in Eq.~(\ref{upper_limit_M_L_15}).

\section{Proof of Eq.~(\ref{upper_limit_S_L_15})}\label{Proof2}

The pseudo-scalar function $F_{S}^l$ in (\ref{Tangent_nu_S}) reads  
\begin{eqnarray}
        F_{S}^l &=& \frac{1}{\left(l-1\right)!}\,\epsilon_{i_l bc}\,\sigma^c\, 
        \delta^{3}_{<{b}}\,\delta^{3}_{i_1} \,\dots \,\delta^{3}_{{i_{l-1}}>} 
        \;{\rm STF}_{i_1 \dots i_l}  
        \left[\sum\limits_{n=0}^{[l/2]} G_n^l\,P_{i_1 i_2}\, \dots \, P_{i_{2 n - 1} i_{2 n}}\,
        \frac{\xi_{i_{2 n +1 }}\,\dots\,\xi_{i_{l}}}{\left|\ve{\xi}\right|^{l - 2 n}}  
        \right], 
	\nonumber\\ 
        \label{Appendix_Proof2}
\end{eqnarray}

\noindent 
where the coefficients $G_n^l$ are given by Eq.~(\ref{Relation_D}). In this Appendix we will show that (\ref{upper_limit_S_L_15}) follows from (\ref{Appendix_Proof2}). 
One may use the same replacement as given by relation (\ref{Appendix_Proof1_1}), 
because $\delta^{3}_{<{b}}\,\delta^{3}_{i_1} \,\dots \,\delta^{3}_{{i_{l-1}}>}$ is an STF tensor up to the index $i_{l-1}$. That means 
\begin{eqnarray}
        \hspace{-0.5cm} P_{i_1 i_2}\, \dots \, P_{i_{2 n - 1} i_{2 n}} \rightarrow \left(-1\right)^n \sigma_{\i_1} \sigma_{i_2} \, \dots \, \sigma_{i_{2 n - 1}} \sigma_{i_{2 n}} \,.
        \label{Appendix_Proof2_1}
\end{eqnarray}

\noindent 
Furthermore, in each term in (\ref{Appendix_Proof2}) the auxiliary three-vector $\ve{\xi}$ can be replaced by the impact vector $\ve{d}_{\sigma}$ (cf. text below Eq.~(\ref{Relation_C})). 
Then, the pseudo-scalar function in (\ref{Appendix_Proof2}) takes the following form,
\begin{eqnarray}
        F_{S}^l &=& \frac{1}{\left(l-1\right)!}\,\epsilon_{i_l bc}\,\sigma^c\, 
        \delta^{3}_{<{b}}\,\delta^{3}_{i_1} \,\dots \,\delta^{3}_{{i_{l-1}}>} 
        \;{\rm STF}_{i_1 \dots i_l} 
	\left[\sum\limits_{n=0}^{[l/2]} G_n^l\,\left(-1\right)^n \sigma_{i_1}\,\dots\,\sigma_{i_{2 n}}\,
	\frac{d_{\sigma}^{i_{2 n +1 }}\,\dots\,d_{\sigma}^{i_{l}}}{\left(d_{\sigma}\right)^{l - 2 n}} 
        \right].  
	\nonumber\\ 
        \label{Appendix_Proof2_5}
\end{eqnarray}

\noindent 
For the STF tensor in (\ref{Appendix_Proof2_5}) we get from (\ref{delta})  
\begin{eqnarray}
        \delta^{\;3}_{<{b}}\,\delta^{3}_{i_1} \,\dots \,\delta^{3}_{{i_{l-1}}>} &=& \delta^3_b\,\delta^{3}_{i_1} \,\dots \,\delta^{3}_{i_{l-1}} 
        \nonumber\\
	&& \hspace{-2.5cm} + \sum\limits_{p=1}^{[l/2]} H^l_p\;\delta_{\{b i_1} \,\dots\, \delta_{i_{2p - 1} i_{2p}} \delta^{3}_{i_{2 p + 1}} \,\dots\, \delta^{3}_{i_{l-1}\}} \;,
        \label{STF_Formula_Proof2}
\end{eqnarray}

\noindent 
where $H^l_p$ are the scalar coefficients in (\ref{Coefficients_G_l_p}). The further evaluation of (\ref{Appendix_Proof2_5}) is considerably be simplified 
by taking account of the fact, that the terms in the second line in (\ref{STF_Formula_Proof2}) do not contribute in (\ref{Appendix_Proof2_5}). Therefore, we get  
\begin{eqnarray}
        F_{S}^l &=& \frac{1}{\left(l-1\right)!}\,\left(\ve{e}_3 \times \ve{\sigma} \right)^{i_l} \,\delta^{3}_{i_1} \,\dots \,\delta^{3}_{i_{l-1}}
        \;{\rm STF}_{i_1 \dots i_l}
        \left[\sum\limits_{n=0}^{[l/2]} G_n^l\,\left(-1\right)^n \sigma_{i_1}\,\dots\,\sigma_{i_{2 n}}\,
        \frac{d_{\sigma}^{i_{2 n +1 }}\,\dots\,d_{\sigma}^{i_{l}}}{\left(d_{\sigma}\right)^{l - 2 n}}
        \right], 
        \nonumber\\
        \label{Appendix_Proof2_10}
\end{eqnarray}

\noindent 
where $\delta^3_b = e_{3}^{b}$ and $\epsilon_{i_l bc}\,e_{3}^{b}\,\sigma^c = \left(\ve{e}_3 \times \ve{\sigma} \right)^{i_l}$ have been used. Now the summation over variable 
$n$ in (\ref{Appendix_Proof2_10}) is considered which, by means of relations (\ref{scalar_products_1}) - (\ref{scalar_products_4}), can be written in the following form,  
\begin{eqnarray}
	&& \delta^{3}_{i_1} \dots \delta^{3}_{i_{l-1}} {\rm STF}_{i_1 \dots i_l} \sum\limits_{n=0}^{[l/2]} G_n^l \left(-1\right)^n 
        \sigma_{i_1} \dots \sigma_{i_{2 n}} 
        \frac{d_{\sigma}^{i_{2 n +1 }}\!\dots d_{\sigma}^{i_{l}}}{\left(d_{\sigma}\right)^{l - 2 n}}
        \nonumber\\
	&& = \frac{d^{i_l}_{\sigma}}{d_{\sigma}} \sum\limits_{n=0}^{[l/2]} G_n^l\,\left(-1\right)^n \frac{l - 2 n}{l} 
        \left(\frac{\ve{d}_{\sigma} \cdot \ve{e}_3}{d_{\sigma}}\right)^{l - 2n - 1}
	\sum\limits_{p=0}^{n} \left(-1\right)^p {n \choose p} \left(\ve{\sigma} \cdot \ve{e}_3\right)^{2n-2p}\;.  
	\label{Appendix_Proof2_20}
\end{eqnarray}

\noindent
In relation (\ref{Appendix_Proof2_20}) all those terms, which are proportional to either $\sigma^{i_l}$ or $e_3^{i_l}$ have been omitted, because they vanish when contracted with 
$\left(\ve{e}_3 \times \ve{\sigma} \right)^{i_l}$. Then, by inserting (\ref{Appendix_Proof2_20}) into (\ref{Appendix_Proof2_10}) and 
using relation (\ref{binomial_theorem_2}), one arrives at (with $\left(-1\right)^n \left(-1\right)^n = 1$)
\begin{eqnarray}
        F_{S}^l &=& + \frac{1}{\left(l-1\right)!}\,\frac{\left(\ve{\sigma} \times \ve{d}_{\sigma}\right) \cdot \ve{e}_3}{d_{\sigma}} 
	\sum\limits_{n=0}^{[l/2]} G_n^l\,\frac{l - 2n}{l} \bigg(1 - \left(\ve{\sigma} \cdot \ve{e}_3\right)^2  \bigg)^{n}  
        \left(\frac{\ve{d}_{\sigma} \cdot \ve{e}_3}{d_{\sigma}}\right)^{l - 2 n -1} ,  
        \label{Appendix_Proof2_45}
\end{eqnarray}

\noindent 
where $\left(\ve{e}_3 \times \ve{\sigma}\right) \cdot \ve{d}_{\sigma} = \left(\ve{\sigma} \times \ve{d}_{\sigma}\right) \cdot \ve{e}_3$ 
has been used. The expression in Eq.~(\ref{Appendix_Proof2_45}) coincides with the expression in Eq.~(\ref{upper_limit_S_L_15}).

\section{Examples of total light deflection}\label{Examples} 

In this Appendix some examples for the total light deflection caused by mass-multipoles (\ref{Total_Light_Deflection_Mass_Chebyshev}) and 
caused by spin-multipoles (\ref{Total_Light_Deflection_Spin_Chebyshev_B}) are given.

\subsection{The mass-monopole}

For completeness, we consider the unit tangent vector of a light ray in the gravitational field of a mass-monopole.  
The monopole term of the unit tangent vector follows from (\ref{vector_mu_M_L}) for $l=0$ to be  
\begin{eqnarray}
	\ve{\nu}_{\rm 1PN}^{M_0} &=& - \frac{4 G M_0}{c^2 d_{\sigma}}\,\frac{\ve{d}_{\sigma}}{d_{\sigma}}\;, 
	\label{Example_tangent_vector_M}
\end{eqnarray}

\noindent
where from (\ref{M}) follows that the mass-monopole equals the Newtonian mass of the body: $M_0 = M$. Eq.~(\ref{Example_tangent_vector_M}) agrees with Eq.~(16) in \cite{Klioner1991} 
for the case of one massive body. 
For the angle of total light deflection caused by the mass-monopole one needs the Chebyshev polynomial of first kind for $l=0$, which follows from (\ref{Chebyshev_Polynomials_0}),
\begin{eqnarray}
        T_0 \left(x\right) &=& + 1\;.
        \label{Example_Monopole_5}
\end{eqnarray}

\noindent 
Inserting (\ref{Example_Monopole_5}) into (\ref{Total_Light_Deflection_Mass_Chebyshev}) yields 
\begin{eqnarray}
        \delta\left(\ve{\sigma}, \ve{\nu}_{\rm 1PN}^{M_0}\right) &=& - \,\frac{4 G M}{c^2}\,\frac{J_0}{d_{\sigma}} \;,
	\label{Example_Monopole_10}
\end{eqnarray}

\noindent
where $J_0 = -1$. Eq.~(\ref{Example_Monopole_10}) is the term in the total light deflection (\ref{Total_Light_Deflection_Mass_Chebyshev}) which is caused by the
mass-monopole structure of the body, which coincides with Eq.~(\ref{upper_limit_M_0_5}).

\subsection{The mass-quadrupole}

For the mass-quadrupole one needs the Chebyshev polynomial of second kind for $l=1$ and $l=2$, which follow from (\ref{Chebyshev_Polynomials_2}), 
\begin{eqnarray}
	U_1 \left(x\right) &=& + 2\,x\;,
        \label{Example_Quadrupole_5}
	\\
	U_2 \left(x\right) &=& - 1 + 4\,x^2\;,
        \label{Example_Quadrupole_6}
\end{eqnarray}

\noindent 
where variable $x$ is defined by Eq.~(\ref{Variable_x_Chebyshev_A}). 
By inserting (\ref{Example_Quadrupole_5}) and (\ref{Example_Quadrupole_6}) into (\ref{Tangent_nu_M_Chebyshev_Final}) we obtain the mass-quadrupole 
term of the unit tangent vector, 
\begin{eqnarray}
	\ve{\nu}_{\rm 1PN}^{M_2} &=& - \frac{4G M}{c^2 d_{\sigma}} J_2 \left(\frac{P}{d_{\sigma}}\right)^2
	\Bigg[ 2 \left(\frac{\ve{d}_{\sigma} \cdot \ve{e}_3}{d_{\sigma}}\right) \ve{\sigma} \times \left(\ve{e}_3 \times \ve{\sigma} \right)
	+ \left(1 - \left(\ve{\sigma} \cdot \ve{e}_3\right)^2 - 4 \left(\frac{\ve{d}_{\sigma} \cdot \ve{e}_3}{d_{\sigma}}\right)^2\right) 
	\frac{\ve{d}_{\sigma}}{d_{\sigma}} \Bigg], 
	\nonumber\\ 
        \label{Example_Quadrupole_7}
\end{eqnarray}

\noindent
which agrees with Eq.~(40) in \cite{Klioner1991} as well as with Eq.~(48) in \cite{Poncin_Lafitte_Teyssandier}. 
For evaluating the angle of total light deflection one may either use the formula (\ref{Total_Light_Deflection_Mass_Chebyshev}) or one may  
insert (\ref{Example_Quadrupole_7}) into (\ref{total_light_deflection_M_L}). One obtains for the quadrupole structure of the massive body, 
\begin{eqnarray}
	\delta\left(\ve{\sigma}, \ve{\nu}_{\rm 1PN}^{M_2}\right) &=&  
        + \frac{4\,G M}{c^2 d_{\sigma}} \,J_2 \left(\frac{P}{d_{\sigma}}\right)^2 
	\left[1 - \left(\ve{\sigma} \cdot \ve{e}_3\right)^2 - 2 \left(\frac{\ve{d}_{\sigma} \cdot \ve{e}_3}{d_{\sigma}}\right)^2\right]. 
        \label{Example_Quadrupole_10}
\end{eqnarray}

\noindent 
Eq.~(\ref{Example_Quadrupole_10}) is the term in the total light deflection (\ref{Total_Light_Deflection_Mass_Chebyshev}) which is caused by the 
mass-quadrupole structure of the body, which coincides with Eq.~(41) in \cite{Klioner1991} as well as with our investigation
in \cite{Zschocke6}; note that $\ve{e}_3 = \left(0,0,1\right)$ hence $\left(\ve{\sigma} \cdot \ve{e}_3\right)^2 = \left(\sigma^3\right)^2$
and $\left(\ve{d}_{\sigma} \cdot \ve{e}_3\right)^2 = \left(d_{\sigma}^3\right)^2$.

\subsection{The mass-octupole}

For the mass-octupole one needs the Chebyshev polynomial of second kind for $l=1$ and $l=2$, which follow from (\ref{Chebyshev_Polynomials_2}),
\begin{eqnarray}
        U_3 \left(x\right) &=& - 4\,x + 8\,x^3\;,
        \label{Example_Octupole_5}
        \\
        U_4 \left(x\right) &=& + 1 - 12\,x^2 + 16\,x^4\;,
        \label{Example_Octupole_6}
\end{eqnarray}

\noindent
where variable $x$ is defined by Eq.~(\ref{Variable_x_Chebyshev_A}).
By inserting (\ref{Example_Octupole_5}) and (\ref{Example_Octupole_6}) into (\ref{Tangent_nu_M_Chebyshev_Final}) we obtain the mass-octupole 
term of the unit tangent vector,
\begin{eqnarray}
        \ve{\nu}_{\rm 1PN}^{M_4} &=& \frac{16\,G M}{c^2 d_{\sigma}} \,J_4 \left(\frac{P}{d_{\sigma}}\right)^4 \left(\frac{\ve{d}_{\sigma} \cdot \ve{e}_3}{d_{\sigma}}\right) 
        \ve{\sigma} \times \left(\ve{e}_3 \times \ve{\sigma} \right)
        \Bigg[1 - \left(\ve{\sigma} \cdot \ve{e}_3\right)^2 - 2 \left(\frac{\ve{d}_{\sigma} \cdot \ve{e}_3}{d_{\sigma}}\right)^2 \Bigg]
        \nonumber\\ 
	&& \hspace{-1.5cm} + \frac{4\,G M}{c^2 d_{\sigma}}\,J_4 \left(\frac{P}{d_{\sigma}}\right)^4 \frac{\ve{d}_{\sigma}}{d_{\sigma}} 
	\Bigg[\left(1 - \left(\ve{\sigma} \cdot \ve{e}_3\right)^2\right)^2 
	- 12 \left(1 - \left(\ve{\sigma} \cdot \ve{e}_3\right)^2\right) 
	\left(\frac{\ve{d}_{\sigma} \cdot \ve{e}_3}{d_{\sigma}}\right)^2 
	+ 16 \left(\frac{\ve{d}_{\sigma} \cdot \ve{e}_3}{d_{\sigma}}\right)^4\Bigg] 
	\nonumber\\
        \label{Example_Octupole_7}
\end{eqnarray}

\noindent
which agrees with Eq.~(50) in \cite{Poncin_Lafitte_Teyssandier}.
For evaluating the angle of total light deflection one may either use the formula (\ref{Total_Light_Deflection_Mass_Chebyshev}) or one may
insert (\ref{Example_Octupole_7}) into (\ref{total_light_deflection_M_L}). One obtains for the octupole structure of the massive body, 
\begin{eqnarray}
        \delta\left(\ve{\sigma}, \ve{\nu}_{\rm 1PN}^{M_4}\right) &=&  
        - \frac{4\,G M}{c^2 d_{\sigma}}\,J_4 \left(\frac{P}{d_{\sigma}}\right)^4
	\nonumber\\ 
	&& \hspace{-0.5cm} \times \Bigg[\left(1 - \left(\ve{\sigma} \cdot \ve{e}_3\right)^2\right)^2 
	- 8 \left(\frac{\ve{d}_{\sigma} \cdot \ve{e}_3}{d_{\sigma}}\right)^2 \left(1 - \left(\ve{\sigma} \cdot \ve{e}_3\right)^2\right)
	+ 8 \left(\frac{\ve{d}_{\sigma} \cdot \ve{e}_3}{d_{\sigma}}\right)^4\Bigg]. 
        \label{Example_Octupole_10}
\end{eqnarray}

\noindent 
Eq.~(\ref{Example_Octupole_10}) is the term in the total light deflection (\ref{Total_Light_Deflection_Mass_Chebyshev}) which is caused by the 
mass-octupole structure of the body.

\subsection{The spin-dipole}

For completeness, we consider the unit tangent vector of a light ray in the gravitational field of a mass-monopole in uniform rotation.
The spin-dipole term of the unit tangent vector follows from (\ref{Tangent_nu_S_Dipole}) for $l=1$ to be
\begin{eqnarray}
        \ve{\nu}_{\rm 1PN}^{S_1} &=& + \frac{4 G M}{c^3}\,\Omega\,\kappa^2\,J_0 \left(\frac{P}{d_{\sigma}}\right)^2
        \bigg[2\,\frac{\left(\ve{\sigma}\times\ve{d}_{\sigma}\right) \cdot \ve{e}_3}{d_{\sigma}}\,\frac{\ve{d}_{\sigma}}{d_{\sigma}} +
        \left(\ve{\sigma} \times \ve{e}_3\right)\bigg], 
        \label{Example_Spin_Dipole_nu}
\end{eqnarray}

\noindent
which is in agreement with Eq.~(60) in \cite{Klioner1991}. 
By inserting (\ref{Example_Spin_Dipole_nu}) or (\ref{Tangent_nu_S_Dipole}) into (\ref{total_light_deflection_S_L}) one obtains the angle of total light deflection 
(cf. Eq.~(\ref{upper_limit_S_1_5})), 
\begin{eqnarray}
        \delta\left(\ve{\sigma}, \ve{\nu}_{\rm 1.5PN}^{S_1}\right) &=& - \,\frac{4 G M}{c^3}\,\Omega\,\kappa^2\,J_0\,\left(\frac{P}{d_{\sigma}}\right)^2\,
	\frac{\left(\ve{\sigma} \times \ve{d}_{\sigma}\right) \cdot \ve{e}_3}{d_{\sigma}}\;,
        \label{Example_Spin_Dipole}
\end{eqnarray}

\noindent
where $J_0 = - 1$ and which is in agreement with Eq.~(61) in \cite{Klioner1991}.

\subsection{The spin-hexapole}

For the unit tangent vector of light trajectory caused by the spin-hexapole of the massive body ($l=3$ in (\ref{Tangent_nu_S_Chebyshev_Final})) 
one needs the following Chebyshev polynomials which are deduced from (\ref{Chebyshev_Polynomials_1}) and (\ref{Chebyshev_Polynomials_2}),
\begin{eqnarray}
	T_3 \left(x\right) &=& - 3\,x + 4\,x^3\;,
        \label{Example_Hexapole_1_A}
	\\
	T_4 \left(x\right) &=& + 1 - 8\,x^2 + 8\,x^4\;,
        \label{Example_Hexapole_2_A}
        \\
        U_2 \left(x\right) &=& - 1 + 4\,x^2\;,
        \label{Example_Hexapole_3_A}
        \\
	U_3 \left(x\right) &=& - 4\,x + 8\,x^3\;,
        \label{Example_Hexapole_4_A}
\end{eqnarray}

\noindent 
where variable $x$ is defined by Eq.~(\ref{Variable_x_Chebyshev_A}). Inserting (\ref{Example_Hexapole_1_A}) - (\ref{Example_Hexapole_4_A}) into (\ref{Tangent_nu_S_Chebyshev_Final}) 
yields for the spin-hexapole term of the unit tangent vector, 
\begin{eqnarray}
	\ve{\nu}_{\rm 1PN}^{S_3} &=& + \frac{8}{7}\,\frac{G M}{c^3}\,\Omega\,J_2  
	\left(\frac{P}{d_{\sigma}}\right)^4 
	\bigg[ \frac{\left(\ve{\sigma} \times \ve{d}_{\sigma} \right) \cdot \ve{e}_3}{d_{\sigma}} 
	\left(24 \left(\frac{\ve{d}_{\sigma} \cdot \ve{e}_3}{d_{\sigma}}\right)^2 - 4 + 4 \left(\ve{\sigma} \cdot \ve{e}_3\right)^2\right) \frac{\ve{d}_{\sigma}}{d_{\sigma}}
	\nonumber\\
	&& \hspace{-1.5cm} - 8 \frac{\left(\ve{\sigma} \times \ve{d}_{\sigma} \right) \cdot \ve{e}_3}{d_{\sigma}} \left(\frac{\ve{d}_{\sigma} \cdot \ve{e}_3}{d_{\sigma}}\right) 
	\ve{\sigma} \times \left(\ve{e}_3 \times \ve{\sigma}\right) 
	+ \left(4 \left(\frac{\ve{d}_{\sigma} \cdot \ve{e}_3}{d_{\sigma}}\right)^2 - 1 + \left(\ve{\sigma} \cdot \ve{e}_3\right)^2 \right) \left(\ve{\sigma} \times \ve{e}_3\right) 
		\bigg].
	\nonumber\\ 
        \label{Example_Spin_Hexapole_nu}
\end{eqnarray}

\noindent 
By inserting (\ref{Example_Spin_Hexapole_nu}) into (\ref{total_light_deflection_S_L}) one obtains the angle of total light deflection, 
\begin{eqnarray}
        && \hspace{-0.5cm} \delta\left(\ve{\sigma}, \ve{\nu}_{\rm 1.5PN}^{S_3}\right) =  
        + \frac{24}{7}\,\frac{G M}{c^3}\,\Omega\,J_{2} \left(\frac{P}{d_{\sigma}}\right)^{4}\,
	\frac{\left(\ve{\sigma} \times \ve{d}_{\sigma}\right) \cdot \ve{e}_3}{d_{\sigma}}  
        \left[1 - \left(\ve{\sigma} \cdot \ve{e}_3\right)^2 - 4 \left(\frac{\ve{d}_{\sigma} \cdot \ve{e}_3}{d_{\sigma}}\right)^2\right]. 
	\nonumber\\ 
        \label{Example_Hexapole_10}
\end{eqnarray}

\noindent
Eq.~(\ref{Example_Hexapole_10}) is the term in the total light deflection (\ref{Total_Light_Deflection_Spin_Chebyshev_B}) which is caused by the 
spin-hexapole structure of the body.


\end{document}